\documentclass[a4paper,11pt]{article}
\usepackage{float}
\usepackage{amsmath}
\usepackage{amssymb}
\usepackage{graphicx}
\usepackage{color,xcolor}
\usepackage{mathrsfs}
\usepackage{tcolorbox}
\usepackage{tikz}
\usepackage{tikz-cd}
\usepackage{physics}
\usepackage{diagbox}
\usepackage{listings}
\usepackage[textsize=scriptsize,textwidth=2.5cm]{todonotes}
\usepackage{multicol}
\usepackage{multirow,makecell}
\usepackage{subfig}
\usepackage{cancel}
\usepackage{jheppub}
\usepackage{upgreek}

\allowdisplaybreaks[2]
\numberwithin{equation}{section}
\definecolor{darkgreen}{RGB}{40,150,60}
\newcommand{\CP}[1]{\textcolor{blue}{ #1}}

\begin{document}
\begin{flushright}
USTC-ICTS/PCFT-24-27
\end{flushright}

\title{Correlators of long strings on AdS$_3\times$S$^3\times$T$^4$ }

\author[a]{Zhe-fei Yu}
\author[a,b]{and Cheng Peng}

\affiliation[a]{Kavli Institute for Theoretical Sciences (KITS), University of Chinese Academy of Sciences, Beijing 100190, China}

\affiliation[b]{Peng Huanwu Center for Fundamental Theory, Hefei, Anhui 230026, China}

\emailAdd{yuzhefei@ucas.ac.cn, pengcheng@ucas.ac.cn}

\abstract{In this work, we calculate correlators of long strings on AdS$_3\times$S$^3\times$T$^4$ with pure NS-NS flux. We first construct  physical vertex operators that correspond to long strings. Due to the GSO projection, 
they depend on  the parity of the spectral flow parameter $w$. For a given $w$, we construct the physical operators that have the lowest space-time weights in both the NS and R sector. Then, we calculate three point correlators for each possible type of parities of  spectral flows. We  find that the recursion relations of correlators in the bosonic SL$(2,\mathbb{R})$ WZW model can be understood from the equivalence of these superstring correlators with different picture choices. Furthermore, after carefully mapping  the vertex operators to appropriate operators in the dual CFT, we find that once the fermionic contributions together with the picture changing effects are correctly taken into account, some mathematical identities of covering maps lead to the matching of the correlators of the two sides. We check this explicitly at the leading order in the conformal perturbation computation and conjecture that this remains correct to all orders.
}

\maketitle

\section{Introduction}

Superstring  on AdS$_3\times$S$^3\times$T$^4$ with pure NS-NS flux is one of the string theories with good computational control.
It can be studied using the RNS formalism (see, e.g. \cite{Giveon:1998ns,Kutasov:1999xu,deBoer:1998gyt,Maldacena:2000hw,Maldacena:2000kv,Maldacena:2001km}), 
and  the special case of minimal tension (referred to as the tensionless limit)  being well-described by the 
hybrid  formalism \cite{Berkovits:1999im,Eberhardt:2018ouy,Dei:2020zui,Gaberdiel:2022oeu}. 
In the context of holographic duality~\cite{Maldacena:1997re},   the CFTs dual to strings with pure NS-NS flux typically lie on the moduli space of the symmetric orbifold of T$^4$~\cite{Seiberg:1999xz}. Remarkably, for the special case of minimal tension, it was shown that the dual theory is exactly the symmetric orbifold CFT itself \cite{Eberhardt:2018ouy,Eberhardt:2019ywk}. Evidence for this duality include  the matching of both the spectrum \cite{Gaberdiel:2018rqv,Eberhardt:2018ouy} and the (structure of) correlators \cite{Eberhardt:2019ywk,Dei:2020zui,Eberhardt:2020akk,Knighton:2020kuh,Gaberdiel:2022oeu,Dei:2023ivl}, see also \cite{Eberhardt:2020bgq,Eberhardt:2021jvj,Aharony:2024fid}. However, for the string theory with non-minimal tension,  
the matching was for a long time only achieved for BPS protected quantities \cite{Strominger:1996sh,deBoer:1998us,Maldacena:1999bp,Gaberdiel:2007vu,Dabholkar:2007ey,Pakman:2007hn,Giribet:2007wp,Cardona:2009hk}. Recently, based on the exact duality between the tensionless string and the symmetric orbifold of T$^4$, a perturbative  dual of the AdS$_3$ (super)string theory with pure NS-NS flux was proposed in \cite{Eberhardt:2021vsx,Dei:2022pkr} (see also \cite{Balthazar:2021xeh,Martinec:2021vpk,Martinec:2022ofs})\footnote{One can also deform the theory  away
from the tensionless point by switching on R-R flux, see \cite{Gaberdiel:2023lco,Fiset:2022erp,Frolov:2023pjw} for some recent progress.}. 
It is  a symmetric orbifold CFT deformed by 
an exactly marginal operator. Thereafter, progress of  matching  correlators beyond the  tensionless limit was made in \cite{Eberhardt:2021vsx,Dei:2022pkr,Knighton:2023mhq,Hikida:2023jyc,Knighton:2024qxd} (in a perturbative sense). Nevertheless, these works focus on the bosonic duality\footnote{See \cite{Eberhardt:2021vsx,Sriprachyakul:2024gyl} for some preliminary discussions for the supersymmetric case.} where higher genus string correlators are not well-defined.  
One of the motivations of this work is to  generalize  the duality to the supersymmetric case.

Spectrum of string theory on AdS$_3\times$S$^3\times$T$^4$ with pure NS-NS flux includes 
short strings and long strings~\cite{Maldacena:2000hw}.
In the RNS formalism, the worldsheet CFT includes a non-compect WZW model with the target space being (the universal covering of) $SL(2,R)$. The short strings and long strings lie in the discrete and continuous representations of the $SL(2,R)$ affine symmetry respectively. To properly characterize the spectrum, one should include the  spectrally flowed 
representations  into the theory \cite{Maldacena:2000hw, Maldacena:2000kv, Maldacena:2001km}. 
Spectrally
flowed vertex operators correspond to winding strings in spacetime and 
capture interesting physics. However, 
much is yet to be understood about the correlators of spectrally flowed operators \cite{Giribet:2000fy,Giribet:2001ft,Giribet:2005ix,Ribault:2005ms,Giribet:2005mc,Minces:2005nb,Iguri:2007af,Baron:2008qf,Iguri:2009cf,Giribet:2011xf,Cagnacci:2013ufa,Giribet:2015oiy,Giribet:2019new,Hikida:2020kil}. Recently, a closed formula for the 3-point and 4-point functions of the bosonic $SL(2,R)$ WZW model on the worldsheet was proposed in \cite{Dei:2021xgh,Dei:2021yom}\footnote{see  \cite{Bufalini:2022toj} for a proof for the 3-point formula and \cite{Iguri:2024yhb} for  highly non-trivial checks for the 4-point formula.}, which is obtained by the 
``local Ward identities" (this method is valid for $SL(2,R)$  WZW models with general levels $k$, in particular, it plays a crucial role in the discussion of the    correlators in the tensionless string \cite{Eberhardt:2019ywk}). For short strings, these results helped to complete the matching of the chiral ring of the two sides \cite{Iguri:2023khc}. For long strings, these results lead to a proposal for a perturbative   CFT dual of the  bosonic string on AdS$_3\times X$ \cite{Eberhardt:2021vsx, Dei:2022pkr}.
In this work, we study long strings in the superstring theory on AdS$_3\times$S$^3\times$T$^4$. We  construct physical vertex operators representing long strings  and calculate their 3-point correlators. We will also match these vertex operators with the ones in the dual CFT side, and compare the 3-point correlators of the two sides.

The rest of the paper is organized as follows. In  section \ref{stringside}, we construct physical vertex operators for long strings on AdS$_3\times$S$^3\times$T$^4$. Because of the GSO projection, their form depends on the parity of the spectral flow. We construct all  physical vertex operators with the lowest space-time weights\footnote{Since we study long strings, we will always construct a continuum of vertex operators with a continuum of lowest space-time weights.} for both odd and even parities
and for both the NS and R sectors. In  section~\ref{superstring correlator}, we calculate the three-point correlators of these physical operators. Since their form again depends on the parities of  the spectral flows of the 3 operators, we calculate one representative for every choice of the parities. The main body of this section is devoted to 
the calculation of various fermionic correlators (coming from the worldsheet fermions and picture changing), which can be done cleanly using the formula obtained in \cite{Dei:2021xgh}.  As a byproduct, we find the recursion relations of correlators in the bosonic $SL(2,R)$ WZW model can be understood from the equivalence of these superstring correlators with different picture choices. In section \ref{theCFTside}, we identify the corresponding operators in the CFT side (following \cite{Eberhardt:2019qcl}), which is proposed to be a deformed symmetric orbifold CFT \cite{Eberhardt:2021vsx}. We also compare the three-point correlators of the two sides at the leading order (where the deformation is  turned off). It turns out that the matching at this order is already non-trivial: due to some interesting mathematical identities for covering maps, 
the correlators of the two sides match precisely. In particular, the fermionic contributions together with the picture changing effects
 guarantee that the dual symmetric orbifold CFT has the  central charge $6k$.  In section \ref{discussion}, we conclude our work and discuss some future directions. 
Some conventions and  backgrounds are described in the appendices.

\section{Physical operators of the superstring}\label{stringside}
In this section, we describe physical vertex operators  of the string theory on AdS$_3\times$S$^3\times$T$^4$. For short strings, the physical chiral operators, including both the spectrally flowed and unflowed sectors,  
are constructed in \cite{Kutasov:1998zh,Dabholkar:2007ey,Gaberdiel:2007vu,Giribet:2007wp}. 
For long strings, physical spectrum  
are discussed in \cite{Ferreira:2017pgt,Eberhardt:2019qcl} (see also \cite{Gaberdiel:2018rqv,Iguri:2022pbp,Maldacena:2000hw}). For our purpose to calculate  string correlators, we need the explicit 
expressions of the physical operators\footnote{Notice that some physical operators of long strings were constructed in literature, see e.g. \cite{Iguri:2022pbp,Eberhardt:2019qcl}. Here we will give a complete construction of  physical operators with the lowest space-time weights (for a given $w$). In particular, we find a special one (\eqref{NSeven2}) in the NS sector for $w$ even and  give a detailed construction for the ones in the R sector. }. 
Therefore in this section we firstly give explicit expressions of all physical vertex operators with the lowest space-time weights for any given spectral flow parameter $w$ that corresponds to long strings.
The construction  depends on the parity of $w$ and the sector we consider (NS or R). We will mostly focus on the left-moving part and omit a similar analysis for the right-movers (and always suppress the anti-holomorphic dependence).

\subsection{Superstring on AdS$_3\times$S$^3\times$T$^4$}
Firstly,  we  review some basic facts about  superstring theory on AdS$_3\times$S$^3\times$T$^4$, see e.g. \cite{Dabholkar:2007ey,Ferreira:2017pgt}. We  discuss this string theory in the RNS formalism, where the worldsheet CFT is described by
\begin{equation}
    sl(2,R)^{(1)}_k\oplus  su(2)^{(1)}_k \oplus U(1)^{4(1)}
\end{equation}
In the above, $sl(2,R)^{(1)}_k$ and $su(2)^{(1)}_k$ represent $\mathcal{N}=1$ supersymmetric WZW model with affine symmetry $sl(2,R)^{(1)}_k$  and $su(2)^{(1)}_k$ respectively. They describe the AdS$_3$ and S$^3$ factor.  $U(1)^{4(1)}$ describes the $\mathcal{N}=1$ supersymmetric version of T$^4$  (the flat
torus directions). 

The $sl(2,R)^{(1)}_k$ WZW model has symmetries generated by  $sl(2,R)$ currents $J^A$ and fermions $\psi^A$ $(A=1,2,3)$, with OPEs:
        \begin{equation}
        \begin{aligned}
            J^A(z)J^B(w)&\sim \frac{\frac{k}{2}\eta^{AB}}{(z-w)^2}+\frac{i\epsilon^{AB}_CJ^C(w)}{z-w}\\
            J^A(z)\psi^B(w)&\sim \frac{i\epsilon^{AB}_C\psi^C(w)}{z-w}\\
            \psi^A(z)\psi^B(w)&\sim \frac{\frac{k}{2}\eta^{AB}}{z-w}\,,
        \end{aligned}
      \end{equation}
      where $\epsilon^{123}=1$ and the indices are raised and lowered with $\eta^{AB}=\eta_{AB}=\text{diag}(++-)$.
     Similarly, the $su(2)^{(1)}_k$ WZW model has $su(2)$ currents $K^a$ and fermions $\chi^a$ $(a=1,2,3)$, with OPEs:
       \begin{equation}
        \begin{aligned}
            K^a(z)K^b(w)&\sim \frac{\frac{k}{2}\delta^{ab}}{(z-w)^2}+\frac{i\epsilon^{ab}_cJ^C(w)}{z-w}\\
            K^a(z)\chi^b(w)&\sim \frac{i\epsilon^{ab}_c\psi^c(w)}{z-w}\\
            \chi^a(z)\chi^b(w)&\sim \frac{\frac{k}{2}\delta^{ab}}{z-w}\ .
        \end{aligned}
      \end{equation}
     The  indices are raised and lowered with $\delta^{ab}=\delta_{ab}=\text{diag}(+++)$. As usual, 
     we define
        \begin{equation}
            J^{\pm}=J^1\pm iJ^2, \quad K^{\pm}=K^1\pm iK^2, \quad \psi^{\pm}=\psi^1\pm i\psi^2, \quad \chi^{\pm}=\chi^1\pm i\chi^2\ .
        \end{equation}
         It is convenient to split the 
         supersymmetric currents into the bosonic and fermionic parts
         \begin{equation}
             J^A=j^A+\hat{j}^A, \qquad K^a=k^a+\hat{k}^a\,,
         \end{equation}
         where $\hat{j}^A$ and $\hat{k}^a$ are the fermionic currents, defined as:
         \begin{equation}
             \hat{j}^A=-\frac{i}{k}\epsilon^A_{BC}\psi^B\psi^C, \quad \hat{k}^a=-\frac{i}{k}\epsilon^a_{bc}\chi^b\chi^c\ .
         \end{equation}
         The currents $j^A, \hat{j}^A$ and $ k^a, \hat{k}^a$ generate 2 bosonic $SL(2, R)$ affine algebras at levels $k + 2, -2$  and 2 bosonic $SU(2)$ affine algebras at levels $k -2, +2$, respectively.  Since $j^A$ and $k^a$ commute with the free fermions,
          the spectrum and interactions of the original level $k$ supersymmetric WZW
            models are then factorized into 2 (decoupled) bosonic WZW models and  free fermions.
            
      In terms of the decoupled WZW currents and free fermions,  one can easily write down the stress tensor and supercurrent of the worldsheet theory
        \begin{equation}\label{TandG}
            \begin{aligned}
                T&=\frac{1}{k}j^Aj_A-\frac{1}{k}\psi^A\partial\psi_A+\frac{1}{k}k^ak_a-\frac{1}{k}\chi^a\partial\chi_a+T(T^4)\\
                G& =\frac{2}{k}(\psi^Aj_A+\frac{2i}{k}\psi^1\psi^2\psi^3)+\frac{2}{k}(\chi^ak_a+\frac{2i}{k}\chi^1\chi^2\chi^3)+G(T^4)   \,,    
            \end{aligned}
        \end{equation}
        where  and in the rest of the paper, normal-ordering is always understood.
        Superstring on AdS$_3\times$S$^3\times$T$^4$ also contains the standard $bc$ and $\beta\gamma$ ghosts. The standard BRST operator of the superstring is then given by
        \begin{equation}
            Q_{\text{BRST}}=\oint dz\left(c(T+\frac{1}{2}T_{\text{gh}})+\gamma(G+\frac{1}{2}G_{\text{gh}})\right)\ .
        \end{equation}
        Physical vertex operators should be BRST invariant and will be discussed in the next section.    Importantly, to obtain all the physical operators, we need the following automorphism $\sigma$ of the current algebra, 
        namely the spectral flow \cite{Maldacena:2000hw,Giribet:2007wp}
        \begin{equation}\label{spectralflow}
        \begin{aligned}
             &\sigma^w(J_m^\pm)=J^{\pm}_{m\mp w}, \quad \sigma^w(J_m^3)=J_m^3+\frac{kw}{2}\delta_{m,0}\,,\\ 
            &\sigma^w(\psi_m^\pm)=\psi^{\pm}_{m\mp w}, \quad
            \sigma^w(\psi_m^3)=\psi_m^3\,,\\
             &\sigma^w(K_m^\pm)=K^{\pm}_{m\mp w}, \quad \sigma^w(K_m^3)=K_m^3-\frac{kw}{2}\delta_{m,0}\,,\\ 
            &\sigma^w(\chi_m^\pm)=\chi^{\pm}_{m\mp w}, \quad
            \sigma^w(\chi_m^3)=\chi_m^3\ .
        \end{aligned}
        \end{equation}
         Notice that spectral flow acts on  the decoupled currents $j^A, k^a$ and the fermionic currents $\hat{j}^A,\hat{k}^a$  the same way as on the full currents $J^A, K^a$. 
         The spectral flows in the (supersymmetric) AdS$_3$ and S$^3$ directions are independent, in the following we only consider  spectral flow in the AdS$_3$ direction. With this restriction, 
        the spectral flow of the $T$ and $G$ are\footnote{Notice that when discussing chiral operators (short strings), it is  convenient to also spectral flow the (supersymmetric) S$^3$ part \cite{Giribet:2007wp}, though it does not give new representations but only reshuffles states. Here we focus on long string so  spectral flow of the (supersymmetric) AdS$_3$ part is enough for us.}
      \begin{equation}\label{actonTandG}
         \sigma^w(L_n)=L_n-w J^3_n-\frac{k}{4}w^2\delta_{n,0}, \qquad \sigma^w (G_m)=G_m-w \psi^3_m\ .
       \end{equation}

\subsection{Vertex operators in the bosonic model}
Since the worldsheet supersymmetric WZW model is factorized into  bosonic WZW models and  free fermions, let's first describe vertex operators in the bosonic $sl(2,R)$ WZW model (we mainly follow the convention in \cite{Dei:2021xgh} for the 
bosonic model). 
There are spectrally flowed operators and unflowed operators. The unflowed operators are simpler. In the 
$x$-basis \cite{Maldacena:2000hw} they are labeled by the spin $j$ and the defining OPEs are\footnote{Notice that we will always use lowercase letter $j^a$ to denote the decoupled currents and capital  letter $J^a$ to denote the full currents. }:
\begin{equation}
            j^A(z)V_j(x,\Bar{x};w,\Bar{w})\sim \frac{D_x^A V_j(x,\Bar{x};w,\Bar{w})}{z-w}\,,
        \end{equation}
        where 
        \begin{equation}
            D_x^+=\partial_x, \quad D_x^3=x\partial_x+j,\quad D_x^-=x^2\partial_x+2jx\ .
        \end{equation}
        The conformal dimension is (recall that  the level for the decoupled $SL(2,R)$ WZW model is shifted to be $k+2$):
        \begin{equation}
            \Delta_h=\Bar{\Delta}=-\frac{j(j-1)}{k}\,,
        \end{equation}
        which can be formally expanded in modes as
        \begin{equation}
            V_j(x,\Bar{x})=\sum_{m,\Bar{m}}V_{j,m,\Bar{m}}x^{-j-m}\Bar{x}^{-\Bar{j}-\Bar{m}}\ .
        \end{equation}
        Thus, the action of the zero modes on $V_{j,m,\Bar{m}}$ is
        \begin{equation}
            j_0^3V_{j,m,\Bar{m}}=mV_{j,m,\Bar{m}}, \qquad j^{\pm}_0V_{j,m,\Bar{m}}=(m\mp (j-1))V_{j,m\pm 1,\Bar{m}}\ .
        \end{equation}
These operators $V_{j,m,\Bar{m}}$ are in the ``$m$-basis". We are ultimately  interested in  
vertex operators in the $x$-basis, since they are local in $x$ and $\Bar{x}$, which are identified with the holomorphic and anti-holomorphic coordinates of the boundary CFT. Nevertheless, the $m$-basis is also useful because the action of spectral flow is more clear in this basis, as 
we show in the following. 

Now we describe the spectrally flowed operators. In the $m$-basis, we write vertex operators with spin $j$ and spectral flow $w$ as $V_{j,m}^w(z)$. Denote their corresponding states as $[|j,m\rangle]^w$, then they form a spectrally flowed representation of the algebra $sl(2,R)_{k+2}$
\begin{equation}
\begin{aligned}
    j^+_w [|j,m\rangle]^w&=(m+1-j)[|j,m+1\rangle]^w, \quad j^+_n[|j,m\rangle]^w=0, \quad n>w\\
    j^3_0 [|j,m\rangle]^w&=(m+\frac{(k+2)m}{2})[|j,m\rangle]^w, \quad j^3_n[|j,m\rangle]^w=0, \quad n>0\\
    j^-_{-w} [|j,m\rangle]^w&=(m+1-j)[|j,m-1\rangle]^w, \quad j^-_n[|j,m\rangle]^w=0, \quad n>-w\ .
\end{aligned}
\end{equation}
Then one can accordingly  write down the OPEs of the currents and $m$-basis operators $V_{j,m}^w(z)$. Notice that a $m$-basis operator $V_{j,m}^w(z)$ is a Virasoro primary but not an affine primary~\cite{Maldacena:2001km,Giribet:2007wp}. When $w>0 $ $(w<0)$ it is  the lowest (highest) weight state of the the global $SL(2,R)$ algebra generated by $j^a_0 (a=3,\pm)$. 
We also need  the $x$-basis operators in the flowed sector. They can be defined from the  spectrally flowed operator in the $m$-basis \cite{Dei:2021xgh}
\begin{equation}\label{xbaisdef}
    V_{j,h}^{w}(x;z)\equiv e^{zL_{-1}}e^{xJ_0^+}V_{j,h}^{w}(0;0)e^{-xJ_0^+}e^{-zL_{-1}}\,,
\end{equation}
where $V_{j,h}^{w}(0;0)\equiv V_{j,m}^w(0)$ with $h=m+\frac{(k+2)w}{2}$. 
A few comment on this definition are in order:
\begin{itemize}
  \item Notice that $L_{-1}$ and $J^+_0$ commute, thus the order of the exponentials in the definition \eqref{xbaisdef} does not matter. 
    \item The definition \eqref{xbaisdef} means  $J_0^+$ is the generator of translation in the $x$-space. Since both the two $m$-basis operators $V_{j,m}^w(z)$ and $V_{j,-m}^{-w}(z)$ contribute to the same $x$-basis operator $V_{j,h}^{w}(x;z)$, one can always label a $x$-basis operator by a positive spectral flow parameter $w$ \cite{Maldacena:2001km,Giribet:2007wp,Iguri:2022eat}. 
    \item Notice that in the above definition we used the modes $L_{-1}$ and $J_0^+$ in the exponentials, which are modes in the full supersymmetric WZW models. Since   the 
bosonic and fermionic contributions decouple, we can write the modes $L_{-1}$ and $J_0^+$ as sums of the corresponding modes in the bosonic and fermionic WZW models: $L_{-1}=l_{-1}+\hat{l}_{-1}$, $J_0^+=j_0^++\hat{j}_0^+$. Then  $l_{-1}, \hat{l}_{-1}, j_0^+, \hat{j}_0^+$  commute with one another and $V_{j,h}^{w}(0;0)$  commute with the fermionic modes $\hat{l}_{-1}, \hat{j}_0^+$. Thus we will
obtain the same result if we replace $L_{-1}$, $J_0^+$ by $l_{-1}$, $j_0^+$ in \eqref{xbaisdef}..
\end{itemize}
 Then we can  write down the OPEs of operators in the $x$-basis \cite{Dei:2021xgh}
\begin{equation}
    \begin{aligned}
        j^+(\xi) V_{j,h}^{w}(x,z)&=\sum_{p=1}^{w+1}\frac{(j_{p-1}^+V_{j,h}^{w})(x,z)}{(\xi-z)^p}+\mathcal{O}(1)\\
        (j^3(\xi)-xj^+(\xi))V_{j,h}^{w}(x,z)&=\frac{hV_{j,h}^{w}(x,z)}{\xi-z}+\mathcal{O}(1)\\
        (j^-(\xi)-2xj^3(\xi)+x^2j^+(\xi)) V_{j,h}^{w}(x,z)&=(\xi-z)^{w-1}(j_{-w}^-V_{j,h}^{w})(x,z)+\mathcal{O}((\xi-z)^w)\ .
    \end{aligned}
\end{equation}
Notice that in the above, we recombined the currents to simplify the expressions. In fact, the recombined currents are just the  currents written in the $x$-basis: 
\begin{equation}
   \begin{aligned}
    j^+(z)&=e^{zL_{-1}}e^{xJ_0^+}j^{+}(0;0)e^{-xJ_0^+}e^{-zL_{-1}}\\
    j^3(z)-xj^+(z)&=e^{zL_{-1}}e^{xJ_0^+}j^{3}(0;0)e^{-xJ_0^+}e^{-zL_{-1}}\\
    j^-(z)-2xj^3(z)+x^2j^+(\xi)&=e^{zL_{-1}}e^{xJ_0^+}j^{-}(0;0)e^{-xJ_0^+}e^{-zL_{-1}}
    \end{aligned}
\end{equation}
where $j^a(0;0)\equiv j^a(0)$, as in the definition \eqref{xbaisdef}.

\subsection{Physical operators of the superstring}
Now we construct physical vertex operators in the superstring. For this, we need to include contributions from the $su(2)_{k-2}$ part, the free fermions, the internal torus  as well as the ghosts. We focus on physical vertex operators of long strings that have the lowest space-time weights in both the NS and R sector (with $w$ given).   Because of the GSO projection, there will be a difference between operators with  odd and even spectral flow parameters, since  the fermion number depends on the parity of the spectral flow parameter \cite{Giribet:2007wp,Ferreira:2017pgt}. This dependence comes from the fact that when one  performs spectral flow in the decoupled bosonic WZW model ($j^a$), one should at the same time  do spectral flow in the fermionic part ($\psi^a$) to respect the $\mathcal{N}=1$ supersymmetry on the worldsheet\footnote{It is completely fine if one does not spectral flow  the operators in the fermionic part. The point here is that one can always describe a vertex operator in this ``supersymmetric spectrally flowed'' frame since spectral flow only reshuffles states in the fermionic part. Besides, the discussion of the string spectrum will be clearer in this ``supersymmetric spectrally flowed'' frame. }. For the fermionic part,  one can write any spectrally flowed states  explicitly and find that doing  spectral flow once change the parity of the fermionic number \cite{Giribet:2007wp}. 
 Notice that for  long strings,  the spectral flow parameter $w$ will be identified with the cycle length of single cycle  twisted sectors  on the CFT side. Thus a similar difference between operators with odd and even single cycle appears on the CFT side \cite{Eberhardt:2019qcl,Lunin:2001pw}, and we will analysis this in detail in section \ref{theCFTside}.

 In practice, we follow the steps of the construction of physical chiral operators (short strings) in \cite{Dabholkar:2007ey,Giribet:2007wp} to construct physical operators representing long strings in this section. Since in the following we  frequently bosonize the worldsheet fermions to  describe spectrally flowed operators in the fermionic part, we recall its form here \cite{Giveon:1998ns}
        \begin{equation}
            \partial \hat{H}_1=\frac{2}{k}\psi^2\psi^1, \quad \partial \hat{H}_2=\frac{2}{k}\chi^2\chi^1, \quad \partial \hat{H}_3=\frac{2}{k}i\psi^3\chi^3,  \quad \partial \hat{H}_4=\eta^2\eta^1,\quad \partial \hat{H}_5=\eta^4\eta^3\,,
        \end{equation}
        where $\hat{H}$ are canonically normalized bosons, including proper cocycles
        \begin{equation}
             \hat{H}_i=H_i+\pi \sum_{j<i}N_j, \qquad  N_i=i\oint \partial H_i, \qquad  H_i(z)H_j(w)\sim -\delta_{ij}\text{log}(z-w)\ .
        \end{equation}
        Then 
        \begin{equation}
        \begin{aligned}
            e^{\pm i\hat{H}_1}&=\frac{\psi^1\pm i\psi^2}{\sqrt{k}}, \quad e^{\pm i\hat{H}_2}=\frac{\chi^1\pm i\chi^2}{\sqrt{k}}, \quad e^{\pm i\hat{H}_3}=\frac{\chi^3\mp \psi^3}{\sqrt{k}}\\
            e^{\pm i\hat{H}_4}&=\frac{\eta^1\pm i\eta^2}{\sqrt{2}}, \quad e^{\pm i\hat{H}_5}=\frac{\eta^3\pm i\eta^4}{\sqrt{2}}\ .
        \end{aligned}
        \end{equation}
         and the fermionic $SL(2,R)$ currents are:
        \begin{equation}
            \hat{j}^3=i\partial \hat{H}_1, \qquad \hat{j}^\pm=\pm e^{\pm i\hat{H}_1}(e^{-i\hat{H}_3}-e^{+i\hat{H}_3})
        \end{equation}
     The final results for the physical operators are summarized in the table \ref{numberofoperators}.
 
\subsubsection{Odd spectral flow parameters}
We first discuss the  case where operators have odd spectral flow parameters $w$. Recall that operators corresponding to local operators on the 
field theory side should be in the $x$-basis. Nevertheless, we will firstly write them in the $m$-basis (then transform them into the $x$-basis), since spectral flow is simpler to perform in the $m$-basis.

In the NS sector, one can construct the following vertex operators (in the $m$-basis):
\begin{equation}\label{mbasisphy}
    O_{j,m}^w(z)\equiv e^{-\phi(z)}\textbf{1}_{\psi}^w(z) V_{j,m}^{w}(z)\,,
\end{equation}
where $\phi$ is the bosonized $\beta\gamma$ ghosts so the  term $e^{-\phi}$ means the operators above are in the standard $(-1)$ picture. The term $\textbf{1}_{\psi}^w(z)$ denote the  $w$ spectrally flowed operator of the identity operator in the free fermion theory of $\psi^A$ $(A=3,\pm)$. It terms of the bosonized field $\hat{H}_i$, it can be written as \cite{Giribet:2007wp}:
\begin{equation}
    \textbf{1}_{\psi}^w=e^{-iw\hat{H}_1}
\end{equation}
 Since we want to obtain the physical operators with lowest spacetime weights, in \eqref{mbasisphy} we 
turned off any excitations in the $su(2)_{k-2}$ WZW model, the free fermions $\chi^a$ and the torus theory T$^4$.
For these operators to be physical, one needs to impose the mass-shell condition in the NS-sector
\begin{equation}
\left[0-\frac{-2w^2}{4}\right]+\left[ -\frac{j(j-1)}{k}-wm-\frac{(k+2)w^2}{4}\right]=\frac{1}{2}\,,
\end{equation}
where terms in the 2 square brackets are conformal weights of $\textbf{1}_{\psi}^w, V_{j,m}^{w}$ respectively. In terms of the full space-time weight 
\begin{equation}
    H=h+\hat{h}=m+\frac{(k+2)w}{2}+0+\frac{-2w}{2}=m+\frac{wk}{2}\,,
\end{equation} the above mass shell condition becomes:
\begin{equation}
    -\frac{j(j-1)}{k}-w H+\frac{kw^2}{4}=\frac{1}{2}\ .
\end{equation}
For long strings, we have $j=\frac{1}{2}+ip$, 
then the  mass shell condition determine the lowest space-time weights as
\begin{equation}\label{NSodd}
    H_{\text{NS,odd}}=\frac{\frac{1}{4}+p^2}{kw}+\frac{kw}{4}-\frac{1}{2w}\ .
\end{equation}
Besides, $\textbf{1}_{\psi}^w(z) V_{j,m}^{w}(z)$ are clearly  super-Virasoro primaries so $O_{h,m}^w(z)$ are indeed BRST invariant. Finally we should demand $O_{h,m}^w(z)$ to survive the GSO projection, which restrict $w$ to be odd \cite{Giribet:2007wp}. Notice that the space-time weights \eqref{NSodd} had  been determined in \cite{Gaberdiel:2018rqv,Eberhardt:2019qcl}. While in \cite{Eberhardt:2019qcl} the ground states of the $su(2)_{k-2}$ WZW model could be an arbitrary affine primary with spin $l$, here we  focus on the case with $l=0$  (since we only concern the operators that have the lowest space-time weights) and construct these physical operators explicitly. 

Now we write the operators with the lowest space-time weights in the $x$-basis:
\begin{equation}\label{oddNS}
    O_{j,h}^w(x;z)=e^{-\phi}(z) \textbf{1}_{\psi}^w(x;z)V_{j,h}^w(x;z).
\end{equation}
Notice  that we have labeled the operator $O_{j,h}^{w}(x;z)$ by the weight from the bosonic WZW $h=m+\frac{(k+2)w}{2}$, while the full space-time weight is $H=h-w$. 
$\textbf{1}_{\psi}^w(x;z)$ is the $x$-basis operators of $\textbf{1}_{\psi}^{w}(z)$. When one expands it as a power series of $x$, only finite terms appear (with each mode being a member in the  $SL(2,R)$ multiplet of $\textbf{1}_{\psi}^w(z)$) and in particular, it contains the $m$-basis operator $\textbf{1}_{\psi}^w=e^{-iw\hat{H}_1}$ and its conjugate $\textbf{1}_{\psi}^{-w}=e^{iw\hat{H}_1}$ \cite{Giribet:2007wp}\footnote{This is different from a general $x$-basis operator $V_{j,h}^w(x;z)$, where there are typically infinite members in the $SL(2,R)$ multiplet of $V_{j,m}^w (V_{j,-m}^{-w})$ (which is true for both the  continuous and discrete  spectrally flowed representations), so $V_{j,h}^w(x;z)$ is generally an infinite power series of $x$.}.
 To calculate  3-point correlators, one also needs  the picture 0 version, which can be obtained by acting the picture rasing operator  $e^{\phi}G$ \cite{Friedan:1985ge}.  Since $e^{\phi}G$ commute with $J_0^+$, we firstly calculate the action of the  picture rasing operator on the $m$-basis physical operators \eqref{mbasisphy}. The result is:
 \begin{equation}\label{mbasispicture0}
     O_{j,m}^{w(0)}(z)=\frac{1}{k}[(m+j-1)\psi^{+,w}V_{j,m-1}^w-2\left(m+\frac{km}{2}\right)\psi^{3,w}V_{j,m}^w+(m-j+1)\psi^{-,w}V_{j,m+1}^w\ ].
 \end{equation}
 where $\psi^{a,w} (a=3,\pm)$ are the   $w$ spectrally flowed operator of the fermions $\psi^a$, which can be written in terms of $\hat{H}_i$ as \cite{Giribet:2007wp,Iguri:2022pbp}:
 \begin{equation}
     \psi^{\pm,w}=\sqrt{k}e^{i(\pm 1-w)\hat{H}_1}, \quad \psi^{3,w}=\psi^3e^{-iw\hat{H}_1}=\frac{\sqrt{k}}{2}(e^{-i\hat{H}_3}-e^{+i\hat{H}_3})e^{-iw\hat{H}_1}
 \end{equation}
 Then from the definition \eqref{xbaisdef}, we can translate these operators into the ones in the $x$-basis:
 \begin{equation}
 \begin{aligned}
      O_{j,h}^{w(0)}(x;z)&=\frac{1}{k}[
      -2(h-w)\psi^{3,w}(x;z)V_{j,h}^w(x;z)+\left(h-\frac{(k+2)w}{2}+j-1\right)\psi^{+,w}(x;z)V_{j,h-1}^w(x;z)\\
      &\qquad\qquad +\left(h-\frac{(k+2)w}{2}-j+1\right)\psi^{-,w}(x;z)V_{j,h+1}^w(x;z)\,],
 \end{aligned}
 \end{equation}
 where $\psi^{a,w}(x;z)$ are the $x$-basis operators of $\psi^{a,w}(z)$ defined as in \eqref{xbaisdef}. Similar to $\textbf{1}_{\psi}^w(x;z)$,  $\psi^{a,w}(x;z)$  are also finite power series of $x$  and contains both the $m$-basis operator $\psi^{a,w}(z)$ itself and its conjugate \cite{Giribet:2007wp}.

Now we turn to the Ramond sector. Firstly, note that the Ramond ground states are created by acting on the vacuum with the spin fields
        \begin{equation}
            \mathbf{S}(z)=e^{\frac{i}{2}\sum_I\epsilon_I\hat{H}_I}\,,
        \end{equation} 
        where $\epsilon_I=\pm 1$,  and the GSO projection imposes the mutual locality condition
        \begin{equation}\label{spinGSO}
            \prod_{I=1}^5\epsilon_I=+1\ .
        \end{equation} 
Besides, the BRST condition demands \cite{Giveon:1998ns}:
         \begin{equation}
            \prod_{I=1}^3\epsilon_I=+1\ .
        \end{equation} 
       Thus, there are in total $2^{5-2}=8$ supercharges $Q$ obtained from these spin fields
       \begin{equation}\label{supercharge}
           Q=\oint dz e^{-\frac{\phi}{2}}\mathbf{S}(z)\ .
       \end{equation}
       They corresponds on the boundary side to the 8 supercharges of the global $\mathcal{N}=4$ superconformal algebra. 

       Now we can write physical vertex operators in the Ramond sector. 
       Writing out explicitly the $\epsilon$ dependence of the spin fields:
       \begin{equation}
          \mathbf{S}(z)=  e^{\frac{i}{2}\sum_I\epsilon_I\hat{H}_I}\equiv \mathbf{S}_{\epsilon_1\epsilon_2\epsilon_3\epsilon_4\epsilon_5}\ .
       \end{equation}
       These fields have $(H,J,\Delta)=(\frac{\epsilon_1}{2},\frac{\epsilon_2}{2},\frac{5}{8})$.
       The spectrally flowed spin fields $\mathbf{S}_{\epsilon_1\epsilon_2\epsilon_3\epsilon_4\epsilon_5}^w$ have
       \begin{equation}
           (H,J,\Delta)=\left(\frac{\epsilon_1}{2}-w,\frac{\epsilon_2}{2},\frac{5}{8}+\frac{w^2}{2}-\frac{w\epsilon_1}{2}\right)\ .
       \end{equation}
       Notice that we only spectral flow the $sl(2,R)^{(1)}$ part, so  $J$ will not change\footnote{Notice that only $\psi^+,\psi^-$ change under the spectral flow ($\psi^3$ does not change),  so the spectral flow only acts on $e^{\frac{i\epsilon_1}{2}\hat{H}_1}$.}.  These charges fix their bosonizations as
       \begin{equation}
          \mathbf{S}_{\epsilon_1\epsilon_2\epsilon_3\epsilon_4\epsilon_5}^w= e^{\frac{i}{2}\sum_I(\epsilon_I-2\delta_{1,I}w)\hat{H}_I}\ .
       \end{equation}
       In the Ramond sector, (the matter part of) physical operators in the picture $(-\frac{1}{2})$ are superconformal primaries and should survive the GSO projection. To construct them, we start form the ones in the picture $(-\frac{3}{2})$, which can be written as (we focus on the states with the lowest weights, thus  turn off all  possible additional excitations)
       \begin{equation}\label{-3/2m}
           \Tilde{O}_{j,m}^w(z)\equiv e^{-\frac{3\phi(z)}{2}} V_{j,m}^{w}(z)\mathbf{S}_{\epsilon_1\epsilon_2\epsilon_3\epsilon_4\epsilon_5}^w(z)\,,
       \end{equation}
       where $\epsilon_i$ should satisfy  $\sum_{I=1}^5\epsilon_I=1$, which comes form the GSO projection. It takes the same form as the one in \eqref{spinGSO} for the spin fields, because we are in the picture $(-\frac{3}{2})$ and we assume that  the spectral flow parameter $w$ is odd. As will be clear in the following,  the remaining $2^{5-1}=16$ operators only give 8  BRST equivalent classes.
       
       Then the picture $(-\frac{1}{2})$ operators  can be constructed as
       \begin{equation}\label{Ramondform}
            O_{j,m}^w(z)=e^{\phi(z)} G_0 \Tilde{O}_{j,m}^w(z)\ .
       \end{equation}
       Notice that in the following, we always omit the labels in the  fermionic parts and only use the labels  of the bosonic operators to label the full supersymmetric physical operators. Besides, we always use ``$O$'' to denote the physical operators in various situations. 
      Now $O_{j,m}^w(z)$ is guaranteed to be BRST invariant  given that  $\Tilde{O}_{j,m}^w(z)$ (thus also $O_{j,m}^w(z)$) is on-shell since
      \begin{equation}
          G_0O_{j,m}^w(z)=e^{\phi(z)}G_0^2 \Tilde{O}_{j,m}^w(z)=e^{\phi(z)}\left(L_0-\frac{c}{24}\right)\Tilde{O}_{j,m}^w(z)=e^{\phi(z)}\left(\frac{5}{8}-\frac{15}{24}\right)\Tilde{O}_{j,m}^w(z)=0\ .
      \end{equation}
      Now we  write down the explicit form of $ O_{j,m}^w(z)$. For this we need the form of the supercurrent in terms of the bosonized fields:
      \begin{equation}\label{bosonizedG}
      \begin{aligned}
            G=\frac{1}{\sqrt{k}}\Big[ &e^{+i\hat{H}_1}j^-+e^{-i\hat{H}_1}j^++(e^{+i\hat{H}_3}-e^{-i\hat{H}_3})j^3+(i\partial \hat{H}_2-i\partial \hat{H}_1)e^{-i\hat{H}_3}+(i\partial \hat{H}_2+i\partial \hat{H}_1)e^{+i\hat{H}_3}\\
            &+e^{+i\hat{H}_2}k^-+e ^{-i\hat{H}_2}k^++(e^{+i\hat{H}_3}+e^{-i\hat{H}_3})k^3
          \Big]+G(T^4)\ .
      \end{aligned}
      \end{equation}
      For the calculation of $O_{j,m}^w(z)$ here, only the first line for $G$ above is needed. Notice that  \eqref{Ramondform} can be generalized to 
      allow additional excitations in $\Tilde{O}_{j,m}^w(z)$, such as those in the  $su(2)_{k-2}$ WZW model or the T$^4$ theory. For these generalizations one will also need the second line of $G$ above. 
      Since the operators we consider are spectrally flowed, we also need to count the effect of spectral flow acting on $G$ in \eqref{actonTandG}. 
      With all these effects included, finally we finds 8  physical operators:
      \begin{equation}\label{-+}
      \begin{aligned}
          O_{j,m}^w(z)&=e^{-\frac{\phi(z)}{2}} \frac{1}{\sqrt{k}}\Big[(m-j+1)V_{j,m+1}^{w}(z)\mathbf{S}_{-\epsilon_2 +\epsilon_4\epsilon_5}^w(z)\\
          &\qquad +\epsilon_2\left(m+\frac{kw}{2}+\frac{1-\epsilon_2}{2}\right)V_{j,m}^{w}(z)\mathbf{S}_{+\epsilon_2 -\epsilon_4\epsilon_5}^w(z)\Big],\qquad 
          (\epsilon_2\epsilon_4\epsilon_5=+1)\,,
      \end{aligned}
      \end{equation}
      or
          \begin{equation}\label{++}
            \begin{aligned}
          O_{j,m}^w(z)&=e^{-\frac{\phi(z)}{2}}\frac{1}{\sqrt{k}} \Big[(m-j+1)V_{j,m+1}^{w}(z)\mathbf{S}_{-\epsilon_2 -\epsilon_4\epsilon_5}^w(z)\\
          &\qquad -\epsilon_2\left(m+\frac{kw}{2}+\frac{1+\epsilon_2}{2}\right)V_{j,m}^{w}(z)\mathbf{S}_{+\epsilon_2 +\epsilon_4\epsilon_5}^w(z)\Big], \qquad
          (\epsilon_2\epsilon_4\epsilon_5=-1)\ .
          \end{aligned}
      \end{equation}
     A few comments on these operators are in order
      \begin{itemize}
          \item The above operators are obtained by letting $\{\epsilon_1,\epsilon_3\}=\{+,+\}$ and $\{+,-\}$ in \eqref{Ramondform} (with \eqref{-3/2m}) respectively.  To obtain them, the cocycles in  $\hat{H}_i$ need to be properly counted. If we instead let $\{\epsilon_1,\epsilon_3\}=\{-,-\}$ and $\{-,+\}$, we will obtain the same set of operators as the above 8 ones. For example, letting $\{\epsilon_1,\epsilon_3\}=\{-,-\}$, one gets:
         \begin{equation}
      \begin{aligned}
          O_{j,m+1}^w(z)&=e^{-\frac{\phi(z)}{2}} \Big[\epsilon_2\left(m+\frac{kw}{2}+\frac{1+\epsilon_2}{2}\right)V_{j,m+1}^{w}(z)\mathbf{S}_{-\epsilon_2 +\epsilon_4\epsilon_5}^w(z)\\
          &\qquad +\left(m+j\right)V_{j,m}^{w}(z)\mathbf{S}_{+\epsilon_2 -\epsilon_4\epsilon_5}^w(z)\Big],\qquad 
          (\epsilon_2\epsilon_4\epsilon_5=+1)\,,
      \end{aligned}
      \end{equation}
      These operators are proportional to those in \eqref{-+}, due to the mass-shell condition \eqref{massshellR}.
          Thus we only have 8 independent physical operators in total.
          \item One can check that these operators are indeed BRST invariant, again using the bosonized form of $G$ in \eqref{bosonizedG} and the mass-shell condition.
          \item All the 4 terms in  \eqref{-+} and \eqref{++} have the same space-time conformal weight 
          $H=m+\frac{kw}{2}+\frac{1}{2}$, and one can check that they  have the same mass shell condition, which can be written as
          \begin{equation}\label{massshellR}
            -\frac{j(j-1)}{k}-w (m+1)-\frac{(k+2)w^2}{4}+\frac{5}{8}+\frac{w^2}{2}+\frac{w}{2}=\frac{5}{8} \ .
          \end{equation}
        \item Form the above equation, we obtain the lowest weight $H_{\text{R,odd}}$ of states in the Ramond secter with $w$ odd:
        \begin{equation}\label{weightRamond}
           H_{\text{R,odd}}=\frac{j(1-j)}{kw}+\frac{kw}{4}=\frac{\frac{1}{4}+p^2}{kw}+\frac{kw}{4}=H_{\text{NS,odd}}+\frac{1}{2w}\ .
          \end{equation}
          One finds that it is larger than~\eqref{NSodd} in the NS sector, which means these 8 operators correspond to excited states in the spacetime theory. The ground state is unique and lies in the NS sector (for a given $w$). 
      \end{itemize}
      We also need the form of the above operators
      in the $x$-basis. For the operators \eqref{-3/2m} in  picture $(-\frac{3}{2})$, in  $x$-basis they are
      \begin{equation}\label{Roddin-3/2}
          \Tilde{O}_{j,h}^w(x;z)=e^{-\frac{3\phi(z)}{2}} V_{j,h}^{w}(x;z)\mathbf{S}_{\epsilon_1\epsilon_2\epsilon_3\epsilon_4\epsilon_5}^w(x;z)\,,
      \end{equation}
      where $h=m+\frac{(k+2)w}{2}$ and
          $\mathbf{S}_{\epsilon_1\epsilon_2\epsilon_3\epsilon_4\epsilon_5}^w(x;z)$ is the $x$-basis operator of $\mathbf{S}_{\epsilon_1\epsilon_2\epsilon_3\epsilon_4\epsilon_5}^w(z)$, which contains finite terms as a power series of $x$, including in particular the $m$-basis operator $\mathbf{S}_{\epsilon_1\epsilon_2\epsilon_3\epsilon_4\epsilon_5}^w(z)$ itself and its conjugate $\mathbf{S}_{(-\epsilon_1)\epsilon_2(-\epsilon_3)\epsilon_4\epsilon_5}^{-w}(z)$\footnote{Notice that the ``conjugate'' here means $w\to -w, m\to -m$, so only  $\epsilon_1$ and $\epsilon_3$ change their signs.} \cite{Giribet:2007wp}. 
      The operators  in  picture $(-\frac{1}{2})$ read
        \begin{equation}\label{Rodd1}
      \begin{aligned}
          O_{j,h}^w(x;z)=e^{-\frac{\phi(z)}{2}}\frac{1}{\sqrt{k}} \Big[(h-\frac{(k+2)w}{2}-j+1)V_{j,h+1}^{w}(x;z)&\mathbf{S}_{-\epsilon_2 +\epsilon_4\epsilon_5}^w(x;z)\\+\epsilon_2\left(h-w+\frac{1-\epsilon_2}{2}\right)V_{j,h}^{w}(x;z)&\mathbf{S}_{+\epsilon_2 -\epsilon_4\epsilon_5}^w(x;z)\Big],\quad (\epsilon_2\epsilon_4\epsilon_5=+1)
      \end{aligned}
      \end{equation}
      or
          \begin{equation}\label{Rodd2}
            \begin{aligned}
          O_{j,h}^w(x;z)=e^{-\frac{\phi(z)}{2}}\frac{1}{\sqrt{k}} \Big[(h-\frac{(k+2)w}{2}-j+1)V_{j,h+1}^{w}(x;z)&\mathbf{S}_{-\epsilon_2 -\epsilon_4\epsilon_5}^w(x;z)\\
          -\epsilon_2\left(h-w+\frac{1+\epsilon_2}{2}\right)V_{j,h}^{w}(z)&\mathbf{S}_{+\epsilon_2 +\epsilon_4\epsilon_5}^w(x;z)\Big], \quad (\epsilon_2\epsilon_4\epsilon_5=-1)
          \end{aligned}
      \end{equation}

\subsubsection{Even spectral flow parameters}
 Now we turn to the case where operators have even spectral flow parameters 
 $w$. Since $w$ is even, comparing to the odd case \eqref{mbasisphy}, an additional fermion should be excited in order to survive the GSO projection. We first consider the NS sector. Somewhat similar to the case of flat space-time, the BRST condition gives a polarization constraint, which reduces the number of states by one. Meanwhile, the action of $G_{-\frac{1}{2}}$ on the ground state (which does not survive the GSO projection) is spurious, which reduces the number of states by one as well. Thus we have in total $10-2=8$ physical operators at the level $\frac{1}{2}$. In the following we will construct these 8 physical operators concretely.

In the NS sector, 
excited fermionic operators in the 7 compact directions are easy to write down and they are of the form (in the $m$-basis)
 \begin{equation}
     O_{j,m}^w(z) \equiv e^{-\phi(z)}\textbf{1}_{\psi}^w(z) V_{j,m}^{w}(z)\mathcal{F}(z)\,,
 \end{equation}
 where $\mathcal{F}(z)$ is the excited fermion and the above operators are in the standard picture $(-1)$.  
 One set of choices of the $\mathcal{F}(z)$ corresponding to these 7 excitations are
 \begin{equation}
    \eta^1, \quad \eta^2, \quad \eta^3, \quad \eta^4, \quad \chi^-, \quad \chi^3, \quad \chi^+.
 \end{equation}
 It is easy to see that the above 7 operators are all BRST invariant, and the mass shell condition is ($H=m+\frac{kw}{2}$)
\begin{equation}\label{massshell}
    -\frac{j(j-1)}{k}-w H+\frac{kw^2}{4}+\frac{1}{2}=\frac{1}{2}\ .
\end{equation}
This leads to
the lowest space-time weights 
 \begin{equation}\label{NSeven}
     H_{\text{NS,even}}=\frac{\frac{1}{4}+p^2}{kw}+\frac{kw}{4}\ .
 \end{equation}
 Finally, we can write these physical operators in the $x$-basis
\begin{equation}\label{NSeven1}
     O_{j,h}^w(z) \equiv e^{-\phi(z)}\textbf{1}_{\psi}^w(x;z) V_{j,h}^{w}(x;z)\mathcal{F}(z)\,,
\end{equation}
with $h=m+\frac{(k+2)w}{2}$. 

The remaining one physical operator is the fermionic excitations 
in the (super-symmetric) AdS$_3$ part. In the $m$-basis, it has
the following form
\begin{equation}\label{adsfermion}
O_{j,m}^w(z)=e^{-\phi(z)}\mathcal{O}_{j,m}^w(z), \quad
     \mathcal{O}_{j,m}^w(z)=\alpha_- V_{j,m+1}^w\psi^{-,w}+\alpha_3 V_{j,m}^w\psi^{3,w}
    +\alpha_+ V_{j,m-1}^w\psi^{+,w}\,,
\end{equation}
where $(\alpha_-,\alpha_3,\alpha_+)$ are the (to be determined) polarization. 
Notice that the mass-shell conditions of the above three operators are respectively
\begin{equation}
\begin{aligned}
    -\frac{j(j-1)}{k}-w(m+1)-\frac{(k+2)w^2}{4}+\frac{(1+w)^2}{2}&=\frac{1}{2}, \\
    -\frac{j(j-1)}{k}-wm-\frac{(k+2)w^2}{4}+\frac{w^2+1}{2}&=\frac{1}{2},\\
    -\frac{j(j-1)}{k}-w(m-1)-\frac{(k+2)w^2}{4}+\frac{(1-w)^2}{2}&=\frac{1}{2}\,,
\end{aligned}    
\end{equation}
which are the same and coincide with \eqref{massshell} and hence consistent (notice that for all the 3 operators we have $H=m+\frac{wk}{2}$).
BRST invariance requires
\begin{equation}\label{superconformal}
    L_n \mathcal{O}_{j,m}^w(0)|0\rangle=0,\quad  G_r \mathcal{O}_{j,m}^w(0)|0\rangle=0, \qquad \text{for $n,r>0$}\,,
\end{equation}
where $L_n, G_r$ are the modes of the stress tensor and supercurrent. 
From the form of the stress tensor  and supercurrent \eqref{TandG}, as well as the action of the spectral flow on them \eqref{actonTandG}, it is clear that the first condition in \eqref{superconformal} is satisfied. As for the second condition, only when $r=\frac{1}{2}$ it 
gives a non-trivial  constraint
\begin{equation}
    \left(G_{\frac{1}{2}}-w \psi^3_{\frac{1}{2}}\right)\left(\alpha_- V_{j,m+1}\psi^-_{-\frac{1}{2}}+\alpha_3 V_{j,m}\psi^3_{-\frac{1}{2}}
    +\alpha_+ V_{j,m-1}\psi^+_{-\frac{1}{2}}\right)|0\rangle=0\ .
\end{equation}
Using the expression of the supercurrent $G_n$, this equation become
\begin{equation}\label{superprimary}
    (m+j)\alpha_-+\left(m+\frac{w k}{2}\right)\alpha_3+(m-j)\alpha_+=0\ .
\end{equation}
This linear equation has 2 independent solutions but only one gives a real  physical state. The other is   spurious and has the form
\begin{equation}
\begin{aligned}
     O_{j,m}^{w,spurious}(z)&\equiv e^{-\phi(z)}\big[G_{-\frac{1}{2}}\textbf{1}_{\psi}^w(z) V_{j,m}^{w}(z)\big]\\
     &\propto (m-j+1) V_{j,m+1}^w\psi^{-,w}-2\left(m+\frac{wk}{2}\right) V_{j,m}^w\psi^{3,w}
    +(m+j-1) V_{j,m-1}^w\psi^{+,w}\ .\label{owspu}
\end{aligned}
\end{equation}
It can be checked that the 
coefficients in~\eqref{owspu} satisfy  equation~\eqref{superprimary}  after using the mass-shell condition \eqref{massshell}. 
This spurious state has the same form as the picture 0 operator \eqref{mbasispicture0} in the case of odd spectral flow parameters. Of course, that operator is not spurious because the mass-shell condition is different there. The remaining physical operators thus has the form (in the $x$-basis)
 \begin{equation}\label{NSeven2}
      O_{j,h}^{w}(x;z)=e^{-\phi(z)}[\alpha_-\psi^{-,w}(x;z)V_{j,h+1}^w(x;z)+ \alpha_3 \psi^{3,w}(x;z)V_{j,h}^w(x;z)+\alpha_+\psi^{+,w}(x;z)V_{j,h-1}^w(x;z)\,],
 \end{equation}
 where $(\alpha_-, \alpha_3, \alpha_+)$ is the real physical solution of \eqref{superprimary} (up to the spurious one). For example, they can be chosen as 
 \begin{equation}
     \alpha_-=j-m, \quad \alpha_3=0, \quad \alpha_+=j+m
 \end{equation}
 
  Now we can write physical vertex operators in the Ramond sector. The result is almost the same as in the case of odd spectral flow parameters. In  picture $(-\frac{3}{2})$, they take the form
       \begin{equation}
           \Tilde{O}_{j,m}^w(z)\equiv e^{-\frac{3\phi(z)}{2}} V_{j,m}^{w}(z)\mathbf{S}_{\epsilon_1\epsilon_2\epsilon_3\epsilon_4\epsilon_5}^w(z)
       \end{equation}
where now $\epsilon_i$ satisfy  $\sum_{I=1}^5\epsilon_I=-1$, which has a sign difference comparing with the case of odd spectral flow parameters. Accordingly, there are 8 physical operators, written in the picture $(-\frac{1}{2})$ as
  \begin{equation}\label{Reven1}
      \begin{aligned}
          O_{j,m}^w(z)&=e^{-\frac{\phi(z)}{2}} \frac{1}{\sqrt{k}}\Big[(m-j+1)V_{j,m+1}^{w}(z)\mathbf{S}_{-\epsilon_2 +\epsilon_4\epsilon_5}^w(z)\\
         &\qquad \qquad  +\epsilon_2\left(m+\frac{kw}{2}+\frac{1-\epsilon_2}{2}\right)V_{j,m}^{w}(z)\mathbf{S}_{+\epsilon_2 -\epsilon_4\epsilon_5}^w(z)\Big]\,,\qquad (\epsilon_2\epsilon_4\epsilon_5=-1)
      \end{aligned}
      \end{equation}
      or
          \begin{equation}\label{Reven2}
            \begin{aligned}
          O_{j,m}^w(z)&=e^{-\frac{\phi(z)}{2}}\frac{1}{\sqrt{k}} \Big[(m-j+1)V_{j,m+1}^{w}(z)\mathbf{S}_{-\epsilon_2 -\epsilon_4\epsilon_5}^w(z)\,,\\
          &\qquad \qquad -\epsilon_2\left(m+\frac{kw}{2}+\frac{1+\epsilon_2}{2}\right)V_{j,m}^{w}(z)\mathbf{S}_{+\epsilon_2 +\epsilon_4\epsilon_5}^w(z)\Big]\,, \qquad (\epsilon_2\epsilon_4\epsilon_5=+1)\ .
          \end{aligned}
      \end{equation}
One can also write down the corresponding operators in the $x$-basis, just as in the NS sector. 
Mass shell condition gives the same  weight $H$ as in \eqref{weightRamond}, namely $H_{\text{R,even}}=H_{\text{R,odd}}$, the difference is that now it is equal to the lowest space-time weights \eqref{NSeven}  in the NS sector:
\begin{equation}
    H_{\text{R,even}}=H_{\text{NS,even}}\ .
\end{equation}
Thus, for $w$ even, there are in total $8+8=16$ ground states in the space-time theory. 

\subsubsection*{Summary}
 The operators constructed in this section are summarized in the following table \ref{numberofoperators}.  
\begin{table}[!ht]
    \centering
     \begin{tabular}{|c|c|c|}\hline
   \diagbox{Parity of $w$}{Sectors}   & NS sector & R sector\\ \hline
  odd & 1 ground (in \eqref{oddNS}) & 8 excited (in \eqref{Rodd1},\eqref{Rodd2})\\  \hline
 even &8 ground (in \eqref{NSeven1},\eqref{NSeven2})& 8 ground (in \eqref{Reven1},\eqref{Reven2})\\  \hline
\end{tabular} 
    \caption{The operators with the lowest space-time weights.}
    \label{numberofoperators}
\end{table}    
For the matching with the CFT side, we want to specify the the representation contents of these operators. Notice that the (small) $\mathcal{N}=4$ superconformal algebra has an  outer automorphism  $SU(2)_{\text{outer}}$. Then the full automorphism group of the algebra is $SU(2)_R\oplus SU(2)_{\text{outer}}$ (here we only consider the left-moving part). It is therefore helpful to organize operators into  representations of  this $SU(2)_R\oplus SU(2)_{\text{outer}}$.

Firstly, notice that among all the generators in the  (small) $\mathcal{N}=4$ superconformal algebra, $SU(2)_{\text{outer}}$ only acts non-trivially on the supercharges (see, e.g. \cite{deBoer:2008ss}). One can write the indices of the supercharge \eqref{supercharge} explicitly as $Q_{\frac{\epsilon_1}{2}}^{\epsilon_2,\epsilon_4}$, and for each choice of $\epsilon_1$, the four supercharges $Q_{\frac{\epsilon_1}{2}}^{\epsilon_2,\epsilon_4}$ transform as a $(\textbf{2,2,1})$ of $SU(2)_R\oplus SO(4)$ ($SO(4)$ is the rotations of the four fermions in T$^4$) \cite{Giveon:1998ns}. Then it is  clear that  $SU(2)_R$ rotates the index $\epsilon_2$ and $SU(2)_{\text{outer}}$ rotates the index $\epsilon_4$. Note that the generators of the rotation $SO(4)$ of the 4 fermions can be constructed   by (the zero modes of) the following 2 $SU(2)$ currents
\begin{equation}\label{outercurrent}
\begin{aligned}
    J^{(1)3}=\frac{1}{2}(i\partial \hat{H}_4+i\partial\hat{H}_5),\qquad
      J^{(1)+}= e^{ i\hat{H_5}}e^{i\hat{H_4}}, \qquad J^{(1)-}= -e^{ -i\hat{H_5}}e^{-i\hat{H_4}}\\
 J^{(2)3}=\frac{1}{2}(i\partial \hat{H}_4-i\partial\hat{H}_5),\qquad
      J^{(2)+}= e^{ i\hat{H_4}}e^{-i\hat{H_5}}, \qquad J^{(1)-}= -e^{ -i\hat{H_4}}e^{i\hat{H_5}}   
\end{aligned}
\end{equation}
Then it is clear that the zero modes of the  currents $J^{(1)a}$ $(a=3,\pm)$ generate the algebra  $SU(2)_{\text{outer}}$.
 Now, one can directly read the  representation contents with respect to $SU(2)_R\oplus SU(2)_{\text{outer}}$, listed in the  table \ref{representation contents}.
\begin{table}[!ht]
    \centering
     \begin{tabular}{|c|c|c|}\hline
   \diagbox{Parity of $w$}{Sectors}   & NS sector & R sector\\ \hline
  odd & $(\textbf{1,1})$ (in \eqref{oddNS}) & $2(\textbf{2,1})\oplus(\textbf{2,2})$ (in \eqref{Rodd1},\eqref{Rodd2})\\  \hline
 even & \makecell[c]{$(\textbf{3,1})\oplus (\textbf{1,1})\oplus 2(\textbf{1,2})$\\ (in \eqref{NSeven1},\eqref{NSeven2})}&  $2(\textbf{2,1})\oplus(\textbf{2,2})$ (in \eqref{Reven1},\eqref{Reven2})\\  \hline
\end{tabular} 
    \caption{The representation contents}
    \label{representation contents}
\end{table}  
In particular, notice that in the case of NS sector and $w$ is even, the $(\textbf{3,1})$ comes from $\mathcal{F}=\chi^{3,\pm}$ in \eqref{NSeven1}, the  $(\textbf{1,1})$ comes from  \eqref{NSeven2}, the $2(\textbf{1,2})$ comes from $\mathcal{F}=\eta^i$ $(i=1,2,3,4)$ in \eqref{NSeven1}.  In the Ramond sector,  those operators with $\epsilon_4=-\epsilon_5$ ($\epsilon_4=\epsilon_5$) from singlets (doublets) of the algebra $SU(2)_{\text{outer}}$ (and they from doublets of the algebra $SU(2)_R$).   We will find their counterparts on the CFT side in section \ref{theCFTside}.

\section{Superstring correlators}\label{superstring correlator}
Now we calculate the superstring three point correlators. There are various cases with different  parities of the spectral flows. We use O and E to denote $w$ odd and even respectively. We also use X-Y-Z, where X, Y, Z can be O or E, to denote the  parities of the three vertex operators. Since the form of the three point functions in the bosonic $SL(2,R)$ WZW model depends on the total parity of $\sum_i w_i$ \cite{Dei:2021xgh}  (see also \eqref{ybasis3pt}), we discuss 
correlators with different total parity separately. When $\sum_i w_i$ is odd,  there are two possible cases:  O-O-O and O-E-E; when $\sum_i w_i$ is even,  
and another two possible cases,  O-O-E and E-E-E. Furthermore, for $w$ even (E), as we had discussed above, there are 16 different choices for the space-time ground states with different fermionic excitations. On the other hand, for $w$ odd (O), the ground state is unique (lies in the NS sector) and we have 8 choices for excited states with lowest excited energy (lie in the R sector). 

We will not calculate all the cases of the three point functions. Instead, for each type of correlators, we will  calculate one representative as an illustration. Besides, these representatives involve operators lying in both the NS sector and the R sector.  Notice that when the total picture number is $-2$, the number of operators in the R sector is 0 or 2; on the other hand,  the form of~\eqref{ybasis3pt}  depends on the total parity $\sum_i w_i$. Thus the 4 representatives we choose in the following include the above 2 choices of the number of the R sector operators for each value of the total parity of $\sum_i w_i$.  In the following,  we again focus on the left-moving part and omit a similar analysis for the right-movers  (we also mostly suppress the anti-holomorphic dependence of the correlators).

\subsection{Parity odd}
    When $\sum_i w_i$ is odd, there are two possible types of correlators, namely O-O-O or O-E-E. 
\subsubsection{O-O-O}
In this case, there are 2 possibilities:
\begin{enumerate}
    \item   All the three operators are in the NS sector (\eqref{oddNS}); 
    \item One operator is in the NS sector (\eqref{oddNS}) and the other two are in the R sector (\eqref{Rodd1},\eqref{Rodd2}). 
\end{enumerate}
We  choose to calculate the  case 1, which is the simplest. In fact, the calculation of the  case 2 is completely analogous with the correlator we will calculate in the type E-O-O in section \ref{typeEOO}, so we omit it here. Thus the correlator we consider is (here we only write its left-moving part,  the final result should also include the right-moving part).
    \begin{equation}
   \mathcal{M}_{OOO}^{\text{left}}= \left\langle c(z_1)O_{j_1,h_1}^{w_1}(x_1;z_1) c(z_2)O_{j_2,h_2}^{w_2}(x_1;z_1)c(z_3)O_{j_3,h_3}^{w_3(0)}(x_3;z_3)\right\rangle\,,
\end{equation}
  where we have included the $c$ ghosts. Notice that we label the correlator simply by the type ``$OOO$'', which is fine since we only choose one representative for each  type of correlators. The result is
\begin{equation}\label{M3OOO}
 \begin{aligned}
&    \mathcal{M}_{OOO}^{\text{left}}=\frac{1}{k}[-2(h_3-w_3)\langle \textbf{1}_{\psi}^{w_1}(x_1) \textbf{1}_{\psi}^{w_2}(x_2) \psi^{3,w_3}(x_3)\rangle\langle V_{j_1,h_1}^{w_1}(x_1)V_{j_2,h_2}^{w_2}(x_2)V_{j_3,h_3}^{w_3}(x_3)\rangle\\
     &+\left(h_3-\frac{(k+2)w_3}{2}+j_3-1\right)\langle \textbf{1}_{\psi}^{w_1}(x_1) \textbf{1}_{\psi}^{w_2}(x_2) \psi^{+,w_3}(x_3)\rangle \langle V_{j_1,h_1}^{w_1}(x_1)V_{j_2,h_2}^{w_2}(x_2)V_{j_3,h_3-1}^{w_3}(x_3)\rangle\\
      &+\left(h_3-\frac{(k+2)w_3}{2}-j_3+1\right)\langle \textbf{1}_{\psi}^{w_1}(x_1) \textbf{1}_{\psi}^{w_2}(x_2) \psi^{-,w_3}(x_3)\rangle \langle V_{j_1,h_1}^{w_1}(x_1)V_{j_2,h_2}^{w_2}(x_2)V_{j_3,h_3+1}^{w_3}(x_3)\rangle]\ .
 \end{aligned}
\end{equation}
Notice that  $\mathcal{M}_{OOO}$ does not have $z_i$ dependence since the ghosts are included. Thus we have omitted all the $z_i$ dependence of all the operators in \eqref{M3OOO}. 
We can use the global $SL(2,R)$ symmetries to fix  $x_i$ and $z_i$ to be: $(z_1,z_2,z_3)=(x_1,x_2,x_3)=(0,1,\infty)$.  
Then, we need to compute the correlators
\begin{equation}
    \langle \textbf{1}_{\psi}^{w_1}(0;0) \textbf{1}_{\psi}^{w_2}(1;1) \psi^{a,w_3}(\infty;\infty)\rangle \quad a=3,\pm, \qquad \langle V_{j_1,h_1}^{w_1}(0;0)V_{j_2,h_2}^{w_2}(1;1)V_{j_3,h_3}^{w_3}(\infty,\infty)\rangle\,,
\end{equation}
A closed formula for the bosonic correlator $\langle V_{j_1,h_1}^{w_1}(0;0)V_{j_2,h_2}^{w_2}(1;1)V_{j_3,h_3}^{w_3}(\infty,\infty)\rangle$ is obtained in~\cite{Dei:2021xgh}. We review the result in the  appendix \ref{3pointWZW}.

Now we turn to the fermionic correlators: $\langle \textbf{1}_{\psi}^{w_1}(0;0) \textbf{1}_{\psi}^{w_2}(1;1) \psi^{a,w_3}(\infty;\infty)\rangle$. The cases with $a=3,-$ were in fact calculated in \cite{Giribet:2007wp} using free field techniques
\footnote{As pointed out in~\cite{Iguri:2023khc}, while the result for $\langle \textbf{1}_{\psi}^{w_1}(0;0) \textbf{1}_{\psi}^{w_2}(1;1) \psi^{-,w_3}(\infty;\infty)\rangle$ in~\cite{Giribet:2007wp} is correct, the expression for $\langle \textbf{1}_{\psi}^{w_1}(0;0) \textbf{1}_{\psi}^{w_2}(1;1) \psi^{3,w_3}(\infty;\infty)\rangle$  seems not correct there.}. 
We do not use these free field techniques in this work. Instead, we treat them as three special cases of the general results~\eqref{ybasis3pt} of $  \langle V_{j_1,h_1}^{w_1}(0;0)V_{j_2,h_2}^{w_2}(1;1)V_{j_3,h_3}^{w_3}(\infty;\infty)\rangle$. In fact, $\psi^a$ can be viewed as states $|j=-1,m=a\rangle$ in the fermionic $SL(2,R)$ WZW model with $k=-2$. This method turns out to be more \CP{}
systematic.
Before doing the computation, notice that there is a convention difference between the basis  $\psi^a$ and $|j=-1,m=a\rangle$ in the  fermionic  WZW model. To get the correct result, we should multiply the formula in \cite{Dei:2021xgh} by a factor $-\frac{1}{2}$ once a $\psi^3$ appears.  

Firstly, since all $w_i$ are odd, we need to use the formula \eqref{ybasis3pt} for $\sum_iw_i\in 2\mathbb{Z}+1$. In the case at hand, we have
\begin{equation}
    j_1=j_2=0, \quad j_3=-1, \quad k=-2\ .
\end{equation}
Then the $y$-basis correlator is simply (here and in the following, when refer to the  $y$-basis correlator, we always omit the overall  factor in \eqref{ybasis3pt})
\begin{equation}\label{OOOybasis}
    X_3^2=(P_{w_1,w_2,w_3+1}+P_{w_1,w_2,w_3-1}y_3)^2\ .
\end{equation}
Since the three operators belong to the (spectral flow of) discrete representations, the integral of \eqref{ybasis3pt}  just gives 
the residue of the integrand at $y_i=0$ \cite{Dei:2021xgh}. For the case $\langle \textbf{1}_{\psi}^{w_1}(x_1) \textbf{1}_{\psi}^{w_2}(x_2) \psi^{-,w_3}(x_3)\rangle$, it can be checked that  $y_i^{\frac{kw_i}{2}+j_i-h_i-1}=y_i^{-1} (i=1,2,3)$ 
and thus the residue can be read off by setting $y_i=0$ in \eqref{OOOybasis}, leading to:
\begin{equation}\label{---}
    \langle \textbf{1}_{\psi}^{w_1}(x_1) \textbf{1}_{\psi}^{w_2}(x_2) \psi^{-,w_3}(x_3)\rangle=\sqrt{k}P^2_{w_1,w_2,w_3+1}\ .
\end{equation}
Notice that in the above we have included the overall  factor $\sqrt{k}$ for the 3-point correlator\footnote{This factor comes from the prefactor in the formula \eqref{ybasis3pt}, which is related to the unflowed 3-point function. We determined it here by letting $w_1=w_2=0, w_3=-1$.}. 
\eqref{---} 
agrees with eq. (4.74) of \cite{Giribet:2007wp}.  For the case $\langle \textbf{1}_{\psi}^{w_1}(x_1) \textbf{1}_{\psi}^{w_2}(x_2) \psi^{3,w_3}(x_3)\rangle$, we have 
\begin{equation}
    y_i^{\frac{kw_i}{2}+j_i-h_i-1}=y_i^{-1} (i=1,2), \qquad y_3^{\frac{kw_3}{2}+j_3-h_3-1}=y_3^{-2}
\end{equation} 
and the residue gives
 \begin{equation}\label{--3}
     \langle \textbf{1}_{\psi}^{w_1}(x_1) \textbf{1}_{\psi}^{w_2}(x_2) \psi^{3,w_3}(x_3)\rangle
     =-\frac{\sqrt{k}}{2}\times 2P_{w_1,w_2,w_3+1}P_{w_1,w_2,w_3-1}=-\sqrt{k}P_{w_1,w_2,w_3+1}P_{w_1,w_2,w_3-1}\ .
 \end{equation}
For the case  $\langle \textbf{1}_{\psi}^{w_1}(x_1) \textbf{1}_{\psi}^{w_2}(x_2) \psi^{+,w_3}(x_3)\rangle$, we have
\begin{equation}
    y_i^{\frac{kw_i}{2}+j_i-h_i-1}=y_i^{-1} ( i=1,2), \qquad y_3^{\frac{kw_3}{2}+j_3-h_3-1}=y_3^{-3}
\end{equation} 
and the residue becomes
\begin{equation}\label{--+}
    \langle \textbf{1}_{\psi}^{w_1}(x_1) \textbf{1}_{\psi}^{w_2}(x_2) \psi^{+,w_3}(x_3)\rangle
    =\sqrt{k}P_{w_1,w_2,w_3-1}^2\ .
\end{equation}
With these expressions 
$\mathcal{M}_{OOO}$ evaluates to (including the right-moving dependence)
\begin{equation}\label{OOO} 
\begin{aligned}
    \mathcal{M}_{OOO}=\frac{C_{S^2}}{k}\Big[& (h_3-
    \frac{(k+2)w_3}{2}+j_3-1)P^2_{w_1,w_2,w_3-1}\langle h_3-1\rangle+
    2(h_3-w_3)P_{w_1,w_2,w_3-1}P_{w_1,w_2,w_3+1}\langle ...\rangle\\
    +&(h_3-\frac{(k+2)w_3 }{2}-j_3+1)P^2_{w_1,w_2,w_3+1}\langle h_3+1\rangle\Big]\times (\text{anti-homomorphic part})\,,
\end{aligned}
\end{equation}
where $C_{S^2}$ is the normalization of the string path integral \cite{Eberhardt:2021vsx,Polchinski:1998rq}. The ``anti-homomorphic part'' above denote the right-moving part.  Here and in the following when we calculate other correlators, we always take the excitations in the right-moving part to be the similar ones as the left-moving part. Then ``anti-homomorphic part'' here is  an expression obtained by replacing all $h$ in the square brackets above by  $\Bar{h}$.
Besides, 
we always use $\langle h_3\pm 1\rangle$, $\langle ...\rangle$ to denote $\langle V_{j_1,h_1}^{w_1}V_{j_2,h_2}^{w_2}V_{j_3,h_3\pm 1}^{w_3}\rangle$, $\langle V_{j_1,h_1}^{w_1}V_{j_2,h_2}^{w_2}V_{j_3,h_3}^{w_3}\rangle$, with the anti-holomorphic part not  specified.  The product of two such terms means specifying  both the holomorphic and 
anti-holomorphic dependence, e,g.
\begin{equation}
    \langle h_3-1\rangle\times \langle \Bar{h}_3+1\rangle\equiv \langle V_{j_1,h_1,\Bar{h}_1}^{w_1}V_{j_2,h_2,\Bar{h}_2}^{w_2}V_{j_3,h_3-1,\Bar{h}_3+1}^{w_3}\rangle
\end{equation}
Then the right-hand-side  can be obtained by the formula \eqref{ybasis3pt}.

\subsubsection{O-E-E}
In this case, there are 3 possibilities:
\begin{enumerate}
    \item   ``O'' is in the NS sector (\eqref{oddNS}) and the  two ``E''  are in the Ramond sector (\eqref{Reven1},\eqref{Reven2}); 
    \item ``O'' is in the NS sector (\eqref{oddNS}) and the  two ``E'' are also in the NS sector (\eqref{NSeven1},\eqref{NSeven2});
    \item ``O'' is in the R sector (\eqref{Rodd1},\eqref{Rodd2}) and one ``E'' is also in the R sector (\eqref{Reven1},\eqref{Reven2}), the other ``E'' is in the NS sector (\eqref{NSeven1},\eqref{NSeven2}).
\end{enumerate}
We choose to calculate the  case 1. \CP{}
The calculation of the second case is analogous with the case 1 in the type of O-O-O, and the calculation of the third case is analogous with the case 1 here. So we omit the calculation for the latter 2 cases.
One choice of the spin fields so that the correlators are non-vanishing is
(other choices for the spin fields can be calculated similarly)
\begin{equation}
      O_{j_1,h_1}^{w_1(-1)}=e^{-\phi}\textbf{1}_{\psi}^{w_1}V_{j_1,h_1}^{w_1},\quad O_{j_2,h_2}^{w_2(-\frac{3}{2})}=e^{-\frac{3\phi}{2}}\mathbf{S}^{w_2}_{++++-}V_{j_2,h_2}^{w_2},\quad O_{j_3,h_3}^{w_3(-\frac{3}{2})}=e^{-\frac{3\phi}{2}}\mathbf{S}^{w_3}_{+---+}V_{j_3,h_3}^{w_3}\ .
\end{equation}
To calculate the correlator, the total picture number should be $-2$. For this, we choose the first operator  to be 
in picture $(-1)$ and the other two \CP{}
in  picture $(-\frac{1}{2})$
\begin{equation}\label{picture-1/2}
\begin{aligned}
    O_{j_2,h_2}^{w_2(-\frac{1}{2})}&=e^{-\frac{\phi}{2}}\frac{1}{\sqrt{k}}\Big[(h_2-\frac{(k+2)w_2}{2}-j_2+1)\mathbf{S}^{w_2}_{-+++-}V_{j_2,h_2+1}^{w_2}
    +\left(h_2-w_2\right)\mathbf{S}^{w_2}_{++-+-}V_{j_2,h_2}^{w_2}\Big]\\
     O_{j_3,h_3}^{w_3(-\frac{1}{2})}&=e^{-\frac{\phi}{2}}\frac{1}{\sqrt{k}}\Big[(h_3-\frac{(k+2)w_3}{2}-j_3+1)\mathbf{S}^{w_3}_{----+}V_{j_3,h_3+1}^{w_3}
    +\left(h_3-w_3\right)\mathbf{S}^{w_3}_{+-+-+}V_{j_3,h_3}^{w_3}\Big]\ .
\end{aligned}
\end{equation}
As in the case of O-O-O, $\langle V_{j_1,h_1}^{w_1}(0;0)V_{j_2,h_2}^{w_2}(1;1)V_{j_3,h_3}^{w_3}(\infty;\infty)\rangle$ is already known \cite{Dei:2021xgh}, thus we need to compute the following 
4 correlators of the fermionic multiplets
\begin{equation}
    \left\langle \textbf{1}_{\psi}^{w_1}(0;0)\mathbf{S}^{w_2}_{a+b+-}(1;1)\mathbf{S}^{w_3}_{c-d-+}(\infty,\infty)\right\rangle,\qquad (a,b)=(\pm,\mp), \quad (c,d)=(\pm,\pm)\ .
\end{equation}
 This time, $\mathbf{S}^{w_2}_{a+b+-}$ ($\mathbf{S}^{w_3}_{c-d-+}$) can be viewed as the state $|j=-\frac{1}{2},m=\frac{a}{2}\rangle$ ($|j=-\frac{1}{2},m=\frac{c}{2}\rangle$) in the fermionic $SL(2,R)$ WZW model with $k=-2$. Then we have $\sum_iw_i\in 2\mathbb{Z}+1$ and $j_1=0, j_2=j_3=-\frac{1}{2}$. Thus the correlator in the $y$-basis 
 is
 \begin{equation}
     X_2X_3=(P_{w_1,w_2+1,w_3}+P_{w_1,w_2-1,w_3}y_2)(P_{w_1,w_2,w_3+1}+P_{w_1,w_2,w_3-1}y_3)\ .
 \end{equation}
Reading out the residue as in the O-O-O case, we get respectively
\begin{itemize}
    \item $(a,b)=(-,+)$, $(c,d)=(-,-)$
    \begin{equation}
         \left\langle \textbf{1}_{\psi}^{w_1}\mathbf{S}^{w_2}_{-+++-}\mathbf{S}^{w_3}_{----+}\right\rangle=(X_2X_3)|_{y_i=0}=P_{w_1,w_2+1,w_3}P_{w_1,w_2,w_3+1}\,,
    \end{equation}
    \item $(a,b)=(-,+)$, $(c,d)=(+,+)$
    \begin{equation}
         \left\langle \textbf{1}_{\psi}^{w_1}\mathbf{S}^{w_2}_{-+++-}\mathbf{S}^{w_3}_{+-+-+}\right\rangle=\partial_{y_3}(X_2X_3)|_{y_i=0}=P_{w_1,w_2+1,w_3}P_{w_1,w_2,w_3-1}\,,
    \end{equation}
    \item $(a,b)=(+,-)$, $(c,d)=(-,-)$
    \begin{equation}
         \left\langle \textbf{1}_{\psi}^{w_1}\mathbf{S}^{w_2}_{-+++-}\mathbf{S}^{w_3}_{+-+-+}\right\rangle=\partial_{y_2}(X_2X_3)|_{y_i=0}=P_{w_1,w_2-1,w_3}P_{w_1,w_2,w_3+1}\,,
    \end{equation}
      \item $(a,b)=(+,-)$, $(c,d)=(+,+)$
    \begin{equation}
         \left\langle \textbf{1}_{\psi}^{w_1}\mathbf{S}^{w_2}_{++-+-}\mathbf{S}^{w_3}_{+-+-+}\right\rangle=\partial_{y_2}\partial_{y_3}(X_2X_3)|_{y_i=0}=P_{w_1,w_2-1,w_3}P_{w_1,w_2,w_3-1}\ .
    \end{equation}
\end{itemize}
The correlator then reads 
\begin{equation}\label{OEE}
\begin{aligned}
        \mathcal{M}_{OEE}=\frac{C_{S^2}}{k^2}\Big[&(h_2-
    \frac{(k+2)w_2}{2}-j_2+1) (h_3-
    \frac{(k+2)w_3}{2}-j_3+1)P_{w_1,w_2+1,w_3}P_{w_1,w_2,w_3+1}\langle 0++\rangle\\
    &+(h_2-
    \frac{(k+2)w_2}{2}-j_2+1) (h_3-w_3)P_{w_1,w_2+1,w_3}P_{w_1,w_2,w_3-1}\langle 0+0 \rangle\\
    &+(h_2-w_2) (h_3-
    \frac{(k+2)w_3}{2}-j_3+1)P_{w_1,w_2-1,w_3}P_{w_1,w_2,w_3+1}\langle 00+\rangle\\
    &+(h_2-w_2) (h_3-w_3)P_{w_1,w_2-1,w_3}P_{w_1,w_2,w_3-1}\langle 000\rangle\Big]\times (\text{anti-homomorphic part})\,,
\end{aligned}
\end{equation}
where the ``anti-homomorphic part'' is again an expression obtained by replacing all $h$ in the square brackets above by  $\Bar{h}$. Here, we use $\langle 0++\rangle$ to denote $\langle V_{j_1,h_1}^{w_1}V_{j_2,h_2+1}^{w_2}V_{j_3,h_3+1}^{w_3}\rangle$, with the anti-holomorphic part not specified.   All other correlators in \eqref{OEE} are similarly defined.  The product of two such terms means specifying  both the holomorphic and 
anti-holomorphic dependence.

\subsection{Parity even}
When $\sum_iw_i$ is even, there are two possible types, namely E-O-O or E-E-E.

\subsubsection{E-O-O}\label{typeEOO}
    In this case, there are 3 possibilities:
\begin{enumerate}
    \item ``E'' is in the NS sector (\eqref{NSeven1},\eqref{NSeven2}) and the  two ``O'' are also in the NS sector (\eqref{oddNS});
    \item   ``E'' is in the NS sector (\eqref{NSeven1},\eqref{NSeven2}) and the  two ``O''  are in the R sector (\eqref{Rodd1},\eqref{Rodd2}); 
    \item ``E'' is in the R sector (\eqref{Reven1},\eqref{Reven2}) and one ``O'' is also in the R sector (\eqref{Rodd1},\eqref{Rodd2}), the other ``O'' is in the NS sector (\eqref{oddNS}).
\end{enumerate}
We  choose to calculate the  case 1 with the ``E'' being the  special physical operator \eqref{NSeven2}. In fact, the calculation of the second case is  similar with the one we will calculate in the type of E-E-E in section \ref{typeEEE}, and the calculation of the third case is similar with the case 1 here. So we omit the calculation for the latter 2 cases.
    Concretely, the three operators we choose are:
    \begin{equation}
    \begin{aligned}
        O_{j_1,h_1}^{w_1(-1)}&=e^{-\phi}( \alpha_- V_{j_1,h_1+1}^{w_1}\psi^{-,w_1}+\alpha_+ V_{j_1,h_1-1}^{w_1}\psi^{+,w_1}
    +\alpha_3 V_{j_1,h_1}^{w_1}\psi^{3,w_1})\,,\\
    O_{j_2,h_2}^{w_2(-1)}&=e^{-\phi}\textbf{1}_{\psi}^{w_2}V_{j_2,h_2}^{w_2},\quad O_{j_3,h_3}^{w_3(-1)}=e^{-\phi}\textbf{1}_{\psi}^{w_3}V_{j_3,h_3}^{w_3}\,,
    \end{aligned}
    \end{equation}
    where $w_1$ is even and $w_2$ and $w_3$ are odd. Notice that if we demand all the three operators corresponding to ground states, the above choice is the only one that makes the  correlator non-vanishing.
    
   For the total picture number being $-2$, we choose the third operator to be 
   in  picture $0$:
   \begin{align}
       O_{j_3,h_3}^{w_3(0)}&=\frac{1}{k}\Big[(h_3-\frac{(k+2)w_3}{2}-j_3+1) V_{j_3,h_3+1}^{w_3}\psi^{-,w_3}-2(h_3-w_3) V_{j_3,h_3-1}^{w_3}\psi^{+,w_3}\\
    &\qquad\qquad +(h_3-\frac{(k+2)w_3}{2}+j_3-1) V_{j_3,h_3}^{w_3}\psi^{3,w_3}\Big]\ .
   \end{align}
   Then there are 9 terms contribute to $\mathcal{M}_{EOO}$. Since $\langle V_{j_1,h_1}^{w_1}(0;0)V_{j_2,h_2}^{w_2}(1;1)V_{j_3,h_3}^{w_3}(\infty;\infty)\rangle$ is already known \cite{Dei:2021xgh}, we 
   simply compute explicitly the following 9 correlators of the fermionic multiplets  
   \begin{equation}
       \left\langle \psi^{a,w_1}(0;0)\textbf{1}_\psi^{w_2}(1;1)\psi^{b,w_3}(\infty;\infty)\right\rangle, \qquad a,b=3,\pm\ .
   \end{equation}
    Now we have $\sum_i w_i\in2\mathbb{Z}$ and $j_1=j_3=-1$, $j_2=0$, so the  $y$-basis correlator is simply
    \begin{equation}
        X_{13}^2=(P_{w_1+1,w_2,w_3+1}+P_{w_1-1,w_2,w_3+1}y_1+P_{w_1+1,w_2,w_3-1}y_3+P_{w_1-1,w_2,w_3-1}y_1y_3)^2\ .
    \end{equation}
    Reading out the residues, we get (in the following, we omit the same overall factor $k$ for all the correlators)
   \begin{itemize}
       \item $(a,b)=(-,-)$\\
       \begin{equation}
           \left\langle \psi^{-,w_1}\textbf{1}_\psi^{w_2}\psi^{-,w_3}\right\rangle=X_{13}^2|_{y_i=0}=P^2_{w_1+1,w_2,w_3+1}\ . 
       \end{equation}
       \item $(a,b)=(-,3)$\\
      \begin{equation}
           \left\langle \psi^{-,w_1}\textbf{1}_\psi^{w_2}\psi^{3,w_3}\right\rangle=-\frac{1}{2}\times \partial_{y_3} X_{13}^2|_{y_i=0}=-P_{w_1+1,w_2,w_3+1}P_{w_1+1,w_2,w_3-1}\ .
       \end{equation}
       \item $(a,b)=(-,+)$\\
      \begin{equation}
           \left\langle \psi^{-,w_1}\textbf{1}_\psi^{w_2}\psi^{+,w_3}\right\rangle=\frac{1}{2}\partial_3^2X_{13}^2|_{y_i=0}=P^2_{w_1+1,w_2,w_3-1}\ . 
       \end{equation}
       \item $(a,b)=(3,3)$
\CP{}
       \begin{equation}
       \begin{aligned}
            \left\langle \psi^{3,w_1}\textbf{1}_\psi^{w_2}\psi^{3,w_3}\right\rangle
            =&\left(-\frac{1}{2}\right)^2\partial_{y_1}\partial_{y_3}X_{13}^2|_{y_i=0}\\
            =&\frac{1}{2}P_{w_1+1,w_2,w_3+1}P_{w_1-1,w_2,w_3-1}+\frac{1}{2}P_{w_1+1,w_2,w_3-1}P_{w_1-1,w_2,w_3+1}\ .
       \end{aligned}
       \end{equation}
       \item $(a,b)=(3,+)$\\
\CP{}
       \begin{equation}
            \left\langle \psi^{3,w_1}\textbf{1}_\psi^{w_2}\psi^{+,w_3}\right\rangle=-\frac{1}{2}\times \frac{1}{2!}\partial_{y_1}\partial^2_{y_3}X_{13}^2|_{y_i=0}
            =-P_{w_1+1,w_2,w_3-1}P_{w_1-1,w_2,w_3-1}\ .
       \end{equation}
       \item $(a,b)=(+,+)$\\
\CP{}
       \begin{equation}
            \left\langle \psi^{+,w_1}\textbf{1}_\psi^{w_2}\psi^{+,w_3}\right\rangle=\left(\frac{1}{2!}\right)^2\partial^2_{y_1}\partial^2_{y_3}X_{13}^2|_{y_i=0}
            =P^2_{w_1-1,w_2,w_3-1}\ .
       \end{equation}
   \end{itemize}
Notice that we omit 
all coordinates in the above expressions.
With all these results, the correlator reads
\begin{equation}\label{EOO}
\begin{aligned}
    \mathcal{M}_{EOO}=C_{S^2}\Big\{\alpha_-\Big[  &P^2_{w_1+1,w_2,w_3+1}(h_3-\frac{(k+2)w_3}{2}-j_3+1)\langle +0+\rangle\\
    +&2P_{w_1+1,w_2,w_3+1}P_{w_1+1,w_2,w_3-1}(h_3-w_3)\langle +00\rangle\\
    +&P^2_{w_1+1,w_2,w_3-1}(h_3-\frac{(k+2)w_3}{2}+j_3-1)\langle +0-\rangle\Big]\\
    -\alpha_3\Big[  &P_{w_1+1,w_2,w_3+1}P_{w_1-1,w_2,w_3+1}(h_3-\frac{(k+2)w_3}{2}-j_3+1)\langle 00+\rangle\\
    +& (P_{w_1+1,w_2,w_3+1}P_{w_1-1,w_2,w_3-1}+P_{w_1+1,w_2,w_3-1}P_{w_1-1,w_2,w_3+1})(h_3-w_3)\langle 000\rangle\\
    +&P_{w_1+1,w_2,w_3-1}P_{w_1-1,w_2,w_3-1}(h_3-\frac{(k+2)w_3}{2}+j_3-1)\langle 00-\rangle\Big]\\
    +\alpha_+\Big[  &P^2_{w_1-1,w_2,w_3+1}(h_3-\frac{(k+2)w_3}{2}-j_3+1)\langle -0+\rangle\\
    +&2P_{w_1-1,w_2,w_3-1}P_{w_1-1,w_2,w_3+1}(h_3-w_3)\langle -00\rangle\\
    +&P^2_{w_1-1,w_2,w_3-1}(h_3-\frac{(k+2)w_3}{2}+j_3-1)\langle -0-\rangle\Big]\Big\}\times \text{(anti-homomorphic part)}.
\end{aligned}
\end{equation}
where $(\alpha_-,\alpha_3,\alpha_+)=(j_1-m_1,0,j_1+m_1)$, up to the spurious polarization  $(\alpha_-,\alpha_3,\alpha_+)_{spurious}=(m_1-j_1+1,-2m_1-w_1k,m_1+j_1-1)$.
Again the ``anti-homomorphic part'' is  an expression obtained by replacing all $h$ in the brace  above by  $\Bar{h}$.

\subsubsection{E-E-E}\label{typeEEE}
  In this case, there are 2 possibilities:
\begin{enumerate}
    \item One ``E'' is in the NS sector (\eqref{NSeven1},\eqref{NSeven2}) and the other  two ``E'' are  in the R sector (\eqref{Reven1},\eqref{Reven2});
    \item  All three ``E'' are in the NS sector (\eqref{NSeven1},\eqref{NSeven2}). 
\end{enumerate}
We  choose to calculate the  case 1 with the ``E'' in the NS sector being the  special physical operator \eqref{NSeven2}. In fact, the calculation of the second case is  similar with the case 1 in the type of E-O-O.\footnote{There is a special case where all the three operators are the special one \eqref{NSeven2}. While in other cases one can always avoid to calculate correlator with the picture 0 version of \eqref{NSeven2} (by using a suitable picture choice), in this case one must calculate such correlators. These correlators are  more complicated because they are correlators of spectrally flowed  operators with descendant insertions. See the last paragraph in section \ref{picturechanging}. } So we omit the calculation for the case 2 here.
    Concretely, the three operators we choose are:
 \begin{equation}
    \begin{aligned}
        O_{j_1,h_1}^{w_1(-1)}&=e^{-\phi}( \alpha_- V_{j_1,h_1+1}^{w_1}\psi^{-,w_1}+\alpha_+ V_{j_1,h_1-1}^{w_1}\psi^{+,w_1}
    +\alpha_3 V_{j_1,h_1}^{w_1}\psi^{3,w_1})\\
   O_{j_2,h_2}^{w_2(-\frac{3}{2})}&=e^{-\frac{3\phi}{2}}\mathbf{S}^{w_2}_{++++-}V_{j_2,h_2}^{w_2},\quad O_{j_3,h_3}^{w_3(-\frac{3}{2})}=e^{-\frac{3\phi}{2}}\mathbf{S}^{w_3}_{+---+}V_{j_3,h_3}^{w_3}\ .
    \end{aligned}
    \end{equation}
Again we choose the first operator to be 
in  picture $(-1)$ and the other two lie in  picture $(-\frac{1}{2})$ (see \eqref{picture-1/2}).  Then there are 12  correlators of the fermionic multiplets we need to calculate
\begin{equation}
    \left\langle \psi^{e,w_1}(0;0)\mathbf{S}^{w_2}_{a+b+-}(1;1)\mathbf{S}^{w_3}_{c-d-+}(\infty,\infty)\right\rangle,\qquad e=\pm,3, \quad (a,b)=(\pm,\mp), \quad (c,d)=(\pm,\pm)\ .
\end{equation}
  Now we have $\sum_i w_i\in2\mathbb{Z}$ and $j_1=-1, j_2=j_3=-\frac{1}{2}$ and the correlator in the $y$-basis is 
    \begin{equation}
    \begin{aligned}
        X_{12}X_{13}=&(P_{w_1+1,w_2+1,w_3}+P_{w_1-1,w_2+1,w_3}y_1+P_{w_1+1,w_2-1,w_3}y_2+P_{w_1-1,w_2-1,w_3}y_1y_2)\\
        &\times(P_{w_1+1,w_2,w_3+1}+P_{w_1-1,w_2,w_3+1}y_1+P_{w_1+1,w_2,w_3-1}y_3+P_{w_1-1,w_2,w_3-1}y_1y_3)\ .
    \end{aligned}
    \end{equation}
We have respectively the following contributions (in the following, we omit the same overall factor $\sqrt{k}$ for all the correlators)
\begin{itemize}
    \item $e=-$, $(a,b)=(-,+)$, $(c,d)=(-,-)$
    \begin{equation}
         \left\langle \psi^{-,w_1}\mathbf{S}^{w_2}_{-+++-}\mathbf{S}^{w_3}_{----+}\right\rangle=(X_{12}X_{13})|_{y_i=0}=P_{w_1+1,w_2+1,w_3}P_{w_1+1,w_2,w_3+1}\ .
    \end{equation}
     \item $e=-$, $(a,b)=(-,+)$, $(c,d)=(+,+)$
    \begin{equation}
         \left\langle \psi^{-,w_1}\mathbf{S}^{w_2}_{-+++-}\mathbf{S}^{w_3}_{+-+-+}\right\rangle=\partial_{y_3}(X_{12}X_{13})|_{y_i=0}=P_{w_1+1,w_2+1,w_3}P_{w_1+1,w_2,w_3-1}\ .
    \end{equation}
    \item $e=-$, $(a,b)=(+,-)$, $(c,d)=(-,-)$
    \begin{equation}
         \left\langle \psi^{-,w_1}\mathbf{S}^{w_2}_{++-+-}\mathbf{S}^{w_3}_{----+}\right\rangle=\partial_{y_2}(X_{12}X_{13})|_{y_i=0}=P_{w_1+1,w_2-1,w_3}P_{w_1+1,w_2,w_3+1}\ .
    \end{equation}
    \item $e=-$, $(a,b)=(+,-)$, $(c,d)=(+,+)$
    \begin{equation}
         \left\langle \psi^{-,w_1}\mathbf{S}^{w_2}_{++-+-}\mathbf{S}^{w_3}_{----+}\right\rangle=\partial_{y_2}\partial_{y_3}(X_{12}X_{13})|_{y_i=0}=P_{w_1+1,w_2-1,w_3}P_{w_1+1,w_2,w_3-1}\ .
    \end{equation}
    \item $e=3$, $(a,b)=(-,+)$, $(c,d)=(-,-)$
    \begin{equation}
    \begin{aligned}
        \left\langle \psi^{3,w_1}\mathbf{S}^{w_2}_{-+++-}\mathbf{S}^{w_3}_{----+}\right\rangle&=\partial_{y_1}(X_{12}X_{13})|_{y_i=0}\\
        &=P_{w_1+1,w_2+1,w_3}P_{w_1-1,w_2,w_3+1}+P_{w_1-1,w_2+1,w_3}P_{w_1+1,w_2,w_3+1}\ .
    \end{aligned}
    \end{equation}
     \item $e=3$, $(a,b)=(-,+)$, $(c,d)=(+,+)$
    \begin{equation}
    \begin{aligned}
              \left\langle \psi^{3,w_1}\mathbf{S}^{w_2}_{-+++-}\mathbf{S}^{w_3}_{+-+-+}\right\rangle&=\partial_{y_1}\partial_{y_3}(X_{12}X_{13})|_{y_i=0}
              \\
              &=P_{w_1-1,w_2+1,w_3}P_{w_1+1,w_2,w_3-1}+P_{w_1+1,w_2+1,w_3}P_{w_1-1,w_2,w_3-1}\ .
    \end{aligned}
    \end{equation}
    \item $e=3$, $(a,b)=(+,-)$, $(c,d)=(-,-)$
    \begin{equation}
    \begin{aligned}
         \left\langle \psi^{3,w_1}\mathbf{S}^{w_2}_{++-+-}\mathbf{S}^{w_3}_{----+}\right\rangle&=\partial_{y_1}\partial_{y_2}(X_{12}X_{13})|_{y_i=0}\\&=P_{w_1+1,w_2-1,w_3}P_{w_1-1,w_2,w_3+1}+P_{w_1-1,w_2-1,w_3}P_{w_1+1,w_2,w_3+1}\ .
    \end{aligned}
    \end{equation}
    \item $e=3$, $(a,b)=(+,-)$, $(c,d)=(+,+)$
    \begin{equation}
    \begin{aligned}
        \left\langle \psi^{3,w_1}\mathbf{S}^{w_2}_{++-+-}\mathbf{S}^{w_3}_{----+}\right\rangle&=\partial_{y_1}\partial_{y_2}\partial_{y_3}(X_{12}X_{13})|_{y_i=0}\\
        &=P_{w_1-1,w_2-1,w_3}P_{w_1+1,w_2,w_3-1}+P_{w_1+1,w_2-1,w_3}P_{w_1-1,w_2,w_3-1}\ .
    \end{aligned}
    \end{equation}
    \item $e=+$, $(a,b)=(-,+)$, $(c,d)=(-,-)$
    \begin{equation}
         \left\langle \psi^{+,w_1}\mathbf{S}^{w_2}_{-+++-}\mathbf{S}^{w_3}_{----+}\right\rangle=\frac{1}{2}\partial^2_{y_1}(X_{12}X_{13})|_{y_i=0}=P_{w_1-1,w_2+1,w_3}P_{w_1-1,w_2,w_3+1}\ .
    \end{equation}
     \item $e=+$, $(a,b)=(-,+)$, $(c,d)=(+,+)$
    \begin{equation}
         \left\langle \psi^{+,w_1}\mathbf{S}^{w_2}_{-+++-}\mathbf{S}^{w_3}_{+-+-+}\right\rangle=\frac{1}{2}\partial^2_{y_1}\partial_{y_3}(X_{12}X_{13})|_{y_i=0}=P_{w_1-1,w_2+1,w_3}P_{w_1-1,w_2,w_3-1}\ .
    \end{equation}
    \item $e=+$, $(a,b)=(+,-)$, $(c,d)=(-,-)$
    \begin{equation}
         \left\langle \psi^{+,w_1}\mathbf{S}^{w_2}_{++-+-}\mathbf{S}^{w_3}_{----+}\right\rangle=\frac{1}{2}\partial^2_{y_1}\partial_{y_2}(X_{12}X_{13})|_{y_i=0}=P_{w_1-1,w_2-1,w_3}P_{w_1-1,w_2,w_3+1}\ .
    \end{equation}
    \item $e=+$, $(a,b)=(+,-)$, $(c,d)=(+,+)$
    \begin{equation}
         \left\langle \psi^{+,w_1}\mathbf{S}^{w_2}_{++-+-}\mathbf{S}^{w_3}_{----+}\right\rangle=\frac{1}{2}\partial^2_{y_1}\partial_{y_2}\partial_{y_3}(X_{12}X_{13})|_{y_i=0}=P_{w_1-1,w_2-1,w_3}P_{w_1-1,w_2,w_3-1}\ .
    \end{equation}
\end{itemize}
Now we can straightforwardly write  $\mathcal{M}_{EEE}$ , similar as the other 3 cases discussed above. Since the expression is lengthy and not instructive, we omit it here.

\subsection{Picture choices and recursion relations}\label{picturechanging}

In this section, we discuss  the equivalence of different choices of pictures  in supersymmetric correlators. In fact, these   equivalences  can be verified by the recursion relations of the bosonic $SL(2,R)$ WZW model found in \cite{Eberhardt:2019ywk}, which relates three point functions with different $h_i (i=1,2,3)$ (we will review their detailed form soon). Conversely, one may understand the recursion relations in \cite{Eberhardt:2019ywk}  from the equivalence of the  different picture  choices. In the following, we will first review and complete the recursion relations in \cite{Eberhardt:2019ywk,Dei:2021xgh,Bufalini:2022toj} by introducing several new ones in the 
``edge" cases (see \eqref{thecondition} for details). 
As we will see later,  in the special case when $w_i$ are all odd and satisfy $w_i+w_j=w_k-1$ for one $(i,j,k)$
(see case $\mathrm{II}$ below), the  recursion relations are the same as the equivalence of different choices of pictures (while generally they are only related but not the same).  We expect the equivalence of different choices of pictures for arbitrary $n$-point correlators are also related to the recursion relations\footnote{Notice that while it is known that  recursion relations  also exist for $n\geq 4$-point correlators, analytic closed forms for them are not known. }, although here we only focus on the case of 3-point correlators.

Firstly, let's describe the recursion relations for all possible configurations of $w_i, i=1,2,3$. Recursion relations generally exist for (depending on the total parity of the spectral parameter $\sum_{i}w_i$)
\begin{align}
\begin{aligned}
    \text{When} \quad \sum_{i=1}^3 w_i&\in2\mathbb{Z}+1: \qquad  \sum_{i\neq j}w_i\geq w_j-1 \qquad  (j=1,2,3)\,,\\
    \text{When} \quad \sum_{i=1}^3 w_i&\in2\mathbb{Z}: \qquad  \sum_{i\neq j}w_i\geq w_j \qquad  (j=1,2,3)\,\label{constn}
\end{aligned}
\end{align}
since correlators that violate this condition simply vanish \cite{Maldacena:2001km, Eberhardt:2019ywk}. For each case of the total parity $\sum_{i}w_i$, we further break our discussion into 2 cases, depending on the saturation of~\eqref{constn} 
\cite{Bufalini:2022toj}:
\begin{equation}\label{thecondition}
\begin{aligned}
    \mathrm{I} :& \quad  \sum_{i=1}^3 w_i\in2\mathbb{Z}+1, \qquad w_i+w_j\geq w_k+1 \quad \text{for all triples  $(i,j,k)$, this is the general cases,} \\
     \mathrm{II} :& \quad  \sum_{i=1}^3 w_i\in2\mathbb{Z}+1, \qquad w_i+w_j= w_k-1 \quad \text{for one triple $(i,j,k)$, this is the edge cases,}\\
      \mathrm{III} :& \quad  \sum_{i=1}^3 w_i\in2\mathbb{Z}, \qquad w_i+w_j\geq w_k+2 \quad \text{for all triples $(i,j,k)$, this is the general cases,}\\
         \mathrm{IV} :& \quad  \sum_{i=1}^3 w_i\in2\mathbb{Z}, \qquad w_i+w_j= w_k \quad \text{for one triple $(i,j,k)$, this is the edge cases.}
\end{aligned}
\end{equation}
For the general cases $\mathrm{I}$ and $\mathrm{III}$,  closed formulas of  differential equations satisfied by correlators in the $y$ basis are obtained in \cite{Bufalini:2022toj} (eq. (3.14), (3.32), (3.33), (3.34) there). One can easily transform them into recursion relations using the following rules \cite{Eberhardt:2019ywk}\footnote{These rules can be read from the definition of the OPEs in $y$-basis,  or by using integration by parts based on the (inverse) $y$-transform \cite{Dei:2021xgh}.}
\begin{equation}\label{replace rule}
\begin{aligned}
      (h_i-\frac{k+2}{2}w_i+j_i-1)\langle h_i-1\rangle &\longleftrightarrow y_i(y_i\partial_{y_i}+2j_i)\langle ...\rangle_y\\   
      h_i\langle ...\rangle&\longleftrightarrow \left(y_i\partial_{y_i}+j_i+\frac{k+2}{2}w_i\right)\langle ...\rangle_y\\
      (h_i-\frac{k+2}{2}w_i-j_i+1)\langle h_i+1\rangle&\longleftrightarrow \partial_{y_i}\langle ...\rangle_y\,,
\end{aligned} 
\end{equation}
where the level of the model is $k+2$ and we use a subscript ``$y$'' to denote the corresponding correlator in the $y$ basis. 
Thus, the  recursion relations for the case $\mathrm{I}$ are\footnote{See  Eq. (3.15) in \cite{Dei:2021xgh}. However there are typos in  Eq. (3.15) in  \cite{Dei:2021xgh}, which is corrected in Eq. (3.14) in \cite{Bufalini:2022toj} (there is a typo even in this equation: $-\frac{k}{2}\to +\frac{k}{2}$).  In the following equation, We have corrected the typos.}
\begin{equation}\label{recursion}
\begin{aligned}
        a_i^{-1}(h_i-\frac{(k+2)w_i}{2}+j_i-1)\langle h_i-1\rangle=\left(\frac{w_i}{N}-1\right)a_i
        (h_i-\frac{(k+2)w_i}{2}-j_i+1)\langle h_i+1\rangle\\
        +\sum_{l=1,2}\frac{w_{i}}{N}a_{i+l}
        (h_{i+l}-\frac{(k+2)w_{i+l}}{2}-j_{i+l}+1)\langle h_{i+l}+1\rangle
        +\left(-\frac{w_i}{N}\sum_{l=1}^3 h_l+2h_i\right)\langle ...\rangle\ .
\end{aligned}
\end{equation}
where indices are understood to be mod 3, $N=\frac{1}{2}\sum_i(w_i-1)+1$ and $a_i$  are related to the functions $P$ and can be written as:
\begin{equation}\label{aiP}
    a_i=-\frac{P_{\boldsymbol{w}+\boldsymbol{e}_i}}{P_{\boldsymbol{w}-\boldsymbol{e}_i}}=\frac{\left(
    \begin{aligned}
        \frac{1}{2}(w_i+w_{i+1}+w_{i+2}-1)\\
        \frac{1}{2}(-w_i+w_{i+1}+w_{i+2}-1)
    \end{aligned}
    \right)}{\left(
    \begin{aligned}
        \frac{1}{2}(-w_i+w_{i+1}-w_{i+2}-1)\\
        \frac{1}{2}(w_i+w_{i+1}-w_{i+2}-1)\ .
    \end{aligned}
    \right)}
\end{equation}
In fact, $a_i$ are the Taylor coefficients (see \eqref{covering}) of the unique covering map  that appears in the three-point function case \cite{Eberhardt:2019ywk,Dei:2021xgh}(see also \cite{Lunin:2000yv} for the explicit construction of the covering map with three ramified points).
Similarly, the recursion relation for the case $\mathrm{III}$ is (for $i=1$)
\begin{equation}\label{case3}
\begin{aligned}
        a_1[\Gamma_3^-]^{-1}(h_1-&\frac{(k+2)w_1}{2}+j_1-1)\langle h_1-1\rangle=\left(\frac{w_1}{N'}-1\right)a_1[\Gamma_3^-]
        (h_1-\frac{(k+2)w_1}{2}-j_1+1)\langle h_1+1\rangle\\
        +&\frac{w_{1}}{N'}a_{2}[\Gamma_3^-]
        (h_{2}-\frac{(k+2)w_{2}}{2}-j_{2}+1)\langle h_{2}+1\rangle\\
        -&\frac{w_{1}}{N'}a_{3}[\Gamma_2^-]
        (h_{3}-\frac{(k+2)w_{3}}{2}-j_{3}+1)\langle h_{3}+1\rangle
        +\left[-\frac{w_1}{N'}(h_1+h_2-h_3)+2h_1\right]\langle ...\rangle \ .
\end{aligned}
\end{equation}
where $N'=\frac{1}{2}(w_1+w_2-w_3)$ and $a_i[\Gamma_j^-]$ is the $a_i$ coefficients of the covering map $\Gamma$ with $w_j$ shifted to be $w_j-1$ \cite{Bufalini:2022toj}. Recursion relations for $i=2,3$ can be obtained by changing all the subscripts as $1\leftrightarrow 2$ and $1\leftrightarrow 3$ respectively. 

Now there remains two edge cases $\mathrm{II}$ and $\mathrm{IV}$. In \cite{Bufalini:2022toj}, differential equations are also obtained for these cases. However, these equations are for the correlators with a  different configuration of $x_i$:
$(x_1,x_2,x_3)=(0,0,\infty)$. Here we will give differential equations or recursion relations for the standard configuration $(x_1,x_2,x_3)=(0,1,\infty)$, just as in the cases $\mathrm{I}$ and $\mathrm{III}$.  Firstly, notice that the reason why the edge cases are special is that the recursion relations for the general cases are not well-defined for the edge cases. Let's consider the case $\mathrm{II}$ for an illustration.  Without loss of generality, we assume $w_1+w_2=w_3-1$, then we have:
\begin{equation}\label{vanishingP}
    P_{w_1-1,w_2,w_3}=P_{w_1,w_2-1,w_3}=P_{w_1,w_2,w_3+1}=0
\end{equation}
Since $a_i=-\frac{P_{\boldsymbol{w}+\boldsymbol{e}_i}}{P_{\boldsymbol{w}-\boldsymbol{e}_i}}$,   the denominators of $a_i$ or $1/a_i$ will be zero, and both of them cause ill-definedness\footnote{This means the corresponding covering map  does not exist. Thus, the method making use of the covering map in \cite{Bufalini:2022toj} does not work and the author of \cite{Bufalini:2022toj} choose to consider a different configuration $(x_1,x_2,x_3)=(0,0,\infty)$.}. Nevertheless,  we observe that the singular terms in each equation are at the same order, which means we can multiply a  denominator to make all the singular terms finite. However, the vanishing denominators could be anyone of the 3 cases in  \eqref{vanishingP}. Thus in practice, we can multiply all terms with a factor: $w_1P_{w_1-1,w_2,w_3}P_{w_1+1,w_2,w_3}$. This works nicely because of the following observed identity (which can be easily proven  by showing the quotient of any two terms in the following is $1$)
\begin{equation}\label{symmetricfactor}
    w_1P_{w_1-1,w_2,w_3}P_{w_1,w_2+1,w_3}= w_2P_{w_1,w_2-1,w_3}P_{w_1,w_2+1,w_3}=w_3P_{w_1,w_2,w_3-3}P_{w_1,w_2,w_3+1}
\end{equation}
which means the multiplied factor is symmetric in the three indices (this fact will also be very useful in other places in this work).
Then we can freely choose the form of this factor to cancel any vanishing denominators.
The resulting recursion relations are
\begin{equation}\label{reIII}
    \begin{aligned}
     &\text{For $i=1$}:\\
     &\quad  0=P_{w_1+1,w_2,w_3}^2(h_1-\frac{k+2}{2}w_1-j_1+1)\langle h_1+1\rangle-P_{w_1,w_2+1,w_3}^2(h_2-\frac{k+2}{2}w_2-j_2+1)\langle h_2+1\rangle\,,\\
   &\text{For $i=2$}:\\
   & \quad 0=P_{w_1,w_2+1,w_3}^2(h_2-\frac{k+2}{2}w_2-j_2+1)\langle h_2+1\rangle-P_{w_1+1,w_2,w_3}^2(h_1-\frac{k+2}{2}w_1-j_1+1)\langle h_1+1\rangle \,, \\
   &\text{For $i=3$}:\\
   &\quad  P_{w_1,w_2,w_3-1}^2(h_3-\frac{k+2}{2}w_3+j_3-1)\langle h_3-1\rangle=\frac{w_1}{w_1+w_2}P_{w_1+1,w_2,w_3}^2(h_1-\frac{k+2}{2}w_1-j_1+1)\langle h_1+1\rangle\\
   &\quad +\frac{w_2}{w_1+w_2}P_{w_1,w_2+1,w_3}^2(h_2-\frac{k+2}{2}w_2-j_2+1)\langle h_2+1\rangle\ .
\end{aligned}
\end{equation}
It is obvious that  $i=1$ and $i=2$ give  the same equation. So there are 2 independent equations. One can also do the same calculation for the case $\mathrm{IV}$. Again without loss of generality, we assume $w_1+w_2=w_3$.  There is a difference in this case: from \eqref{case3} (for $i=1$) one can see that  the divergence on the right hand side comes form the factor $N'$, rather than  $a_1[\Gamma^-_3]$, $a_2[\Gamma^-_3]$ or $a_3[\Gamma^-_2]$ (they are finite, in particular, $a_3[\Gamma^-_2]=0$).  The same type of divergence happens for $i=2$, while  for  $i=3$, the divergences come from $a_3[\Gamma_1^-]^{-1}$, $a_2[\Gamma_3^-]$ and $a_3[\Gamma_2^-]$ (similar with the one in the edge case $\mathrm{II}$). Again with the help of the identity \eqref{symmetricfactor}, we can multiply a factor to cancel  all the divergences in the recursion relations. The result are
\begin{equation}
    \begin{aligned}
     &\text{For $i=1$}:\\
     &\quad  0=w_1a_1[\Gamma_3^-](h_1-\frac{k+2}{2}w_1 -j_1+1)\langle h_1+1\rangle-w_1a_2[\Gamma_3^-](h_2-\frac{k+2}{2}w_2-j_2+1)\langle h_2+1\rangle\\
     &\qquad +w_1(h_1+h_2-h_3)\langle ...\rangle\,,\\
   &\text{For $i=2$}:\\
   &\quad  0=w_2a_2[\Gamma_3^-](h_2-\frac{k+2}{2}w_2 -j_2+1)\langle h_2+1\rangle-w_2a_1[\Gamma_3^-](h_1-\frac{k+2}{2}w_1-j_1+1)\langle h_1+1\rangle\\
     &\qquad +w_2(h_1+h_2-h_3)\langle ...\rangle\,,\\
   &\text{For $i=3$}:\\
   &\quad  w_3P_{w_1-1,w_2,w_3-1}^2(h_3-\frac{k+2}{2}w_3+j_3-1)\langle h_3-1\rangle=w_3P_{w_1-1,w_2+1,w_3}^2(h_2-\frac{k+2}{2}w_2-j_2+1)\langle h_2+1\rangle\\
   &\quad -w_3P_{w_1+1,w_2-1,w_3}P_{w_1-1,w_2+1,w_3}(h_1-\frac{k+2}{2}w_1-j_1+1)\langle h_1+1\rangle\ .
\end{aligned}
\end{equation}
Similar with the edge case  $\mathrm{II}$, the first two equations are equivalent and we only obtain two independent recursion relations for the edge case  $\mathrm{IV}$.
For both the two edge cases $\mathrm{II}$ and $\mathrm{IV}$, we have checked that the above  recursion relations or their corresponding differential equations are indeed satisfied by the 3-point function \eqref{ybasis3pt}. Our analysis here shows that the edge cases  can be seen as the limiting cases of the general cases\footnote{This should be expected because the 3-point functions in the edge cases, being solutions of the recursion relations, have the same form as the one in the general cases.}.  

Now we discuss the relation between the choices of pictures in 
the superstring correlators and the recursion relations of correlators in the bosonic $SL(2, R)$ WZW model.
Firstly, consider the case of O-O-O.   
Substitute \eqref{recursion} into \eqref{OOO} (with $i=3$), we can write the three point correlator O-O-O as
\begin{equation}
\begin{aligned}
    \mathcal{M}_{OOO}=&\frac{C_{S^2}}{k} \Big\{P^2_{w_1,w_2,w_3-1}a_3\Bigg[\left(\frac{w_3}{N}-1\right)a_3
        (h_3-\frac{(k+2)w_3}{2}-j_3+1)\langle h_3+1\rangle_3\\
        &+\frac{w_{3}}{N}a_{1}
        (h_{1}-\frac{(k+2)w_1}{2}-j_{1}+1)\langle h_{1}+1\rangle_3
        +\frac{w_{3}}{N}a_{2}
        (h_{2}-\frac{(k+2)w_{2}}{2}-j_{2}+1)\langle h_{2}+1\rangle_3\\
        &+\left(-\frac{w_3}{N}\sum_{l=1}^3 h_l+2h_3\right)\langle ...\rangle_3  \Bigg]\\
        &+(h_3-\frac{(k+2)w_3 }{2}-j_3+1)P^2_{w_1,w_2,w_3+1}\langle h_3+1\rangle_3
    +2(h_3-w_3)P_{w_1,w_2,w_3-1}P_{w_1,w_2,w_3+1} .\langle ...\rangle_3\ \Big\}\\
    &\times (\text{anti-homomorphic part}).
\end{aligned}
\end{equation}
Using the relation \eqref{aiP}, one finds
\begin{equation}\label{OOO2}
\begin{aligned}
    \mathcal{M}_{OOO}=\frac{C_{S^2}}{k}&\Big\{\frac{w_3}{N}P_{w_1,w_2,w_3-1}P_{w_1,w_2,w_3+1}\Big[-\sum_{i=1}^3 a_{i}
        (h_{i}-\frac{kw_{i}}{2}-j_{i}+1)\langle h_{i}+1\rangle_3
        \\&+(h_1+h_2+h_3-2N)\langle ...\rangle_3\Big] \Big\}\times (\text{anti-homomorphic part}).
\end{aligned}
\end{equation}
The  above expression  is clearly symmetric in the 3 subscripts $1,2,3$ (recall the identities \eqref{symmetricfactor}).   So no matter which of the  
first, second or third operator to be in its picture 0 version, we will get the same result for the correlator.

One can  try to reverse the above discussion. In the following, we will keep the right-moving dependence to be arbitrary (as long as satisfying the level-matching), so that we can obtain equations that only depend on the left-moving part. In fact, for the general case $\mathrm{I}$, equivalence of different picture choices is simply the difference of two  recursion relations. Explicitly, by this we mean that the equality of the correlator $\mathcal{M}_{OOO}$ \eqref{OOO} (where the third operator is in the picture 0) and the one that with the second  operator in the picture 0  can be written as:
\begin{equation}
\begin{aligned}
    & (h_2-
    \frac{(k+2)w_2}{2}+j_2-1)P^2_{w_1,w_2-1,w_3}\langle h_2-1\rangle-(h_3-
    \frac{(k+2)w_3}{2}+j_3-1)P^2_{w_1,w_2,w_3-1}\langle h_3-1\rangle\\
    =&(h_3-\frac{(k+2)w_3 }{2}-j_3+1)P^2_{w_1,w_2,w_3+1}\langle h_3+1\rangle-(h_2-\frac{(k+2)w_2 }{2}-j_2+1)P^2_{w_1,w_2+1,w_3}\langle h_2+1\rangle\\
    &+2(h_3P_{w_1,w_2,w_3-1}P_{w_1,w_2,w_3+1}-h_2P_{w_1,w_2-1,w_3}P_{w_1,w_2+1,w_3})\langle ...\rangle
\end{aligned}
\end{equation}
where we have used \eqref{symmetricfactor}. Then after dividing the two sides by the factor \eqref{symmetricfactor} (notice that it is symmetric in the 3 indices) and use the relation between $a_i$ and $P$ \eqref{aiP}, the above equation just becomes
\begin{equation}
    \frac{1}{w_3}\times \eqref{recursion}(\text{for $i=3$})- \frac{1}{w_2}\times \eqref{recursion}(\text{for $i=2$})\ .
\end{equation}
A similar relation holds if we choose the first operator in picture 0. 
This is reasonable that reversely one cannot obtain the recursion relations but only their difference:  from equality of different picture choices one get 2 independent equations, while there are 3 independent recursion relations for the case $\mathrm{I}$. 
 Since we are considering the case of O-O-O, the edge case $\mathrm{III}$ is also possible. Superstring correlators of  this edge case is  simpler: for each picture choice, only one term in \eqref{OOO} does not vanish. Since there are only 2 independent recursion relations in the case $\mathrm{III}$, one can check that the equalities of different picture choices are precisely the recursion relations \eqref{reIII} in the edge case $\mathrm{III}$.

We  expect that similar relations between recursion relations and equalities of different picture choices hold for the cases E-E-O, E-O-O, E-E-E, and even for arbitrary $n$-point functions as well. In these cases, however, their relation will not be as clear as in the case of O-O-O. The reason is two-folded; on the one hand, correlators in these three cases include more bosonic correlators (of the $SL(2,R)$ WZW model) than in the case of O-O-O, 
which indicates that more sophisticated use of recursion relations is required to get them straight. As an illustration, we demonstrate this analysis with an example in appendix \ref{pictureEOO};  on the other hand, picture changing  sometimes leads to correlators of spectrally flowed operators (of long strings) with descendant insertions. For example, the picture 0 version of the physical  operator \eqref{NSeven2} is (written in the $m$-basis and omitting the $z$ coordinate)
\begin{equation}
\begin{aligned}
    O_{j,m}^{w(0)}(z)=\frac{1}{k}\Big\{\alpha_-[(m+j)V^w_{j,m}\psi^{+,w}\psi^{-,w}+k(j^{-,w}V^w_{j,m+1})-2(m+1+\frac{wk}{2})V^w_{m+1}\psi^{3,w}\psi^{-,w}]\\
    + \alpha_3 [(m-j+1)V_{j,m+1}^w\psi^{-,w}\psi^{3,w}+(m+j-1)V_{j,m-1}\psi^{-,w}\psi^{3,w}+k(j^{3,w}V^w_{j,m})+\frac{k^2w}{2}V^w_{j,m}]\\
      +\alpha_+[(m-j)V^w_{j,m}\psi^{-,w}\psi^{+,w}+k(j^{+,w}V^w_{j,m-1})-2(m-1+\frac{kw}{2})V^w_{m-1}\psi^{3,w}\psi^{+,w}]\Big\}\,,
\end{aligned}
 \end{equation}
 where operators $j^{a,w}V^w_{j,m-a}$ appear. Three point correlators involving such operators cannot be read off directly from the closed formula in \cite{Dei:2021xgh}. The computation for them is probably not easy: for short strings, see  \cite{Iguri:2022pbp,Iguri:2023khc} for the calculation of such correlators  using the series identifications. In the case at hand for long strings, the equivalence conditions of  different picture choices give some relations or constraints among correlators of this type. 
 For example, other than \eqref{EOO} we can alternatively choose the first operator in  picture 0 and the other two in  picture $(-1)$. The fact that these two choices of picture numbers give identical correlators indicates the existence of non-trivial relations 
 among correlators in the $SL(2,R)$ WZW model that cannot be derived from the recursion relations in \cite{Eberhardt:2019ywk}. To summarize, equivalence of supersymmetric correlators with different  picture choices will give various linear relations among correlators in the bosonic  $SL(2,R)$ WZW model. The explicit form of these linear relations depends on the surperstring correlator that one consider. In particular, some of these relations are closely related to the recursion relations found in \cite{Eberhardt:2019ywk}.

\subsection{In the $y$-basis}
We have seen that a superstring correlator can be expressed in  different but equivalent forms, which can be related by the recursion relations  (for example, \eqref{OOO} and \eqref{OOO2} are two equivalent forms for one correlator). Using \eqref{ybasis3pt}, one can always write them as  integrals over $y_i$ $(i=1,2,3)$, with  different integrands. Generally, this integrands depends 
on (one or some of) $h_i$ $(i=1,2,3)$. Nevertheless, one can always act the $y$-transform \cite{Dei:2021xgh} on the correlator in the $h$-basis  to obtain the corresponding one in the   $y$-basis, which by definition will not depend 
on any $h_i$. Then one can write the correlator in the $h$-basis as  integrals over $y_i$ (that is, the inverse $y$-transform), where now the integrand become the correlator in the   $y$-basis and is unique.

Next we  show that starting from any 
choice of pictures of the various operators, one can obtain this unique integrand without really doing any 
integral transformations\footnote{Both the $y$-transform and inverse $y$-transform are hard to perform. For the inverse $y$-transform, see~\cite{Dei:2021xgh} (appendix D) for some examples of the calculation. The final results are complicated.}.  Thus this procedure gives an alternative (and perhaps simpler) way to show the equivalence of different picture choices.  However, we stress that the notion of ``(super)string correlator in the $y$-basis'' is improper because  physical operators in  string theory  should be on-shell (thus $h$ is fixed) while in the $y$-basis one needs to  sum and/or integrate over all $h$.  Nevertheless, there is no problem to express a string correlator as  $y_i$ integrals of the corresponding  $y$-basis correlator (in particular, $h_i$ are all fixed by the on-shell condition).

The idea to proceed is simply to use the rules \eqref{replace rule}. As an illustration, consider the correlator \eqref{OOO} for the case of O-O-O. It can be written as
\begin{equation}
\begin{aligned}
      \mathcal{M}_{OOO}=\frac{C_{S^2}}{k}\mathcal{N}(j_1)D\left(\frac{k+2}{2}-j_1,j_2,j_3\right)&\\
      \times\int \prod_{i=1}^3 \frac{d^2y_i}{\pi}
      &\left| \prod_{i=1}^3  y_i^{\frac{(k+2)w_i}{2}+j_i-h_i-1}\mathfrak{F}(y_1,y_2,y_3)\mathfrak{B}_y(y_1,y_2,y_3)\right|^2\,,
\end{aligned}
\end{equation}
where $\mathfrak{B}_y(y_1,y_2,y_3)$ is the  bosonic correlator in the $y$-basis without any normalization factor (see \cite{Dei:2021xgh} or \eqref{ybasis3pt}),
\begin{equation}
    \mathfrak{B}_y(y_1,y_2,y_3)=X_{123}^{\frac{k+2}{2}-j_1-j_2-j_3}\prod_{i=1}^3X_i^{-\frac{k+2}{2}+j_1+j_2+j_3-2j_i}\,,
\end{equation}
and  $\mathfrak{F}(y_1,y_2,y_3)$ is
\begin{equation}
\begin{aligned}
    \mathfrak{F}(y_1,y_2,y_3)=P_{w_1,w_2,w_3+1}P_{w_1,w_2,w_3-1}\Bigg[\left(2-\frac{y_3}{a_3}-\frac{a_3}{y_3}\right)h_3+\left(\frac{k+2}{2}w_3-j_3+1\right)\frac{y_3}{a_3}\\
    +\left(\frac{k+2}{2}w_3+j_3-1\right)\frac{a_3}{y_3}-2w_3\Bigg]\ .
\end{aligned}
\end{equation}
Thus, the integrand $\mathfrak{F}(y_1,y_2,y_3)\mathfrak{B}_y(y_1,y_2,y_3)$ depends on $h_3$ so it is not yet
a correlator in the $y$-basis. However, using~\eqref{replace rule} we can eliminate the $h_3$ dependence and change $\mathfrak{F}$ to
\begin{equation}
    \mathfrak{F}_y(y_1,y_2,y_3)=P_{w_1,w_2,w_3+1}P_{w_1,w_2,w_3-1}w_3k+\left(j_1+j_2+j_3-\frac{k+2}{2}\right)\frac{X_1X_2X_3}{X_{123}}\ .
\end{equation}
Then $\mathfrak{F}_y(y_1,y_2,y_3)\mathfrak{B}_y(y_1,y_2,y_3)$ has no  $h_i$ dependence so can be the correlator in the $y$-basis up to normalization. 
Moreover, it is symmetric in the index $1,2,3$ (again thanks to the identities \eqref{symmetricfactor}). This makes it clear that different picture choices lead to the same correlator. Finally, we stress that writing a correlator into this form depends on the normalization (which could have $h$ dependence) of the operators. Thus it is unique provided that the normalization of every operators are fixed.

\subsection{Two point correlators and the normalization}\label{2ptnormalization}
In this section, we calculate the  string two point function, which will determine the normalization of the vertex operators. Firstly, notice that to obtain the string two point function, one should divide the worldsheet two point function by the volume of the 
subgroup of the M$\ddot{\text{o}}$bius transformation that fixes two-points (similar to the flat space case \cite{Erbin:2019uiz}). While this volume is infinite,  the two point function of the worldsheet $SL(2,R)$ WZW model is also divergent under the mass-shell condition. 
These two divergent quantities cancel with each other and leaves a finite result, denoted as $\mathcal{M}_B$
\begin{equation}\label{MB}
\begin{aligned}
    \mathcal{M}_B\equiv &C_{S^2}\times \frac{\text{Two point functions in the bosonic $SL(2,R)$ WZW model}}{\text{Volume of the  M$\ddot{\text{o}}$bius (sub)group}}\\
    =& C_{S^2}\times \frac{N_B(w,j)N_B(w,1-j)}{C_{S^2,B}}\delta_{w_1,w_2}\left(R(j_1,h_1,\Bar{h}_1)\delta(j_1-j_2)+\delta(j_1+j_2-1)\right)\\
    =& \frac{wC_{S^2}}{8\pi}\delta_{w_1,w_2}\left(R(j_1,h_1,\Bar{h}_1)\delta(j_1-j_2)+\delta(j_1+j_2-1)\right)\,,
\end{aligned}
\end{equation}
where $R(j_1,h_1,\Bar{h}_1)$ is the reflection coefficient \eqref{reflection coefficient} and  $N_B(w,j)$ and $C_{S^2,B}$ are the normalizations of the vertex operators and the string path integral constant in the case of bosonic string respectively. They are determined in eq. (5.18) in \cite{Dei:2022pkr}, using the matching of the 3 and 4 point correlators.

Notice that in \eqref{MB}, there are two terms, including $\delta(j_1+j_2-1)$ and $\delta(j_1-j_2)$ respectively with relative normalization denoted by the reflection coefficient  $R(j_1,h_1,\Bar{h}_1)$. When discussing the superstring two point functions in the following, we 
 normalize the term proportional to $\delta(j_1+j_2-1)$. 
This is not only because~\eqref{MB} canonically normalizes this term (the coefficient is a constant) but also the fact that on the dual CFT side, only this term is unchanged in the conformal perturbation theory. This means
\begin{equation}
    b\delta(\alpha_1+\alpha_2-Q)=\delta(j_1+j_2-1)\,,
\end{equation}
where the left-hand-side (LHS) is the charge conservation of the two point function in the linear dilaton theory (see appendix \ref{TheproposedCFT} for more details) and we have used the map \eqref{jalpha} here. In fact, the matching of the term proportional to $\delta(j_1-j_2)$ with the CFT side is a non-trivial  test of the proposed CFT dual of the bosonic string theory on AdS$_3\times X$ \cite{Eberhardt:2021vsx}.

For the full string two point function, we should additionally calculate the fermionic contribution and count the effect of picture changing (the ghost part is always canonically normalized). We now calculate the two point functions of the physical vertex operators one by one.
\begin{itemize}
    \item  \textbf{$w$ odd, NS sector:}  the physical operator (with picture number           $-1$) is \eqref{NSodd}. The fermionic two point function is  $\langle               \textbf{1}_\psi^{w_1}(x_1;z_1)\textbf{1}_\psi^{w_2}(x_2;z_2)\rangle$, which         is unit normalized. Thus, omit the coordinate dependence
          the string two point function is simply $\langle O_{j_1,h_1}^{w_1}O_{j_2,h_2}^{w_2}\rangle= \mathcal{M}_B$. Assume the normalization to be $N(w,j)$, then we have 
         \begin{equation}\label{Nj1-j}
         N(w,j)N(w,1-j)= \frac{wC_{S^2}}{8\pi}\ .
         \end{equation}
         Notice that this condition cannot uniquely determine $N(w,j)$\footnote{This is not strange. Notice that the reflection coefficient  $R(j_1,h_1,\Bar{h}_1)$ itself is not fixed uniquely but with a free parameter $\nu$, which can be viewed as the worldsheet cosmological constant. See e.g. \cite{Dei:2021xgh}.}, one can always multiply it by an extra factor $f_j$ satisfying $f_jf_{1-j}=1$ to get another solution. Just as in the bosonic case \cite{Eberhardt:2021vsx}, the  normalization $N(w,j)$ can be fixed only after we identify the operator \eqref{NSodd} with the canonically normalized operator in the CFT side (see section \ref{matching}).
    \item \textbf{$w$  odd, R sector:}  the physical operators (with picture number $-        \frac{1}{2}$) are \eqref{Rodd1} and \eqref{Rodd2}.  To have the total picture       number $-2$, we need one operator in  picture $(-\frac{1}{2})$ and the other       in  picture $(-\frac{3}{2})$. As an illustration, we choose  the first           operator  $O_{j_1,h_1}^{w_1}$, in picture $(-\frac{3}{2})$, to be a specific one in    \eqref{Roddin-3/2}
         \begin{equation}
          \Tilde{O}_{j_1,h_1}^{w_1}(x;z)=e^{-\frac{3\phi(z)}{2}} V_{j_1,h_1}^{w_1}(x;z)\mathbf{S}_{+++++}^{w_1}(x;z)\ .
      \end{equation}
     Accordingly, we  choose  the second operator to be the conjugate of the first one, in the $m$-basis, it is (we use $-w_2$ instead of $w_2$ to label it and keep $j_2, m_2$ not specified)
     \begin{equation}
         \Tilde{O}_{j_2,-m_2}^{w_2=-w_1}(z)=e^{-\frac{3\phi(z)}{2}} V_{j_2,-m_2}^{-w_1}(z)\mathbf{S}_{--+--}^{-w_1}(z)\ .
     \end{equation}
     Notice that $\epsilon_3$ is not changed since $H_i^\dagger=H_i$ for  $i=1,2,4,5$ but  $H_3^\dagger=-H_3$   \cite{Giveon:1998ns}.   Recall that we always label   $x$-basis operators with  positive $w$ (so $w_1>0$), thus the above $m$-basis operator is in fact collected into the following $x$-basis operators with positive $w_2=w_1$
    \begin{equation}
          \Tilde{O}_{j_2,m_2}^{w_2=w_1}(x;z)=e^{-\frac{3\phi(z)}{2}} V_{j_2,m_2}^{w_1}(x;z)\mathbf{S}_{+----}^{w_1}(x;z)\ .
      \end{equation}
     Then we let the first operator lie in the picture ($-\frac{1}{2}$), that is, it becomes the one in          \eqref{Rodd1} with           $\epsilon_2=\epsilon_4=\epsilon_5=1$.    Then the worldsheet two point function is
      \begin{equation}
       \begin{aligned}
&        \langle O_{j_1,h_1}^{w_1}(x_1;z_1)O_{j_2,h_2}^{w_2}(x_2;z_2)\rangle\\
        &\qquad=\frac{h_1-w_1}{\sqrt{k}}
      \langle \textbf{S}^{w_1}_{++-++}(x_1;z_1)\textbf{S}^{w_1}_{+----}(x_2;z_2)\rangle\langle V_{j_1,h_1}^{w_1}(x_1;z_1)V_{j_2,h_2}^{w_2}(x_2;z_2)\rangle\ .
       \end{aligned}
        \end{equation}
        Notice that only one term in \eqref{Rodd1} contribute to the two point function.
      Thus,
       the string two point function contains an additional factor $\frac{h_1-w_1}{\sqrt{k}}$ (and the corresponding one $\frac{\Bar{h}_1-w_1}{\sqrt{k}}$ in the right moving part)
      \begin{equation}\label{twopoint N'}
        \langle O_{j_1,h_1}^{w_1}O_{j_2,h_2}^{w_2}\rangle=\frac{(h_1-w_1)(\Bar{h}_1-w_1)}{k}\mathcal{M}_B\ .
        \end{equation}
       Notice that the mass-shell condition \eqref{weightRamond} implies
      \begin{equation}
        h-w=\frac{1}{kw}(j+\frac{kw}{2}-1)(1-j+\frac{kw}{2}-1)\ .
       \end{equation}
        Then again  normalizing  the term proportional to $\delta(j_1+j_2-1)$, one finds that the normalization factor, denoted by $N'(w,j)$, of \eqref{Rodd1} is $N'(w,j)=\frac{wk\sqrt{k}}{(j+\frac{kw}{2}-1)^2}N(w,j)$. Notice that this solution is not uniquely determined, similar to the situation of \eqref{Nj1-j}. Nevertheless, we will show in section \ref{matching} that it is indeed the correct normalization by comparing with the CFT side. 
    \item \textbf{$w$  even, NS sector:} the physical operator (with picture number          $-1$) are \eqref{NSeven1} or \eqref{NSeven2}. For the case \eqref{NSeven1},      the calculation of the two point function is almost the same as the operator         \eqref{NSodd}, except for an additional fermionic contraction coming from the             fermion $\mathcal{F}(z)$. For the case \eqref{NSeven2}, the two point function     is
        \begin{equation}
          (j_1-m_1)(j_2-m_2)\langle \psi^{-,\omega_1}\psi^{-,\omega_2}\rangle \mathcal{M}_B^{(m_1+1)}+ (j_1+m_1)(j_2+m_2)\langle \psi^{+,\omega_1}\psi^{+,\omega_2}\rangle \mathcal{M}_B^{(m_1-1)}\,,
        \end{equation}
        where we add a superscript to $\mathcal{M}_B$ indicating the $m_1$ dependence of the reflection coefficient $R(j_1,h_1,\Bar{h}_1)$ in  \eqref{MB}. The fermionic two point function are 
       \begin{equation}
        \langle \psi^{-,\omega_1}\psi^{-,\omega_2}\rangle =\langle \psi^{+,\omega_1}\psi^{+,\omega_2}\rangle=k \ .
        \end{equation}
        Again  normalizing  the term proportional to  $\delta(j_1+j_2-1)$, the two point coefficient is
       \begin{equation}
        k(j_1-m_1)(1-j_1-m_1)+k(j_1+m_1)(1-j_1+m_1)=2k\left(m_1+\frac{kw_1}{2}\right)^2=2kH_1^2\,,
       \end{equation}
        where we have used the mass shell condition \eqref{massshell} to simplify the answer. Then the normalization of \eqref{NSeven2}, denoted as $N''(w,j)$, could be $N''(w,j)=\frac{1}{\sqrt{2k}H}N(w,j)$. Notice that this normalization is also not uniquely determined, which can also be fixed when specifying the corresponding operator in the CFT side (though we will not do such a computation in this work).
    \item \textbf{$w$  even, R sector:} the physical operators (with picture number $-        \frac{1}{2}$) are \eqref{Reven1} and \eqref{Reven2}.
         The calculation in this case is completely analogous with the case above when $w$ is odd and in the R sector. Thus the normalization is also  
         \begin{equation}\label{N'}
             N'(w,j)=\frac{wk\sqrt{k}}{(j+\frac{kw}{2}-1)^2}N(w,j)\ .
         \end{equation}
\end{itemize}
To complete the discussion, and also give a cross check, let us calculate the two point functions involving BRST-exact operators. They are expected to vanish and we will show this explicitly in the following. Consider the  spurious operator \eqref{owspu}. The two point function of \eqref{owspu} and  
a physical operator  \eqref{NSeven2} is (all other physical operators are clearly orthogonal with the spurious operator)
\begin{equation}
\begin{aligned}
     -2(m_1+\frac{w_1k}{2})\alpha_3^{(2)}\langle \psi^{3,\omega_1}\psi^{3,\omega_2}\rangle \mathcal{M}_B^{(m_1)}+&(m_1-j_1+1)\alpha_-^{(2)}\langle \psi^{-,\omega_1}\psi^{-,\omega_2}\rangle \mathcal{M}_B^{(m_1+1)}\\
     + &(m_1+j_1-1)\alpha_+^{(2)}\langle \psi^{+,\omega_1}\psi^{+,\omega_2}\rangle \mathcal{M}_B^{(m_1-1)}\,,
\end{aligned}
\end{equation}
where $\alpha_3^{(2)}, \alpha_-^{(2)}, \alpha_+^{(2)}$ is the physical polarization for the second operator (so with a superscript ``$(2)$'').
Since
\begin{equation}
     \langle \psi^{3,\omega_1}\psi^{3,\omega_2}\rangle=-\frac{1}{2}\langle \psi^{-,\omega_1}\psi^{-,\omega_2}\rangle =-\frac{1}{2}\langle \psi^{+,\omega_1}\psi^{+,\omega_2}\rangle=-\frac{k}{2}\ .
\end{equation}
Then two point coefficient of the terms including $\delta(j_1+j_2-1)$ is proportional to
\begin{equation}
    (m_2+\frac{w_2k}{2})\alpha_3^{(2)}+(m_2+j_2)\alpha_-^{(2)}+(m_2-j_2)\alpha_+^{(2)}\,,
\end{equation}
which is zero because of the physical states condition \eqref{superprimary}. One can also show that again due to the physical states condition, the terms proportional to $\delta(j_1-j_2)$  vanish as well. This result also imply that the spurious operator has a zero two point function with itself, since spurious operator is  a solution of the condition \eqref{superprimary} as well.

\section{Match with the CFT side}\label{theCFTside}
In this section, we discuss the matching of the physical operators and their correlators calculated in the previous sections with the dual CFT side. The dual CFT was proposed to be a deformed symmetric orbifold CFT \cite{Eberhardt:2021vsx,Eberhardt:2019qcl}. We  review this proposal in the appendix \ref{TheproposedCFT}. A crucial point is that for long strings, the spectrum will not be affected by the marginal deformation \cite{Balthazar:2021xeh,Eberhardt:2021vsx}. Thus the main aim here is to find operators in the symmetric orbifold CFT that (after the marginal deformation) correspond to the physical vertex operators we found in  section \ref{stringside}\footnote{The matching of the spectrum of long strings were discussed in \cite{Eberhardt:2019qcl}, by finding all the DDF operators (see also \cite{Sriprachyakul:2024xih}). However, there the discussion of matching the ground states (see section (6.3) in \cite{Eberhardt:2019qcl}) seems not complete.    Our discussion here  (matching the operators with the lowest space-time weights of the two sides) can be viewed as a complement of \cite{Eberhardt:2019qcl}. }, and then compare the three point correlators of the two sides at the leading order of the conformal perturbation theory.  We find the matching at this order is already non-trivial: it predicts an interesting mathematical identity for covering maps, which can be checked (or proved) to be correct. More physically, this means that at the level of correlators, the picture changing effect  is essential to reproduce the correct central charge of the boundary CFT.

\subsection{The seed theory}
Since the undeformed theory is a symmetric orbifold CFT, we firstly describe the seed theory, that is,  $\mathbb{R}_Q\times \mathfrak{su}(2)_{k-2}\times \text{four free fermions} \times \left(\text{U}(1)^{(1)}\right)^4$ \cite{Eberhardt:2021vsx,Eberhardt:2019qcl} (see appendix \ref{conventionforseed} for our conventions for the seed theory). As in the string side, here   we  mainly  focus on the holomorphic part.
Denote the generating fields of this seed theory  as:
\begin{equation}
   \partial\phi, \qquad  J^a , \qquad \psi^{\alpha\beta}, \qquad  X^i, \qquad X^{i\dagger},\qquad \lambda^j, \qquad \lambda^{j\dagger}.
\end{equation}
where the first three kinds of fields generate the $\mathcal{N}=4$ linear dilaton theory; $\partial\phi$ is the linear dilaton with background charge $Q=\frac{k-1}{\sqrt{k}}$; $J^a (a=\pm,3)$ generate the affine algebra $\mathfrak{su}(2)_{k-2}$  and $\psi^{\alpha\beta}$ are the 4 free fermions, with $\alpha, \beta=\pm$. These indices are in fact  spinor indices of the $SU(2)$ R-symmetry and the $SU(2)$ automorphism (see the next paragraph for more explanation on $SU(2)_R\oplus SU(2)_{\text{outer}}$). The remaining fields generate the torus theory; 
$X^i, X^{i\dagger} (i=1,2)$ are two complex bosons and their conjugates, $\lambda^j, \lambda^{j\dagger} (j=1,2)$ are two complex fermions and their conjugates. For convenience, we can also relabel the fermions  $\lambda^{\alpha\beta}$ by 2  superscript $\alpha, \beta=\pm$:  
\begin{equation}
    \lambda^{++}\equiv\lambda^1, \quad \lambda^{--}\equiv\lambda^{1\dagger}, \quad \lambda^{+-}\equiv i\lambda^2, \quad \lambda^{-+}\equiv i\lambda^{2\dagger}\ .
\end{equation}
Then $\alpha$ is a spinor index of  $SU(2)_R$ but $\beta$ is not a spinor index of $SU(2)_{\text{outer}}$ (see the next paragraph).
Both the $\mathcal{N}=4$ linear dilaton and torus theory have (small) $\mathcal{N}=4$ superconformal symmetries, see appendix \ref{conventionforseed} for constructions of the $\mathcal{N}=4$ superconformal generators of the two theories.

Before proceeding, let us comment on the algebra $SU(2)_{\text{outer}}$. Generally, as an outer automorphism of the small $\mathcal{N}=4$ superconformal algebra, $SU(2)_{\text{outer}}$ is not a symmetry of the theory. That means there is no corresponding conserved currents. 
However, in ether case of the $\mathcal{N}=4$ linear dilaton or the ($\mathcal{N}=4$) T$^4$ theory, one can construct  $SU(2)_{\text{outer}}$ using the free fields. In the $\mathcal{N}=4$ linear dilaton theory, there are 2 global $SU(2)$ symmetries which correspond to the $SO(4)$
rotations of the 4 fermions. In fact,  one of these 2 $SU(2)$ algebras could be the  algebra $SU(2)_{\text{outer}}$ (see appendix \ref{conventionforseed} for its explicit form). Thus, the 4 fermions  $\psi^{\alpha\beta}$ form a $\textbf{(2,2)}$ of $SU(2)_R\oplus SU(2)_{\text{outer}}$ and two indices $\alpha,\beta=\pm$ are just the spinor indices. On the contrary, in the theory T$^4$,  the  algebra $SU(2)_{\text{outer}}$ is constructed by the bosons: it is one $SU(2)$ component of the $SO(4)$
rotations of the 4 bosons (see \cite{David:2002wn} for the construction). Consequently, the 4 fermions $\lambda^{\alpha\beta}$ form  2 $\textbf{(2,1)}$'s of $SU(2)_R\oplus SU(2)_{\text{outer}}$ and the first index $\alpha$ is  the spinor index of $SU(2)_R$. The full algebra $SU(2)_R\oplus SU(2)_{\text{outer}}$ is the sum of the ones in  the $\mathcal{N}=4$ linear dilaton and the T$^4$ theory. In the following, we will organize states according to this $SU(2)_R\oplus SU(2)_{\text{outer}}$.  

Now we describe primary operators that have the lowest conformal weights. 
In the NS sector, it is unique and is constructed in \cite{Eberhardt:2019qcl} (rf e.g. eq. (6.7) there)\footnote{Since we focus on the operator with lowest weight, we set $l=0$  in Eq. (6.7) in \cite{Eberhardt:2019qcl}. Besides, \eqref{theoperator} including an extra factor $i$ in the exponent because we use a different convention of the free boson $\phi$.  },  
\begin{equation}\label{theoperator}
    \mathbb{V}_{\alpha}\equiv e^{\sqrt{2}\alpha\phi}=e^{\frac{i}{\sqrt{2k}}(2p-ik+i)\phi}\,,
\end{equation}
where $p\in \mathbb{R}$ and the momenta $\alpha$ is
\begin{equation}
    \alpha=\frac{ip+\frac{k-1}{2}}{\sqrt{k}}=\frac{\frac{1}{2}+ip+\frac{k}{2}-1}{\sqrt{k}}\ .
\end{equation}
Notice that we are considering the whole seed theory of the  symmetric orbifold theory (not only the $\mathcal{N}=4$ linear dilaton), so when writing \eqref{theoperator}, we have set the operator in the torus theory to be the identity.  This operator is a singlet $(\textbf{1},\textbf{1})$ of $SU(2)_R\oplus SU(2)_{\text{outer}}$ and has  conformal weight
\begin{equation}\label{groundNS}
    h=\frac{\frac{1}{4}+p^2}{k}+\frac{k-2}{4}\ .
\end{equation}
For our purpose to find the corresponding operators of the ones in the string side, we also needs the lowest excited states in the NS sector. They are:
\begin{equation}\label{onefermionex}
   \mathbb{V}_{\alpha,\mathbb{F}}\equiv e^{\frac{i}{\sqrt{2k}}(2p-ik+i)\phi}\mathbb{F}\,,
\end{equation}
where $\mathbb{F}$ represents an excited fermion and we have 8 different choices for it: $\psi^{\alpha\beta}, \lambda^{\alpha\beta}$. These 8 excited states has conformal weight
\begin{equation}\label{excitedNS}
    h=\frac{\frac{1}{4}+p^2}{k}+\frac{k-2}{4}+\frac{1}{2}\,,
\end{equation}
and they form a representation $(\textbf{2},\textbf{2})\oplus2(\textbf{2},\textbf{1})$  of $SU(2)_R\oplus SU(2)_{\text{outer}}$.

Now we turn to the Ramond sector\footnote{Notice that we will always concern the NS sector of the symmetric orbifold theory, since states in the Ramond sector cannot be treat perturbatively on the string side \cite{Lunin:2001jy}.  Here we need the Ramond sector because when the cycle  length (of a single cycle-twisted sector) is even, states (in the NS sector) will effectively lie in the Ramond sector when lift up to the covering surface \cite{Lunin:2001pw}.}. There will be an additional contribution from the Ramond ground states  of the 8 fermions. Zero modes of the 4 fermions $\psi^{\alpha\beta}_0$ in the $\mathcal{N}=4$ linear dilaton theory result in ground states which form a $(\textbf{2},\textbf{1})\oplus (\textbf{1},\textbf{2})$ representation of $SU(2)_R\oplus SU(2)_{\text{outer}}$. In contrast, zero modes of the 4 fermions $\lambda^{\alpha\beta}_0$ in the torus theory result in ground states which form a  $(\textbf{2},\textbf{1})\oplus 2(\textbf{1},\textbf{1})$ representation of $SU(2)_R\oplus SU(2)_{\text{outer}}$, since the algebra $SU(2)_{\text{outer}}$ is constructed by bosons in the theory T$^4$. Then we have in total $4\times 4=16$ ground states, in the representation 
\begin{equation}
\left[\textbf{(2,1)}\oplus\textbf{(1,2)}\right]\otimes\left[\textbf{(2,1)}\oplus2\textbf{(1,1)}\right]=\textbf{(3,1)}\oplus\textbf{(1,1)}\oplus\textbf{(2,2)}\oplus 2\textbf{(2,1)}\oplus2\textbf{(1,2)}
\end{equation}
All these 16 states have conformal weight $h=\frac{1}{4}\times 2=\frac{1}{2}$. To write down the  spin fields which generate the above 16 Ramond ground states, we firstly   bosonize the 8 fermions\footnote{As usual, we use a hat to denote the bosons with cocycles: $\hat{B}_i=B_i+\pi\sum_{j<i}N_j$, $B_i(z)B_j(w)\sim -\delta_{ij}\text{log}(z-w)$, $N_i=i\oint \partial H_i$.}
\begin{equation}
    i\partial \hat{B}_1=\psi^{++}\psi^{--},\quad  i\partial \hat{B}_2=-\psi^{+-}\psi^{-+}, \quad i\partial \hat{B}_3=\lambda^{++}\lambda^{--}, \quad i\partial \hat{B}_4=-\lambda^{+-}\lambda^{-+}\ .
\end{equation}
Accordingly, the fermions can be written as
\begin{equation}
    \begin{aligned}
        \psi^{++}=e^{i\hat{B}_1},\quad \psi^{--}=e^{-i\hat{B}_1}, \quad \psi^{+-}=ie^{i\hat{B}_2}, \quad \psi^{-+}=ie^{-i\hat{B}_2}\\
         \lambda^{++}=e^{i\hat{B}_3},\quad \lambda^{--}=e^{-i\hat{B}_3}, \quad \lambda^{+-}=ie^{i\hat{B}_4}, \quad \lambda^{-+}=ie^{-i\hat{B}_4}\ .
    \end{aligned}
\end{equation}
Then one can easily write down the 16 spin fields as in Table \ref{spinfield}  where the spin field $S^{\epsilon_1\epsilon_2\epsilon_3\epsilon_4}$ are defined as
\begin{equation}
    S^{\epsilon_1\epsilon_2\epsilon_3\epsilon_4}=e^{\frac{i}{2}\sum_{i=1}^4\epsilon_i \hat{B}_i}\ .
\end{equation}
\begin{table}[!ht]
    \centering
     \begin{tabular}{|c|c|c|c|c|c|c|}\hline
  (\textbf{3},\textbf{1})   & (\textbf{1},\textbf{1}) & (\textbf{1},\textbf{2}) & (\textbf{1},\textbf{2}) & (\textbf{2},\textbf{1}) & (\textbf{2},\textbf{1}) & (\textbf{2},\textbf{2})\\ \hline
\makecell[c]{ $S^{++++}$ \\$S^{++--}+S^{--++}$\\$S^{----}$} & $S^{++--}-S^{--++}$& \makecell[c]{ $S^{+-+-}$ \\$S^{-++-}$}  & \makecell[c]{ $S^{+--+}$ \\$S^{-+-+}$} &\makecell[c]{ $S^{+++-}$ \\$S^{--+-}$} &\makecell[c]{ $S^{+++-}$\\ $S^{---+}$} &\makecell[c]{$S^{+-++}$ \\$S^{+---}$\\$S^{-+++}$ \\ $S^{-+--}$ }   \\  \hline
\end{tabular} 
    \caption{The spin fields}
    \label{spinfield}
\end{table}

Note that the $U(1)_R$ and $U(1)_{\text{outer}}$ charges of a spin field $S^{\epsilon_1\epsilon_2\epsilon_3\epsilon_4}$ are $\frac{1}{4}(\epsilon_1+\epsilon_2+\epsilon_3+\epsilon_4)$ and $\frac{1}{4}(\epsilon_1-\epsilon_2)$ respectively. One can easily check that fields in Table \ref{spinfield} have the correct charges, and to verify the full representations in Table \ref{spinfield}, the cocycles should be counted carefully.  
The vertex operators that have the lowest conformal weight in the Ramond sector are:
\begin{equation}\label{CFTRamond}
    \mathbb{V}_{\alpha,\mathcal{S}}\equiv e^{\frac{i}{2k}(2p-ik+i)\phi} \mathcal{S}\,,
\end{equation}
where $\mathcal{S}$ can be any of the 16 spin fields in Table \ref{spinfield}. Then the conformal weight of these operators are:
\begin{equation}\label{groundR}
    h=\frac{\frac{1}{4}+p^2}{k}+\frac{k-2}{4}+\frac{1}{2}\ .
\end{equation}
Notice that it coincides with  \eqref{excitedNS} of excited states in the NS sector.

\subsubsection*{Summary}
The operators constructed in the seed theory  are summarized in the following table \ref{seed}.  
The numbers and  representation contents of operators in this table are the same as the ones in the table \ref{numberofoperators} and table \ref{representation contents}. In the symmetric orbifold theory discussed in the following section, we will make this agreements more precise. In particular, the ``NS sector'' and ``R sector'' of the seed theory in the table \ref{seed} are related to the  ``odd'' and  ``even'' parities of $w$ in the table \ref{numberofoperators} (and the table \ref{representation contents}) respectively.

\begin{table}[!ht]
    \centering
     \begin{tabular}{|c|c|c|}\hline
      \diagbox{Sectors}{}   & Ground & Excited \\ \hline
  NS sector & $(\textbf{1},\textbf{1})$ (in \eqref{theoperator}) &$2 (\textbf{2},\textbf{1})\oplus(\textbf{2},\textbf{2})$ (in \eqref{onefermionex})\\  \hline
 R sector & \makecell[c]{$\textbf{(3,1)}\oplus\textbf{(1,1)}\oplus\textbf{(2,2)}\oplus 2\textbf{(2,1)}\oplus2\textbf{(1,2)}$ \\(in \eqref{CFTRamond})} & $\cdots$
 \\  \hline
\end{tabular} 
    \caption{The operators constructed in the seed theory}
    \label{seed}
\end{table}

\subsection{The symmetric orbifold}
Now we describe the symmetric orbifold theory
\begin{equation}
    \text{Sym}^N(\text{Seed CFT})
\end{equation}
where  ``seed CFT" is the one described in the precious section.
In general, the Hilbert space of a symmetric orbifold CFT is a direct sum of twisted sectors, with each sector 
labeled by a conjugacy class of $S_N$.  We will be interested in the large $N$ limit of the symmetric orbifold CFT, since they  describes perturbative string theory on AdS$_3$ backgrounds. Furthermore, we will  restrict to twisted sectors described by conjugacy classes of single cycles (which are interpreted as single string states on the string side), labeled by the cycle  lengths $w$. 

In the following, we describe operators in  twisted sectors as in the bosonic case \cite{Eberhardt:2021vsx}.
For every vertex operator $V_{h,\Bar{h}}$ with weights $(h,\Bar{h})$ satisfying $h-\Bar{h}\in wZ$ (which is the physical condition comes from the orbifold invariance) in the seed theory, there is a corresponding vertex operator $\mathcal{V}_{h_w,\Bar{h}_w}$ in the $w$-twisted sector  with weight
\begin{equation}\label{weightforw}
    h_w=\frac{c(w^2-1)}{24w}+\frac{h}{w}, \qquad \Bar{h}_w=\frac{c(n^2-1)}{24w}+\frac{\Bar{h}}{w}\,,
\end{equation}
where $c$ is the central charge of the seed theory. In fact, lift $\mathcal{V}_{h_w,\Bar{h}_w}$ up to a covering surface (which is  locally a $w$-folded cover at the insertion point) we will get $V_{h,\Bar{h}}$. For a supersymmetric symmetric orbifold CFT, there is a difference between the two parities of $w$ \cite{Lunin:2001pw} (due to the fermions in the seed theory); $V_{h,\Bar{h}}$ should be in the NS sector when $w$ is odd and in the R sector when $w$ is even.
In the present case, we  consider $V_{h,\Bar{h}}$ to be the ones in the table \ref{seed}, that is,  operators \eqref{theoperator}, \eqref{onefermionex} in the NS sector for $w$ odd and operators \eqref{CFTRamond} in the R sector for $w$ even. Then we denote the corresponding  vertex operators in the $w$-twisted  sector as
\begin{equation}\label{CFToperators}
\begin{aligned}
   \text{$w$ odd}: \qquad  \mathcal{V}_{p}^{(w)}&\equiv e^{\frac{i}{2k}(2p-ik+i)\phi} \Sigma_w, \qquad \mathcal{V}_{p,\mathbb{F}}^{(w)}\equiv e^{\frac{i}{2k}(2p-ik+i)\phi}\mathbb{F} \Sigma_w \\
  \text{$w$ even}: \qquad  \mathcal{V}_{p,\mathcal{S}}^{(w)}&\equiv e^{\frac{i}{2k}(2p-ik+i)\phi} \mathcal{S} \Sigma_w, 
\end{aligned}
\end{equation}
where $\Sigma_n$ is the twist fields of the $w$-twisted sector.  Now using \eqref{weightforw}, \eqref{groundNS}, \eqref{excitedNS}, \eqref{groundR} and $c=6k$, one can check that  their conformal weights agree with the corresponding ones in the string side (anti-holomorphic part is similar):
\begin{equation}\label{weightVW}
\begin{aligned}
    h(\mathcal{V}_{p}^{(w)})&=H_{\text{NS,odd}}, \quad h(\mathcal{V}_{p,\mathbb{F}}^{(w)})=H_{\text{R,odd}}\\
    h(\mathcal{V}_{p}^{(w)})&=H_{\text{NS,even}}=H_{\text{R,even}}
\end{aligned}    
\end{equation}
Notice that this matching of  weights was found in \cite{Eberhardt:2019qcl} for  some operators in \eqref{weightVW}. The matching of weights \eqref{weightVW},
combined with the agreement between the representation contents of  operators in the table \ref{representation contents} and \ref{seed}, show the matching of the operators on the two sides.

\subsection{Matching the correlators at the leading order}\label{matching}

It was shown in \cite{Eberhardt:2021vsx} that in the bosonic case, the three point function in the string side has the same poles as one in the CFT side\footnote{These includs the ``bulk" poles and the ``LSZ" poles. For our propose, we focus on the (residue of) bulk poles.}. Besides, the  corresponding residues of the two sides are also remarkably matched (up to the forth order).
This  matching of poles also extend to the superstring case (just shift the level of the bosonic model as $k\to k+2$). Therefore to further match 
the correlators,  we need to compare the  corresponding residues of the poles, that is,  verify the following equation (eq. (3.1) in \cite{Eberhardt:2021vsx}):
\begin{equation}\label{Thematching}
    \mathop{\text{Res}}\limits_{\sum_i j_i=2-\frac{k}{2}+\frac{mk}{2}}\mathcal{M}_3\mathop{=}\limits^{?}
     \mathop{\text{Res}}\limits_{2b(\sum_i \alpha_i-Q)=m}\mathbb{M}_3\,,
\end{equation}
where $\mathcal{M}_3$ on the left hand side is a string correlator and $\mathbb{M}_3$ on the right hand side is the corresponding correlator on the CFT side. The residues for the RHS can be calculated using the conformal perturbation theory \cite{Eberhardt:2021vsx} and  $m\in \mathbb{N}$ is the perturbation order.  The positions of the poles are the same as in the bosonic case \cite{Eberhardt:2021vsx} (with the shift $k\to k+2$, also see the appendix \ref{TheproposedCFT}). 

In this section, we will match the two sides of \eqref{Thematching} at the leading order i.e. $m=0$. Before  that, we should mention that for correlators in the symmetric orbifold (we review the covering map method in the appendix \ref{coveringmapmethod}), 
there is a qualitative difference depending on the parity of $\sum_i(w_i-1)$. The Riemann-Hurwitz formula \eqref{RH} implies that a covering map only exists when $\sum_i(w_i-1)$ is even.
Since the marginal operator lies in the 2-twisted  sector, every insertion of it change the parity of $\sum_i(w_i-1)$. Therefore, we refer to the three operators as X-Y-Z according to the parity of their twists, 
where $X\,,Y\,,Z$ can be E(even) or O(odd). There are then the following cases 
\begin{itemize}
    \item For $\sum_{i=1}^3(w_i-1)$ even, only even orders in conformal perturbation
           theory can be non-zero. There are two cases: O-O-O and O-E-E.
    \item For $\sum_{i=1}^3(w_i-1)$ odd, only odd orders in conformal perturbation
           theory can be non-zero. There are two cases: E-O-O and E-E-E.
\end{itemize}
These different cases are just the counterparts of the string correlators we have discussed in the previous sections. Since we focus on the order $m=0$, there are thus two possibilities: O-O-O or O-E-E. 

Firstly,  we consider the case of O-O-O, where the left hand side of \eqref{Thematching} being the string correlator $\mathcal{M}_{OOO}$ in \eqref{OOO}.
For the residue of the LHS (string side), using the result for $m=0$ in the bosonic case \cite{Eberhardt:2021vsx} (eq. (3.20)), we have:
\begin{equation}\label{resbosonic}
\begin{aligned}
       \mathop{\text{Res}}\limits_{\sum_i j_i=2-\frac{k}{2}}&\langle V_{j_1,h_1,\Bar{h}_1}^{w_1}(0;0)V_{j_2,h_2,\Bar{h}_2}^{w_2}(1;1)V_{j_3,h_3,\Bar{h}_3}^{w_3}(\infty;\infty)\rangle\\
    &=\frac{\nu^{\frac{k}{2}-1}}{2\pi^2k^2\upgamma(\frac{k+1}{k})}\left|\prod_{i=1}^3 a_i^{\frac{k+2}{4}(w_i-1)-h_i}w_i^{-\frac{k+2}{4}(w_i+1)+1-j_i}\Pi^{-\frac{k+2}{2}}\right|^2\ .
\end{aligned}
\end{equation}
where $\upgamma(x)=\Gamma(x)/\Gamma(1-x)$ and $\Pi$ is the product of residues of the relevant covering map (see \eqref{defPi}) and its explicit form is complicated (see \eqref{Pi}).
Notice that in the above the level is shifted to be $k+2$. Using this result and \eqref{OOO}, we obtain
\begin{equation}\label{LHS}
\begin{aligned}
    \text{LHS}=& \frac{\nu^{\frac{k}{2}-1}C_{S^2}}{2\pi^2k^3\upgamma(\frac{k+1}{k})}\Bigg|\Big[ (h_3-
    \frac{(k+2)w_3}{2}+j_3-1)P^2_{w_1,w_2,w_3-1}a_3+
    (h_3-\frac{(k+2)w_3 }{2}-j_3+1)P^2_{w_1,w_2,w_3+1}a_3^{-1}\\
    &+2(h_3-w_3)P_{w_1,w_2,w_3-1}P_{w_1,w_2,w_3+1}\Big]
     \prod_{i=1}^3 a_i^{\frac{k+2}{4}(w_i-1)-h_i}w_i^{-\frac{k+2}{4}(w_i+1)+1-j_i}\Pi^{-\frac{k+2}{2}}\Bigg|^2\\
     =&\frac{\nu^{\frac{k}{2}-1}C_{S^2}}{2\pi^2k\upgamma(\frac{k+1}{k})}\Bigg|w_3P_{w_1,w_2,w_3-1}P_{w_1,w_2,w_3+1}
     \prod_{i=1}^3 a_i^{\frac{k+2}{4}(w_i-1)-h_i}w_i^{-\frac{k+2}{4}(w_i+1)+1-j_i}\Pi^{-\frac{k+2}{2}}\Bigg|^2\ .
\end{aligned}
\end{equation}
As a cross check, one can check that this result is symmetric in the three index $1,2,3$. Besides, if one were starting with
the form~\eqref{OOO2} for $\mathcal{M}_{OOO}$, one will find the same result.

As for the CFT side, since  the deformation is turned off for $m=0$,  the result can be easily written down as in the bosonic case \cite{Eberhardt:2021vsx}:
\begin{equation}\label{TheRHS}
    \text{RHS}=\frac{1}{\pi\sqrt{N}}\prod_{i=1}^3N(w_i,j_i)w_i^{\frac{1}{2}}\Bigg|\prod_{i=1}^3 a_i^{\frac{k}{4}(w_i-1)-H_i}w_i^{-\frac{k}{4}(w_i+1)}\Pi^{-\frac{k}{2}}\Bigg|^2\,,
\end{equation}
where $N(w_i,j_i)$ are the  normalization factors  of the  vertex operators on the string side (see \eqref{Nj1-j}),  since in the CFT side, vertex operators are already canonically normalized. 
Notice that the above equation is $not$ obtained by replacing $k$ by $k+2$ in equation (3.21) in \cite{Eberhardt:2021vsx}, because the central charge of the seed theory is $6k$ instead of $6(k+2)$ \cite{Argurio:2000tb,Eberhardt:2019qcl} (see \eqref{centralcharge}). However, the corresponding formula on the LHS in \eqref{LHS} is based on the decoupled  bosonic WZW level $k+2$. This causes a disagreement in the power of  $a_i, w_i$ and $\Pi$ between the LHS and RHS, with the difference being\footnote{In the following, we do not include the term $w_i^{1-j_i}$ in \eqref{LHS} since this term is not caused by the difference of  levels, and can be compensated by modifying the normalization $N(w_i,j_i)$ as in the bosonic case \cite{Eberhardt:2021vsx}. }:
\begin{equation}\label{difference in power}
    \Pi^2\prod_{i=1}^3(w_ia_i)^{w_i+1}
\end{equation}
where we have used $H_i=h_i-w_i$.
Crucially, \eqref{difference in power} cannot be compensated by just modifying the normalization $N(w_i,j_i)$ of each vertex operators because $\Pi(w_1,w_2,w_3)$ and $a_i(w_1,w_2,w_3)$ cannot be factorized into products of factors that 
only depends on one of $w_i$. Thus, to make the two sides match, additional factors $|w_3P_{w_1,w_2,w_3-1}P_{w_1,w_2,w_3+1}|^2$ in \eqref{LHS}, coming from the fermionic parts and  picture changing, should be taken into account and cure 
the non-factorizing behaviour  in~\eqref{difference in power}. This is indeed the case, as we  show below.

Recall that covering maps for 3 ramified points can be explicitly constructed by Jacobi
polynomials \cite{Lunin:2000yv}. Accordingly, the associated quantity $\Pi$ can also be   explicitly written down\footnote{It is very hard to give  closed formulas for   general $n (n>3)$ points ramified covering maps. Subsequently,  the associated $\Pi$ is not known.}. The formula is (see eq. (5.30) in \cite{Lunin:2000yv})\footnote{Notice that comparing with eq. (5.30) in \cite{Lunin:2000yv}, we have included an additional factor $w_3^{-w_3-1}$ in the following equation. This factor comes from the difference between  treating the point at infinity symmetrically or not (see footnote 6 in \cite{Eberhardt:2021vsx}).}:
\begin{equation}\label{Pi}
\begin{aligned}
    \Pi=&2^{-2d_2(d_2-1)}w_1^{d_2}\mathcal{D}^2(d_2!)^{-3d_2+4}\left(\frac{d_1!}{w_1!(d_1-w_1)!}\right)^{d_2}\left(\frac{(w_1-1)!}{(w_1-d_2-1)!}\right)^{w_1+d_2-1}\\
    &\times \left(\frac{(d_1-d_2)!}{d_1!}\right)^{d_1-d_2+3}\left(\frac{(d_!+d_2-w_1)!}{(d_1-w_1)!}\right)^{d_1+d_2-w_1}w_3^{-w_3-1}\,,
\end{aligned}
\end{equation}
where $d_1=\frac{1}{2}(w_1+w_2+w_3-1), d_2=\frac{1}{2}(w_1+w_2-w_3-1)$ and $\mathcal{D}$ is the discriminant of Jacobi polynomials
\begin{equation}
    \mathcal{D}=2^{-d_2(d_2-1)}\prod_{j=1}^{d_2}j^{j+2-2d_2}(j-w_1)^{j-1}(j-d_1-d_2+w_1-1)^{j-1}(j-d_1-1)^{d_2-j}\ .
\end{equation}
With this expression for $\Pi$, one can find the following somewhat surprising mathematical identities for covering maps
\begin{equation}\label{idnetity}
    (w_3P_{w_1,w_2,w_3-1}P_{w_1,w_2,w_3+1})^2=\Pi^2\prod_{i=1}^3(w_ia_i)^{w_i+1}\ .
\end{equation}
The above identities can be verified by comparing the total power of every integers on the two sides. 
\CP{}\eqref{idnetity} gives a concise  way to express the covering map data $\Pi$. One can also rewrite \eqref{idnetity} in terms of $X_i$ (recall the identities \eqref{symmetricfactor}):
\begin{equation}
    (w_j\partial_{y_j}(X_j^2))^2=\Pi^2\prod_{i=1}^3(w_ia_i)^{w_i+1}, \qquad j=1,2,3\ .
\end{equation}
Then these identities are analogous to 
various identities found in matching the bosonic correlators in \cite{Eberhardt:2021vsx}, that is, eq. (3.26a), (3.29a), (3.31a) and (3.34b) in  \cite{Eberhardt:2021vsx}. In \cite{Eberhardt:2021vsx}, there is another identity (3.19b) that plays a role in the matching of the bosonic correlators at the leading order. The identity \eqref{idnetity} is in fact a refined version of the identity (3.19b).
While in \cite{Eberhardt:2021vsx} these identities are only verified numerically since the relevant covering map has more than 3 ramified points. Here we can directly prove the identity \eqref{idnetity} using the explicit expression for $\Pi$.

Because of \eqref{idnetity}, the RHS 
agrees with the LHS provided that
\begin{equation}
    N(w_i,j_i)=N_0w_i^{\frac{3}{2}-2j_i} \,,
\end{equation}
and 
\begin{equation}
    C_{S^2}=2\pi k\nu^{1-\frac{k}{2}}\upgamma\left(\frac{k+1}{k}\right)N_0^3N^{-\frac{1}{2}}\,,
\end{equation}
where $N_0$ is an undetermined $k$-dependent function. The above relations, together with \eqref{Nj1-j}, determine $N_0$, which leads to
\begin{equation}\label{NCS^2}
    \begin{aligned}
        N(w,j)&=\frac{4\sqrt{N}\nu^{\frac{k}{2}-1}w^{\frac{3}{2}-2j}}{ k\upgamma\left(\frac{k+1}{k}\right)}\\
        C_{S^2}&=\frac{128\pi N\nu^{k-2}}{k^2\upgamma\left(\frac{k+1}{k}\right)^2}\ .
    \end{aligned}
\end{equation}
Notice that the above $N(w,j)$ and $C_{S^2}$ are not the same as the corresponding constants in the bosonic string case (eq. (5.18) in \cite{Dei:2022pkr}) with a simple shifted level $k\to k+2$.  Note that the above analysis for the matching of residues  is valid for the general cases $\mathrm{I}$ of $w_1, w_2,w_3$ (see \eqref{thecondition}), where the relevant covering map exists on the CFT side. Another possibility are the edge cases $\mathrm{II}$ in \eqref{thecondition}, where the relevant covering map does not exist on the CFT side. Thus, the RHS of \eqref{Thematching} for these edge cases will vanish. One can see  that the result of the LHS in \eqref{LHS}  also vanishes, since  in the edge cases one of the two functions $P_{w_1,w_2,w_3-1}$ and $P_{w_1,w_2,w_3+1}$ will vanish. Consequently, the matching holds for both of the general and edge cases.

Next we discuss the case  O-E-E, where the correlator on the LHS is the $\mathcal{M}_{OEE}$ in  \eqref{OEE}. As shown below, the matching in this case is guaranteed by the same covering map identities, 
which 
also gives us a cross-check for the normalization factors. With the help of \eqref{resbosonic} and \eqref{aiP}, the LHS becomes
\begin{equation}\label{LHS2}
\begin{aligned}
    \text{LHS}=
    \frac{\nu^{\frac{k}{2}-1}C_{S^2}}{2\pi^2k^2\upgamma(\frac{k+1}{k})}\Bigg|\left(\frac{kw_2}{2}+j_2-1\right)\left(\frac{kw_3}{2}+j_3-1\right)P_{w_1,w_2-1,w_3}P_{w_1,w_2,w_3-1}&\\\prod_{i=1}^3 a_i^{\frac{k+2}{4}(w_i-1)-h_i}w_i^{-\frac{k+2}{4}(w_i+1)+1-j_i}\Pi^{-\frac{k+2}{2}}\Bigg|^2&\ .
\end{aligned}
\end{equation}
The RHS is almost the same as \eqref{TheRHS}, except the full space-time weight becomes $H_i=h_i-w_1+\frac{1}{2}$ and the  normalization 
(denoted as $N'(w_i,j_i)$ in \eqref{N'})
\begin{equation}
    \text{RHS}=\frac{1}{\pi\sqrt{N}}\prod_{i=1}^3N'(w_i,j_i)w_i^{\frac{1}{2}}\Bigg|\prod_{i=1}^3 a_i^{\frac{k}{4}(w_i-1)-H_i}w_i^{-\frac{k}{4}(w_i+1)}\Pi^{-\frac{k}{2}}\Bigg|^2\ .
\end{equation}
Since the constant $C_{S^2}$ is already determined in \eqref{NCS^2}, then the two sides match provided that
\begin{equation}\label{nor N'}
    N'(w_i,j_i)=\frac{k\sqrt{k}w_i}{(j_i+\frac{kw_i}{2}-1)^2}N(w_i,j_i)\,,
\end{equation}
and 
\begin{equation}\label{iden}
    (P_{w_1,w_2-1,w_3}P_{w_1,w_2,w_3-1})^2=\Pi^2\prod_{i=1}^3(w_ia_i)^{w_i+\delta_{1,i}}\ .
\end{equation}
 The normalization \eqref{nor N'} is just the one that we have already announced in \eqref{N'}, and it uniquely determines the normalization that satisfies the string two point function \eqref{twopoint N'}. The equation \eqref{iden} is also an identity for  covering maps. In fact,  
with the help of \eqref{symmetricfactor} and \eqref{aiP},  it is easy to see that it coincides with the identities \eqref{idnetity}. As in the case of O-O-O, the above analysis for the matching is valid for the general case $\mathrm{I}$ in \eqref{thecondition}, where  the relevant covering map exists. For the edge case $\mathrm{II}$ in \eqref{thecondition} (where the CFT result vanish), one can see that the string result in \eqref{LHS2} also vanishes since one of the two functions $P_{w_1,w_2-1,w_3}$ and $P_{w_1,w_2,w_3-1}$ will vanish in the edge cases.

In conclusion, the lessons one learns from matching the leading ordering correlators is that the fermionic part and picture changing  are essential for the dual CFT to be a (deformed) symmetric orbifold CFT with the correct central charge $6k$\footnote{The issue of how the central charge of the dual CFT should be the correct one $6k$ instead of $6(k+2)$ was also recently studied in \cite{Sriprachyakul:2024gyl} using the near-boundary approximation (and for some simple cases of $w_i$). Here we directly calculate the three point superstring correlator with general $w_i$ to fix this  issue.}. Besides, comparing the matching in the case O-O-O and O-E-E also gives a cross-check of the difference between the 2 normalization factors  $N(w,j)$ and $N'(w,j)$. Notice that this cross-check crucially depends on the mass-shell condition \eqref{massshellR}. This means, even though we only consider the 3-point correlators (where no moduli integral is needed to do),  the matching of the two sides  makes it clear that the bulk side is a bona fide string theory (instead of being simply the worldsheet CFT), since we really need to count the picture changing and use the mass-shell condition\footnote{Note that the mass-shell condition seems not crucial in the matching of correlators  in the bosonic case \cite{Eberhardt:2021vsx}. However, we believe it could be crucial when considering correlators of more general operators, e.g. some descendants. }. We believe they will also be important for the matching of the two sides at higher orders \cite{ZFYCP}.

\section{Discussion}\label{discussion}
In this work, we calculate the superstring correlators of long strings on  AdS$_3\times$S$^3\times$T$^4$. Firstly, we  construct the relevant physical vertex operators. To avoid complexity from extra worldsheet excitations, we choose the physical operators to be the ones  that represent (a continuum of) long strings with the lowest space-time weights for a given $w$, in both the NS  and R sectors. Because of the GSO projection, the construction depends on the parity of $w$ so we discuss the cases with parity even and odd separately.   The final result for the space-time theory is: for $w$ odd,  there is a unique ground state comes from the NS sector and 8 excited states come from the R sector; for $w$ odd,  there are in total 16 ground states, with 8 come from the NS sector and the other 8 come from the R sector.

Then, we calculate correlators of these physical operators. Since a closed formula for the three point functions in the bosonic $SL(2,R)$ WZW model is derived in \cite{Dei:2021xgh,Bufalini:2022toj}, we only need to calculate the fermionic correlators (together with the picture changing effects). Though they are simply correlators in the free fermion theory $\psi^a (a=\pm,3)$, the calculation could be very complicated if one use the free field technique, since the construction of spectrally flowed operators is not simple \cite{Giribet:2007wp}. A simpler and systemic method is to view the fermion theroy $\psi^a$ as a special $SL(2,R)$ WZW model, then the fermionic correlator can be obtained by the closed formula in \cite{Dei:2021xgh}. Since the formula depends on the total parity $\sum_i w_i$, we calculate 4 representatives  of correlators with different parities $w_i$. As a byproduct, we find the equivalence of different picture choices gives relations among correlators in the bosonic  $SL(2,R)$ WZW model, some of which are  related to the recursion relations found in \cite{Dei:2021xgh}. 

In the discussion of the dual CFT, which is a deformed symmetric orbifold CFT, we found the ground states of the $w$-twisted sector (and the lowest excited states when $w$ is odd) match precisely with the results obtained from the string side. For the correlators, we show that at the leading order in the conformal perturbation, the fermionic contributions, together with the picture changing effects, modify the  central charge on the boundary side to be the correct one, i.e. $c=6k$. This matching  is guaranteed by  interesting identities of covering maps with three ramified points. As a cross-check, we also find the  normalizations of 2 string vertex operators determined holographically from the CFT side agree with the results from the two point string correlators.   

There are several interesting questions and open problems for future studies. Firstly, since our leading order matching of the correlators does not involve the 
 marginal deformation operator, it is rewarding to test the proposed deformation in \cite{Eberhardt:2021vsx} by matching the correlators at higher orders.  One can do this either using  the exact 3 or 4 point functions of the bosonic $SL(2,R)$ WZW model as in  \cite{Eberhardt:2021vsx,Dei:2022pkr}, or using the near-boundary approximation as in \cite{Knighton:2023mhq,Knighton:2024qxd} (see also \cite{Hikida:2023jyc}), where the residues can be obtained for general $n$-point functions (see \cite{Sriprachyakul:2024gyl} for some related discussions). With the results of this work, there are various correlators one can choose to compare with the CFT side at higher orders. A natural one is to choose \eqref{OOO} for $\sum_i w_i$ odd and \eqref{EOO} for $\sum_i w_i$ even, both of which in fact only depend on the $\mathcal{N}=1$ supersymmetric AdS$_3$ part. It is very likely that the matching at  higher orders are also related to some mathematical identities of covering maps. This is currently under investigation \cite{ZFYCP}.

Another interesting but more difficult problem is to test this duality for  higher genus correlators. For this, one can try to firstly calculate the string correlators at higher genus. However  solving the higher genus correlators of the worldsheet  bosonic $SL(2,R)$ WZW model, let alone doing the  moduli space integral, is already a difficult task, which could be related to covering maps from a higher genus surface to a sphere. Nevertheless, regardless of the holographic matching, this calculation for string correlators itself is meaningful and worth pursuing. Returning to the problem of matching the higher genus correlators of the two sides,  perhaps an easier way is to employ the near-boundary approximation, where one bypasses both the problems of solving the worldsheet CFT as well as doing the  moduli space integral \cite{Knighton:2023mhq,Hikida:2023jyc,Knighton:2024qxd}. This is especially hopeful given the fact that in the tensionless limit the localization of moduli space integral holds  also for higher genus correlators \cite{Eberhardt:2020akk,Knighton:2020kuh}.
 
Besides, one can try to generalize the calculation here to the cases of other supersymmetric AdS$_3$ string background, such as AdS$_3\times$S$^3\times$K$3$ or AdS$_3\times$S$^3\times$S$^3\times$S$^1$. Various properties of the related CFT with the appropriate chiral algebra have been studied previously~\cite{Baggio:2015jxa,Gaberdiel:2014yla}. It will also be interesting to    explore the full consequences of  the equivalence of the superstring correlators with different picture choices. A more ambitious goal is to find a non-perturbative definition of the dual CFT, or test the proposed duality beyond the perturbative analysis. Furthermore, it is argued in \cite{Eberhardt:2021vsx} that the dual theory could be a grand canonical ensemble of CFTs (rather than a theory with fixed N), where N is no longer an independent parameter of the theory. It will be very interesting to further explore in the supersymmetric setup  whether the dual theory is a CFT with fixed N or a grand canonical ensemble of CFTs.  Finally, in a wider context, an analogue of the tensionless limit is observed in high dimensional covariant disordered models~\cite{Peng:2018zap,Chang:2021fmd,Chang:2021wbx} where emergent higher spin symmetries are observed. It is thus interesting to relate the long string correlators to correlators in those disordered models.

\section*{Acknowledgments}
    We are grateful to Bin Chen and Matthias Gaberdiel for valuable discussions. We also thank Lorenz Eberhardt, Matthias Gaberdiel and Vit Sriprachyakul for comments on a draft of this paper. CP is supported by NSFC NO.~12175237, and NO.~12247103, the Fundamental Research Funds for the Central Universities, and funds from the Chinese Academy of Sciences. Zhe-fei Yu is supported by the Postdoctoral Fellowship Program of CPSF under Grant Number GZC20241685, and partly supported by NSFC NO.~12175237 and  funds from the UCAS program of Special Research Assistant.

\appendix

\section{Three and two point functions in the $SL(2,R)$ WZW model}\label{3pointWZW}

In this section, we review the closed formula for the three point function of spectrally flowed operators in the   $SL(2,R)$ WZW model in \cite{Dei:2021xgh}, as well as the form of the two point function \cite{Maldacena:2001km,Dei:2021xgh}.  Making use of the local Ward identities, the three point function can be written as an integral of the correlators in the
``$y$-basis"
\begin{equation}\label{ybasis3pt}
\begin{aligned}
    &\langle V_{j_1,h_1,\Bar{h}_1}^{w_1}(0;0)V_{j_2,h_2,\Bar{h}_2}^{w_2}(1;1)V_{j_3,h_3,\Bar{h}_3}^{w_3}(\infty;\infty)\rangle=\int \prod_{i=1}^3 \frac{d^2y_i}{\pi}\prod_{i=1}^3 \left|y_i^{\frac{kw_i}{2}+j_i-h_i-1}\right|^2\\
   &\times\left\{ \begin{aligned}
       &D(j_1,j_2,j_3)\left|X_{\varnothing}^{j_1+j_2+j_3-k}\prod_{i<l}^3X_{il}^{j_1+j_2+j_3-2j_i-2j_l}\right|^2\,, \qquad\qquad\quad\quad\,~\sum_iw_i\in 2Z\\
      & \mathcal{N}(j_1)D\left(\frac{k}{2}-j_1,j_2,j_3\right)\left|X_{123}^{\frac{k}{2}-j_1-j_2-j_3}\prod_{i=1}^3X_i^{-\frac{k}{2}+j_1+j_2+j_3-2j_i}\right|^2, \quad\sum_iw_i\in 2Z+1
   \end{aligned}
   \right.
\end{aligned}
\end{equation}
where
\begin{itemize}
    \item 
    Both $(z_1,z_2,z_3)$ and $(x_1,x_2,x_3)$ are set to $(0,1,\infty)$ 
    and it is easy to get their expressions at generic $z_i$ and $x_i$. 
    \item $D(j_1,j_2,j_3)$ is the three-point function of three unflowed vertex operators \cite{Teschner:1997ft}
    \begin{equation}
        D(j_1,j_2,j_3)=-\frac{G_k(1-j_1-j_2-j_3)}{2\pi^2 \nu^{j_1+j_2+j_3-1}\upgamma\left(\frac{k-1}{k-2}\right)}\prod_{i=1}^3\frac{G_k(2j_i-j_1-j_2-j_3)}{G_k(1-2k_i)}\,,
    \end{equation}
    where $G_k(x)$ is the Barnes double Gamma function. The normalization factor $\mathcal{N}(j)$ is given by
    \begin{equation}
        \mathcal{N}(j)=\frac{\nu^{\frac{k}{2}-2j}}{\upgamma(\frac{2j-1}{k-2})}\ .
    \end{equation}
    where $\upgamma(x)=\Gamma(x)/\Gamma(1-x)$.
    \item For $I\subset \{1,2,3\}$, $X_I$ is defined as
         \begin{equation}
               X_I(y_1,y_2,y_3)=\sum_{i\in I, \epsilon_i=\pm 1}P_{w+\sum_{i\in I}\epsilon_i e_i}
                 \prod_{i\in I}y_i^{\frac{1-\epsilon_i}{2}}\,,
           \end{equation}
           where $P$ is defined by
           \begin{equation}
          P_{w}=\left\{
          \begin{aligned}
              &0, \qquad\qquad~~ \qquad{\text{for }  \sum_{j}w_j<2\text{max}_{i=1,2,3} w_i \quad \text{or} \quad \sum_{i}w_i\in 2Z+1}\\
             &S_{w}G(\frac{w_1+w_2+w_3}{2}+1)\prod_{i=1}^3\frac{G(\frac{w_1+w_2+w_3}{2}-w_i+1)}{G(w_i+1)}, \qquad \text{otherwise}\ .
          \end{aligned}\right.
           \end{equation}
        In the above, $G(n)$ is the Barnes G function 
        \begin{equation}
            G(n)=\prod_{i=1}^{n-1}\Gamma(i)\,,
        \end{equation}
        and the  function $S_{w}$ is a phase depending on $w$ mod 2
      \begin{equation}
          S_w=(-1)^{\frac{1}{2}x(x+1)},\qquad x=\frac{1}{2}\sum_{i=1}^3(-1)^{w_iw_{i+1}}w_i\ .
        \end{equation}
    \end{itemize}
    We choose to normalize the vertex operators in the bosonic $SL(2,R)$ WZW model as in \cite{Eberhardt:2021vsx}, thus the two point function is:
    \begin{equation}\label{TwopointWZW}
        \langle V_{j_1,h_1}^{w_1}(0;0)V_{j_2,h_2}^{w_2}(\infty;\infty)\rangle=4i\delta^{(2)}(h_1-h_2)\delta_{w_1,w_2}\left(R(j_1,h_1,\Bar{h}_1)\delta(j_1-j_2)+\delta(j_1+j_2-1)\right)
    \end{equation}
    where
    \begin{equation}\label{reflection coefficient}
        R(j,h,\Bar{h})=\frac{(k-2)\nu^{1-2j}\upgamma (h-\frac{kw}{2}+j)}{\upgamma(\frac{2j-1}{k-2})\upgamma(h-\frac{kw}{2}+1-j)\upgamma(2j)}
    \end{equation}
    is the reflection coefficient and $\delta^{(2)}(h)\equiv \delta(h+\Bar{h})\delta_{h,\Bar{h}}$.

\section{Another example of different picture choices}\label{pictureEOO}
In this section,  we give a further example to demonstrate the relation between the picture choices and recursion relations in \cite{Eberhardt:2019ywk}. Since we have discussed the correlator $\mathcal{M}_{EOO}$ in section \ref{picturechanging} (whose total parity $\sum_i w_i$ is odd), here we discuss the correlator $\mathcal{M}_{EOO}$  (whose total parity $\sum_i w_i$ is even). We will not do the calcualtion concretely  but only show how to relate the correlators with different picture choices by the recursion relations in \cite{Eberhardt:2019ywk}. In \eqref{EOO}, we choose the third operator in the picture 0.  At the end of the section \ref{picturechanging}, we also comment on the case where  the first operator is in the picture 0, which turns out to be complicated. Here, we  show how to relate the two correlators with the second and third operators in the  picture 0 respectively  by the recursion relations \eqref{case3}.

Since the second operator is the same as the third one, when it is in the picture 0, the resulting correlator will be   \eqref{EOO} with the exchange $2\leftrightarrow 3$.
Since the form of \eqref{EOO} is clearly not symmetric under this exchange, our strategy is to use the recursion relations \eqref{case3} to transform \eqref{EOO} into a form that is symmetric in the index 2 and 3.   We transform all the terms as follows.\\
For the terms proportional to $\alpha_-$:
\begin{equation}
  \langle+0+\rangle\to \langle+0+\rangle, \quad \langle+00\rangle\to \langle+00\rangle, \quad \langle+0-\rangle\xrightarrow{i=3} \langle+00\rangle+\langle200\rangle+\langle++0\rangle+\langle+0+\rangle
\end{equation}
For the terms proportional to $\alpha_3$:
\begin{equation}
  \langle 00+\rangle\to \langle 00+\rangle, \quad \langle000\rangle\to \langle000\rangle, \quad \langle00-\rangle\xrightarrow{i=3} \langle000\rangle+\langle+00\rangle+\langle0+0\rangle+ \langle00+\rangle
\end{equation}
For the terms proportional to $\alpha_+$:
\begin{equation}
\begin{aligned}
  \langle -0+\rangle\xrightarrow{i=1}& \langle 00+\rangle+\langle +0+\rangle+\langle 0++\rangle+\langle 002\rangle, \quad \langle-00\rangle\xrightarrow{i=1} \langle000\rangle+\langle+00\rangle+\langle0+0\rangle+\langle00+\rangle\\
  \langle-0-\rangle\xrightarrow{i=3}& \langle-00\rangle+\langle000\rangle+\langle-+0\rangle+ \langle-0+\rangle\xrightarrow{i=1}
  [\langle000\rangle+\langle+00\rangle+\langle0+0\rangle+\langle00+\rangle]+[\langle000\rangle]\\
  &+[\langle0+0\rangle+\langle++0\rangle+\langle020\rangle+\langle00+\rangle]+[\langle 00+\rangle+\langle +0+\rangle+\langle 0++\rangle+\langle 002\rangle]
\end{aligned}
\end{equation}
In the above,  ``$A\xrightarrow{i=a} B+C+...$'' means using the recursion relation \eqref{case3} for $i=a$, $A$ can be represented by a sum of $B, C, ...$    without specifying the coefficients. $\langle 200\rangle$ means $\langle V_{j_1,h_1+2}^{w_1}V_{j_2,h_2}^{w_2}V_{j_3,h_3}^{w_3}\rangle$, and $\langle 020\rangle, \langle 002\rangle$ are similarly defined. After these transformations, there are 4 terms containing a ``2'', thus should be replaced once more as:   
\begin{equation}
\begin{aligned}
    \langle200\rangle&\xrightarrow{i=1} \langle000\rangle+\langle+00\rangle+\langle++0\rangle+\langle+0+\rangle\\
   \langle020\rangle&\xrightarrow{i=2} \langle000\rangle+\langle0+0\rangle+\langle++0\rangle+\langle0++\rangle\\
    \langle002\rangle&\xrightarrow{i=3} \langle000\rangle+\langle00+\rangle+\langle+0+\rangle+\langle0++\rangle
\end{aligned}
\end{equation}
With all these replacements, one can write \eqref{EOO} as a linear combination of:
\begin{equation}\label{7terms}
    \langle000\rangle, \quad \langle+00\rangle,\quad \langle0+0\rangle, \quad \langle00+\rangle,\quad \langle++0\rangle, \quad \langle+0+\rangle, \quad \langle0++\rangle
\end{equation}
These 7 terms are linear independent with respect to the recursion relation \eqref{case3} and they transform into each other (or invariant) under the exchange $2\leftrightarrow 3$. Thus,  transforming \eqref{EOO} into a linear combination of these 7 terms is similar to transforming \eqref{OOO} into \eqref{OOO2} in the case of O-O-O. We have   checked the  coefficients of the 7 terms in \eqref{7terms} and find that the transformed correlator is indeed  invariant under the exchange  $2\leftrightarrow 3$.  While in  case of O-O-O we only need one recursion relation to make the transformation,  here we need  many recursion relations. So the equivalence of the two picture choices will give an equation which is a linear combination of   more than two  recursion relations.

\section{The proposed CFT dual}\label{TheproposedCFT}
In this section, we   review the dualities proposed in \cite{Eberhardt:2021vsx}. We describe both the perturbative CFT dual of the bosonic string theory on on AdS$_3\times X$ and a similar proposal for the superstring on AdS$_3\times$S$^3\times$T$^4$. 

\subsection*{The bosonic proposal}
Firstly, we briefly review the bosonic duality. The perturbative CFT dual of the bosonic string theory on AdS$_3\times X$ is proposed to be:
\begin{equation}
    \text{Sym}^N(R_Q\times X)
\end{equation}
deformed by a non-normalizable marginal operator
\begin{equation}
    \Phi(x)\equiv \sigma_{2,\alpha=-\frac{1}{2b}}(x)
\end{equation}
Let's explain the two sides more concretely. On the string side, AdS$_3$ is described by a $SL(2,R)$ WZW model at level $k$. $X$ is an arbitrary internal CFT of the compactification, with central charge   
\begin{equation}
    c_X=26-\frac{3k}{k-2}
\end{equation}
On the CFT side, $R_Q$ is a linear dilaton theory with background charge $Q$\footnote{See Appendix \ref{conventionforseed} for our conventions for the linear dilaton theroy with background charge Q. }, defined by
\begin{equation}
    Q\equiv b^{-1}-b=\frac{k-3}{\sqrt{k-2}}, \qquad b\equiv \frac{1}{\sqrt{k-2}}
\end{equation}
$X$ is the same CFT as in the string side. Then the central charge of the seed theory is:
\begin{equation}
    c=1+6Q^2+c_X=6k,
\end{equation}
as expected. The marginal operator  is in the twist-2  sector and has the following dressing
\begin{equation}\label{bosonicdress}
    \sigma_{2,\alpha=-\frac{1}{2b}}(x)=e^{\sqrt{2}\alpha\phi}\sigma_2=e^{-\sqrt{\frac{k-2}{2}}\phi}\sigma_2
\end{equation}
where $\sigma_2$ is the spin field generating the ground state of of the twist-2 sector. One can easily check that this operator is of dimension one so is indeed marginal. To match the spectrum of long strings with  vertex operators in the symmetric orbifold, there is also a map between the $sl(2,R)$ spin $j$ on the string side and the linear dilaton momenta $\alpha$:
\begin{equation}
    \alpha=\frac{j+\frac{k}{2}-2}{\sqrt{k-2}}
\end{equation}
Notice that the marginal deformation does not affect the spectrum of long strings, thus this matching of  spectrum holds no matter whether one deforms the theory or not \cite{Balthazar:2021xeh,Eberhardt:2021vsx}.  

This proposal is confirmed by matching the  three-point functions of the two sides (up to 4th order) in \cite{Eberhardt:2021vsx}. This matching is remarkable since the calculation of the two sides are quite different and both are complicated.

\subsection*{The supersymmetric proposal}
Now we move to the supersymmetric setting. The CFT dual of the superstring was also proposed in \cite{Eberhardt:2021vsx}. 

On the string side, we have the superstring theory on AdS$_3\times$S$^3\times$T$^4$. In the RNS formalism,  the worldsheet CFT is described by
\begin{equation}
    sl(2,R)^{(1)}_k\oplus  su(2)^{(1)}_k \oplus \left(U(1)^{(1)}\right)^4
\end{equation}
where $sl(2,R)^{(1)}_k$ and $su(2)^{(1)}_k$ represent $N=1$ supersymmetric WZW model with affine symmetry $sl(2,R)^{(1)}_k$  and $su(2)^{(1)}_k$ respectively. They describe the AdS$_3$ and S$^3$ factors.  $ \left(U(1)^{(1)}\right)^4$ represents the $\mathcal{N}=1$ supersymmetric version of T$^4$, describing the flat torus directions. 

The candidate CFT dual is again a deformed symmetric orbifold theory, similar to the bosonic case. The theory before deformation is the following symmetric orbifold theory:
\begin{equation}\label{undeformed}
    \text{Sym}^N\left(R_Q\times su(2)_{k-2}\times \text{four free fermions} \times \left(U(1)^{(1)}\right)^4\right)
\end{equation}
where $R_Q$ is the linear dilaton direction with background charge Q:
\begin{equation}\label{Qk}
    Q=b-b^{-1}=\frac{k-1}{\sqrt{k}}, \qquad b=\frac{1}{\sqrt{k}} 
\end{equation}
Then the central charge of the seed theory is:
\begin{equation}\label{centralcharge}
    c=1+6Q^2+\frac{3(k-2)}{k-2+2}+4\times\frac{1}{2}+6=6k,
\end{equation}
as expected.
The map from the $sl(2,R)$ spin $j$ on the worldsheet to the momenta $\alpha$ in the linear dilaton factor is
\begin{equation}\label{jalpha}
    \alpha=\frac{j+\frac{k}{2}-1}{\sqrt{k}}
\end{equation}
Notice that \eqref{Qk} and \eqref{jalpha}
are simply the corresponding ones in the bosonic proposal with the replacement $k\to k+2$.
The first three factors in the  seed theory of the symmetric orbifold \eqref{undeformed} should be thought of as an $\mathcal{N}=4$  linear dilaton theory. The spectrum of this undeformed theory was matched with the spectrum of long strings in the superstring theory on AdS$_3\times$S$^3\times$T$^4$ \cite{Eberhardt:2019qcl}.

As in the bosonic case, one needs to deform this symmetric orbifold theory by a (non-normalizable) marginal operator. In any $\mathcal{N}=4$ theory, marginal operators are obtained as descendants of BPS operators with $h=\Bar{h}=\frac{1}{2}$. These BPS operators can be obtained by dressing some ground states of the twist-2 sector with vertex operators in the linear dialton theory. 
 It was proposed in \cite{Eberhardt:2021vsx} that the  marginal operator should  lie in a singlet
 $(\textbf{1},\textbf{1})$ of $SU(2)_{R}\oplus SU(2)_{\text{outer}}$. Notice that this should hold for two $SU(2)_{\text{outer}}$s in the left and right moving parts respectively. 
 
Thus, one can write the deformation as:
\begin{equation}
    \Phi(x,\Bar{x})\equiv G^{\alpha A}_{-\frac{1}{2}}\Bar{G}^{\beta B}_{-\frac{1}{2}}\Psi_{\alpha\beta AB}(x,\Bar{x})=\epsilon_{\alpha\gamma}\epsilon_{\beta\delta}\epsilon_{AC}\epsilon_{BD}G^{\alpha A}_{-\frac{1}{2}}\Bar{G}^{\beta B}_{-\frac{1}{2}}\Psi^{\gamma\delta CD}(x,\Bar{x})
\end{equation}
where $G^{\alpha A}$ and $\Bar{G}^{\beta B}$ are holomorphic and anti-holomorphic supercurrents respectively. $\alpha,\beta, \gamma, \delta=\pm $ are the  spinor indices of the R-symmetry $SU(2)_{R}$, while $A,B,C,D=\pm $ are the  spinor indices of the outer automorphism group $SU(2)_{\text{outer}}$. $\Psi^{\alpha\beta,AB}$ are non-normalizable BPS operators in the twist-2 sector, obtained by dressing the ground states as (we only write the left moving part, thus only the indices $\alpha$ and  $A$ remain):
\begin{equation}\label{BPSoper}
    \Psi^{\alpha A}=e^{\sqrt{2}\alpha\phi}S^{\epsilon_1\epsilon_2\epsilon_3\epsilon_4}\Sigma_{2}=e^{-\sqrt{\frac{k}{2}}\phi}S^{\epsilon_1\epsilon_2\epsilon_3\epsilon_4}\Sigma_{2}
\end{equation}
 where $S^{\epsilon_1\epsilon_2\epsilon_3\epsilon_4}$ are the spin fields lie in $(\textbf{2},\textbf{2})$ in the Table \ref{spinfield}. The superscripts of the two sides are related as: $\alpha=\frac{1}{2}(\epsilon_1+\epsilon_2+\epsilon_3+\epsilon_4)$, $A=\frac{1}{2}(\epsilon_1-\epsilon_2)$.   Notice that the dressing in the linear dialton direction (the momenta $\alpha$ in \eqref{BPSoper}) is the same as in the bosonic case \eqref{bosonicdress}  (again with the shift $k\to k+2$):
 \begin{equation}
     \alpha=-\frac{1}{2b}=-\frac{\sqrt{k}}{2}
 \end{equation}
 Thus, this deformation is also non-normalizable. This operator creates an exponential wall and is hence  similar to the exponential operator in Liouville theory. Thus it is very different from the deformation corresponding to RR deformation on the string side, which is a singlet  with respect to the   global $so(4)=su(2)_R\oplus su(2)_B$ symmetry (where $su(2)_B$  is the residual torus symmetry that acts on the bosonic modes) \cite{Gaberdiel:2015uca,Fiset:2022erp,Gaberdiel:2023lco}.
 One can check that $\Psi_{\alpha\beta,AB}$ are indeed BPS:
 \begin{equation}
     h=\frac{c}{24}\left(2-\frac{1}{2}\right)+\frac{1}{2}\times \frac{1}{2}+\frac{\alpha(Q-\alpha)}{2}=\frac{3k}{8}+\frac{1}{4}-\frac{\sqrt{k}}{4}\left(\frac{k-1}{\sqrt{k}}+\frac{\sqrt{k}}{2}\right)=\frac{1}{2}=|q|
 \end{equation}

\section{Correlators of symmetric orbifold CFTs}\label{coveringmapmethod}
For the correlators (in the large N limit), there is a algorithm that can reduce the calculation to the one in the seed theory. This is a method making use of the covering map, developed by Lunin and Mathur \cite{Lunin:2000yv,Lunin:2001pw}.  Firstly, note that the operators in the twist-$n$ sector discussed above are not invariant under the action of $S_n$. The real gauge invariant  operators in the twist-$n$ sector can be obtained by summing over elements in the conjugacy class of the permutation $(1,2,...,n)$ as follows:
\begin{equation}\label{Fulltwist}
    \mathcal{O}_n(x)=\frac{\sqrt{(N-n)!n}}{\sqrt{N!}}\sum_{\tau\in [(1,2,...,n)]}O_\tau(x)
\end{equation}
Notice that the prefactor comes form the standard normalization. So the correlator we concerned are of these gauge invariant  operators, which can be written as \cite{Lunin:2000yv,Lunin:2001pw,Dei:2019iym}:
\begin{equation}
    \left\langle \prod_{j=1}^m \mathcal{O}_{n_j}(x_j)\right\rangle=\left(\begin{matrix}
        N\\
        d
    \end{matrix} \right)\left(\prod_{j=1}^m\frac{\sqrt{(N-n_j)!n_j}}{\sqrt{N!}}\right)\sum_{\text{covering map $\Gamma$}}f(\Gamma)\left\langle \prod_{j=1}^m O_{\tau_j}(z_j)\right\rangle\Bigg|_{\Gamma(z_i)=x_i}.
\end{equation}
where
\begin{itemize}
    \item $d$ is the number of elements that $(\tau_1,\tau_2,...,\tau_m)$ truly act on.
    \item  The summation is over all covering map $\Gamma$ with ramification indices $n_j$ at the respective insertion points $x_i$, that is, around  $z_i$  ($z$ is the coordinate of the covering surface) we have:
\begin{equation}\label{covering}
    \Gamma(z)=x_i+a_i(z-z_i)^{n_i}+...
\end{equation}
   \item $f(\Gamma)$ is a factor determined by the covering map $\Gamma$. This is in fact a  Weyl factor that accounts for the non-trivial (induced) metric on the covering space. If the covering surface has genus 0 (we will focus on this simplest case), it can explicitly be
   computed as:
   \begin{equation}
       f(\Gamma)=\left|\prod_{i=1}^m w_i^{-\frac{c(w_i+1)}{24}}a_i^{\frac{c(w_i-1)}{24}-h_i}\Pi^{-\frac{c}{12}}\right|^2
   \end{equation}
   where $a_i$ is the coefficient determined in \eqref{covering} and $\Pi$ is the product of the residues of the covering map:
   \begin{equation}\label{defPi}
       \Pi=\prod_a\Pi_a, \qquad \Gamma(z)\sim \frac{\Pi_a}{z-z_a}+O(1) 
   \end{equation}
   \item  $\left\langle \prod_{j=1}^m O_{\tau_j}(z_j)\right\rangle$ is the correlator of gauge dependent operators, lifted up to the covering surface. 
\end{itemize}
The covering surfaces in the summation can have higher genus (and even be disconnected) and
its genus $g$ can be determined by the Riemann-Hurwitz formula:
\begin{equation}\label{RH}
    g\equiv 1-n+\frac{1}{2}\sum_{j=1}^m(n_j-1)
\end{equation}
Then in the large N limit, the power of N is determined as:
\begin{equation}
    \left(\begin{matrix}
        N\\
        d
    \end{matrix} \right)\left(\prod_{j=1}^m\frac{\sqrt{(N-n_j)!n_j}}{\sqrt{N!}}\right)\sim
    N^{1-g-\frac{m}{2}}
\end{equation}
Thus the normalization factor results in a large N expansion controlled by the genus of the covering surface.

\section{Conventions for the seed theory}\label{conventionforseed}
In this section, we set our conventions for the seed theory:
\begin{equation}
    R_Q\times su(2)_{k-2}\times \text{four free fermions} \times \left(U(1)^{(1)}\right)^4
\end{equation}
which is a product of an $\mathcal{N}=4$ linear dilaton theory and  T$^4$.

\subsection*{Bosonic linear dilaton} For a bosonic   linear dilaton $\phi$ with background charge Q, the defining OPE of the $U(1)$ current $i\partial\phi$ is:
\begin{equation}
    i\partial\phi(z)i\partial\phi(w)\sim \frac{1}{(z-w)^2} 
\end{equation}
With background charge $Q$, the stress-energy tensor is modified to be:
\begin{equation}
    T(z)=-\frac{1}{2}:\partial\phi\partial\phi:(z)+\frac{1}{\sqrt{2}}Q\partial^2\phi(z)
\end{equation}
as a consequence, the central charge is also modified:
\begin{equation}
    c=1+6Q^2
\end{equation}
A vertex operator $e^{\sqrt{2}\alpha\phi}$ has conformal weight:
\begin{equation}
    h(e^{\sqrt{2}\alpha\phi})=\alpha(Q-\alpha)
\end{equation}

\subsection*{$\mathcal{N}=4$ linear dilaton} The OPEs among the generating fields: $\partial\phi,   J^a , \psi^{\alpha\beta}$ of a $\mathcal{N}=4$  linear dilaton theory is:
\begin{equation}
\begin{aligned}
       i\partial\phi(z)i\partial\phi(w)&\sim \frac{1}{(z-w)^2}, \\
       \psi^{\alpha\beta}(z)\psi^{\gamma\delta}(w)&\sim \frac{\epsilon^{\alpha\gamma}\epsilon^{\beta\delta}}{z-w},\\
       J^3(z)J^3(w)&\sim \frac{k-2}{2(z-w)^2},\\
       J^3(z)J^\pm(w)&\sim \frac{J^\pm(w)}{z-w},\\
       J^+(z)J^-(w)&\sim \frac{k-2}{(z-w)^2}+\frac{2J^3(w)}{z-w}.
\end{aligned}
\end{equation}
The generators of the small $\mathcal{N}=4$ superconformal algebra are:
\begin{equation}\label{generatorLD}
\begin{aligned}
    T&=-\frac{1}{2}\partial\phi\partial\phi+\frac{k-1}{\sqrt{2k}}\partial^2\phi+\frac{1}{k}\left( J^3J^3+\frac{1}{2}(J^+J^-+J^-J^+)\right)+\frac{1}{2}\epsilon_{\alpha\gamma}\epsilon_{\beta\delta}\partial\psi^{\alpha\beta}\psi^{\gamma\delta}  \\
    G^{\alpha\beta}&=\frac{i}{\sqrt{2}}(\partial\phi\psi^{\alpha\beta})+\frac{i}{\sqrt{k}}\left( (\sigma_a)^\alpha_\gamma\left(J^a+\frac{1}{3}J^{(f,+)a}\right)\psi^{\gamma\beta}-(\sigma_a)^\beta_\gamma J^{(f,-)a}\psi^{\alpha\gamma}-(k-1)\partial\psi^{\alpha\beta}\right)\\
    K^a&=J^a+J^{(f,+)a}
\end{aligned}
\end{equation}
with the fermionic currents defined as:
\begin{equation}
    J^{(f,+)a}=\frac{1}{4}(\sigma^a)_{\alpha\gamma}\epsilon_{\beta\delta}(\psi^{\alpha\beta}\psi^{\gamma\delta}), \qquad J^{(f,-)a}=\frac{1}{4}\epsilon_{\alpha\gamma}(\sigma^a)_{\beta\delta}(\psi^{\alpha\beta}\psi^{\gamma\delta})
\end{equation}
Then the zero modes of $J^{(f,-)a}$ generate the algebra $SU(2)_{\text{outer}}$.

\subsection*{The torus theory}
For the torus theory, the OPEs among its generators are:
\begin{equation}
\begin{aligned}
    X^a(z)X^{b\dagger}(w)&\sim \delta^{ab}\text{log}(z-w),\\
    \lambda^a(z)\lambda^{b\dagger}(w)&\sim \frac{\delta^{ab}}{z-w}, \qquad a,b=1,2.
\end{aligned}
\end{equation}
This theory has a small $\mathcal{N}=4$ superconformal symmetry with $c=6$, whose generators are $T, G^a, G^{a\dagger}, J^i$ $(a=1,2, i=1,2,3)$ \cite{David:2002wn}:
\begin{equation}\label{generatorT4}
\begin{aligned}
     T&=\sum_{i=1,2}\partial X^{i}\partial X^{i\dagger}+\frac{1}{2}\sum_{a=1,2}(\lambda^a\partial \lambda^{a\dagger}-\partial \lambda^a\lambda^{a\dagger})\\
     G^a&=\left[
     \begin{aligned}
         G^1\\
        G^2
     \end{aligned}
     \right]=\sqrt{2}\left[
     \begin{aligned}
         \lambda^1\\
         \lambda^{2}
     \end{aligned}
     \right]\partial X^{2}+
     \sqrt{2}\left[
     \begin{aligned}
         -\lambda^{2\dagger}\\
         \lambda^{1\dagger}
     \end{aligned}
     \right]\partial X^{1}\\
     J^i&=\frac{1}{2}\left[\lambda_1,\lambda_2\right]\sigma^i\left[
     \begin{aligned}
         \lambda^{1\dagger}\\
         \lambda^{2\dagger}
     \end{aligned}
     \right].
\end{aligned}
\end{equation}
The 4 supercurrents are $G^a, G^{a\dagger}$ $(a=1,2)$.
The small $\mathcal{N}=4$ generators of the full seed theory will be the sum of the corresponding ones in \eqref{generatorLD} and \eqref{generatorT4} (an appropriate scaling is also needed to have a standard normalization of the algebra.).

\bibliographystyle{JHEP}
\bibliography{refs}

\end{document}